\documentclass[preprint,prd,preprintnumbers,amssymb,superscriptaddress,nofootinbib,floatfix]{revtex4}
\pdfoutput=1
\usepackage[english]{babel}
\usepackage{slashed}
\usepackage{amsmath}
\usepackage[]{units}
\usepackage{multirow}
\usepackage{color}
\usepackage{enumitem}
\usepackage{hyperref}
\usepackage[all]{hypcap}
\usepackage{array}
\usepackage{float}
\usepackage{graphicx}

\newcommand{\DMp}{{\Phi}}
\newcommand{\DMm}{{\Phi^*}}
\newcommand{\DM}{\text{DM}}
\newcommand{\med}{\phi}
\newcommand{\medp}{{\phi}}
\newcommand{\medm}{{\phi^*}}
\newcommand{\medz}{{\varphi}}
\newcommand{\medpVcov}[1]{\medp^*_{#1}}

\newcommand{\mDMp}{m_\DMp}
\newcommand{\mDM}{m_\DM}
\newcommand{\mmed}{m_\med}

\newcommand{\rDMp}{r_\DMp}
\newcommand{\rmed}{r_\med}

\newcommand{\cond}{\hyperref[condition:N]{(i)-(v)}}

\newcommand{\dguno}[1]{
\raisebox{-.5\height}{\includegraphics[height=#1cm]{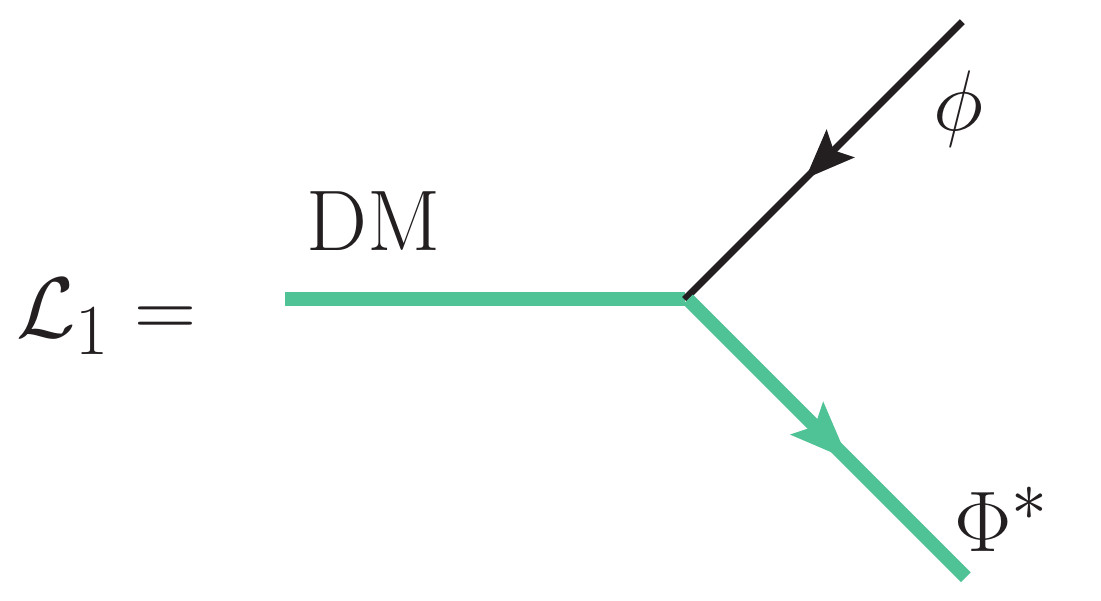}}
}

\newcommand{\dgunop}[1]{
\raisebox{-.5\height}{\includegraphics[height=#1cm]{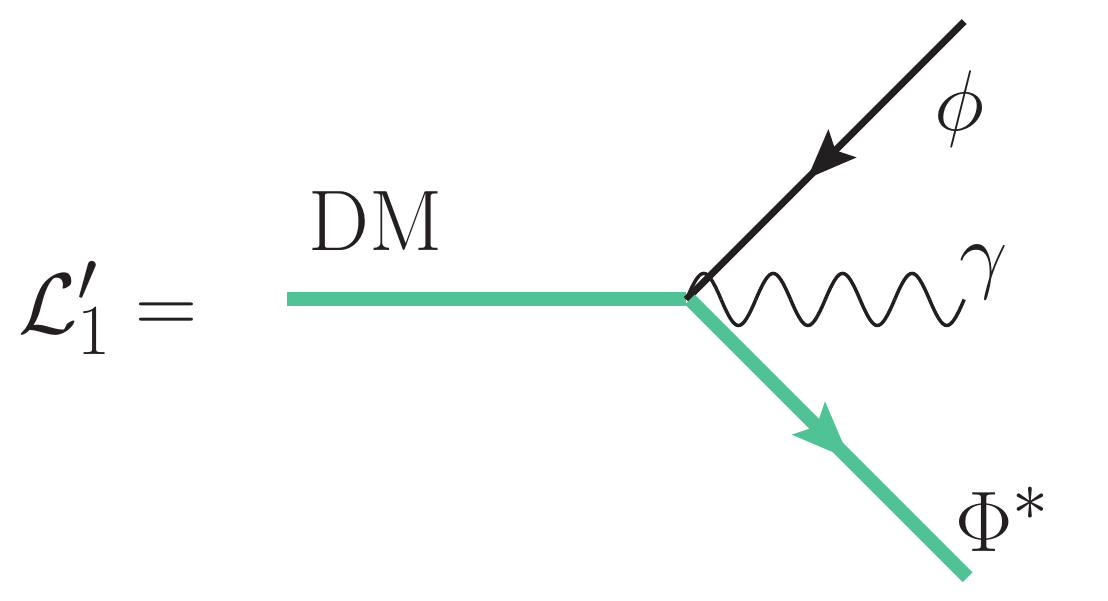}}
}

\newcommand{\dgdos}[1]{
\raisebox{-.5\height}{\includegraphics[height=#1cm]{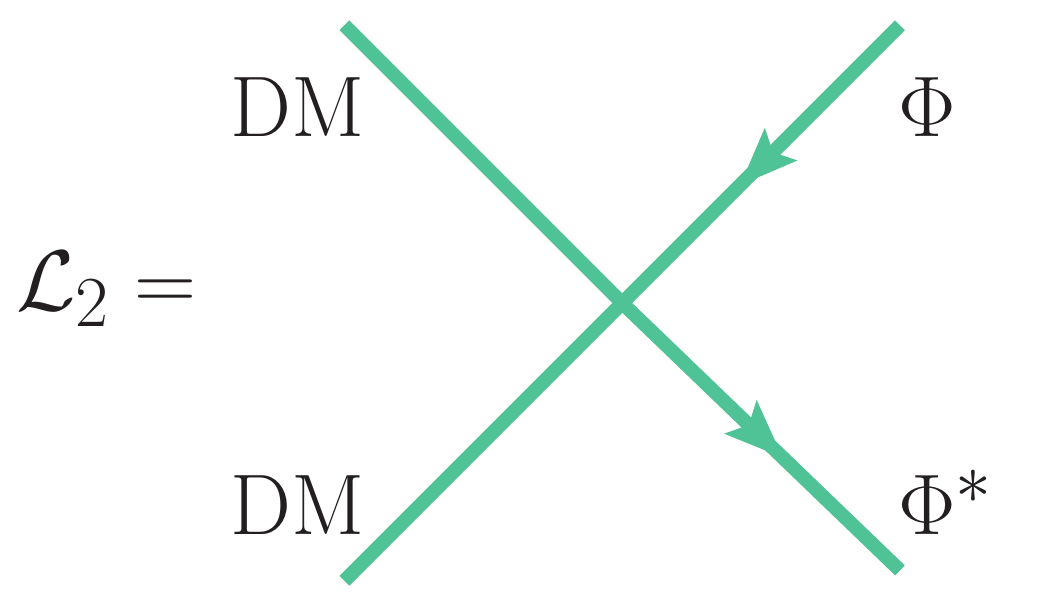}}
}
\newcommand{\dgtres}[1]{
\raisebox{-.5\height}{\includegraphics[height=#1cm]{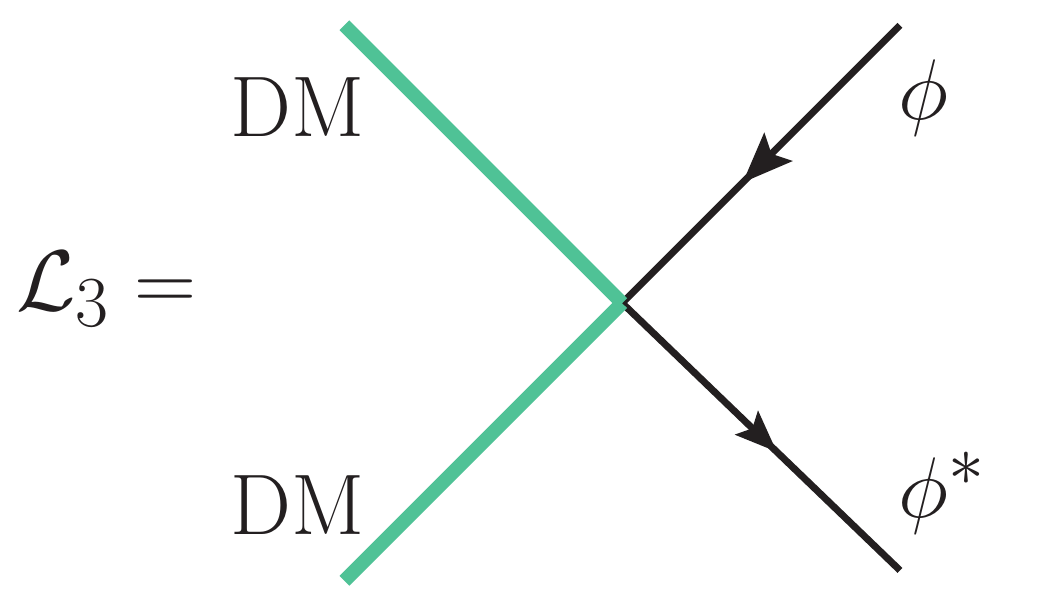}}
}
\newcommand{\dgcuatro}[1]{
\raisebox{-.5\height}{\includegraphics[height=#1cm]{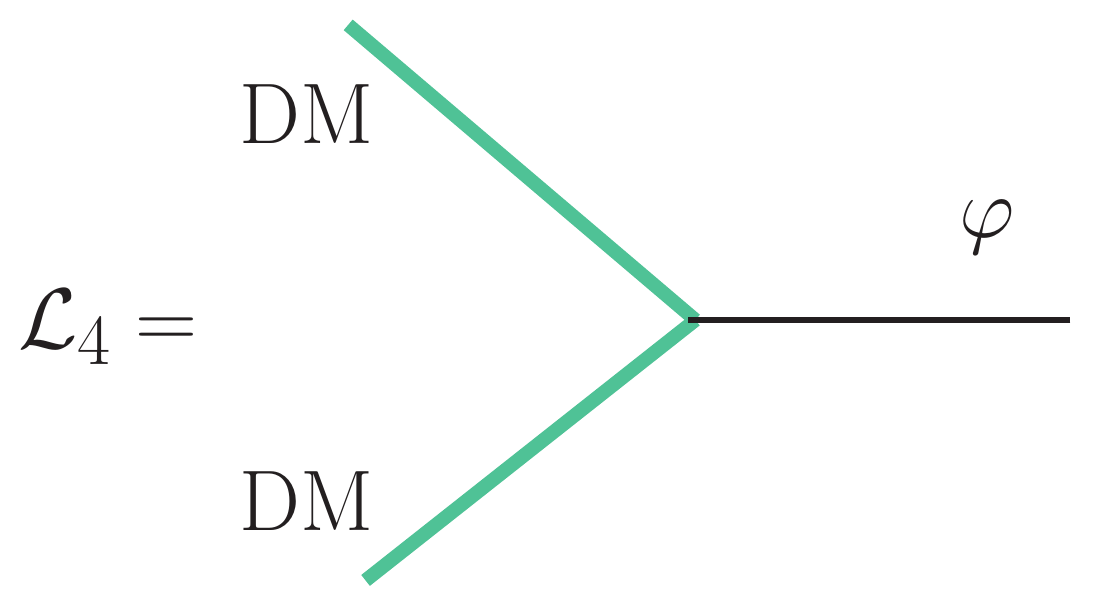}}
}
\newcommand{\dgcinco}[1]{
\raisebox{-.5\height}{\includegraphics[height=#1cm]{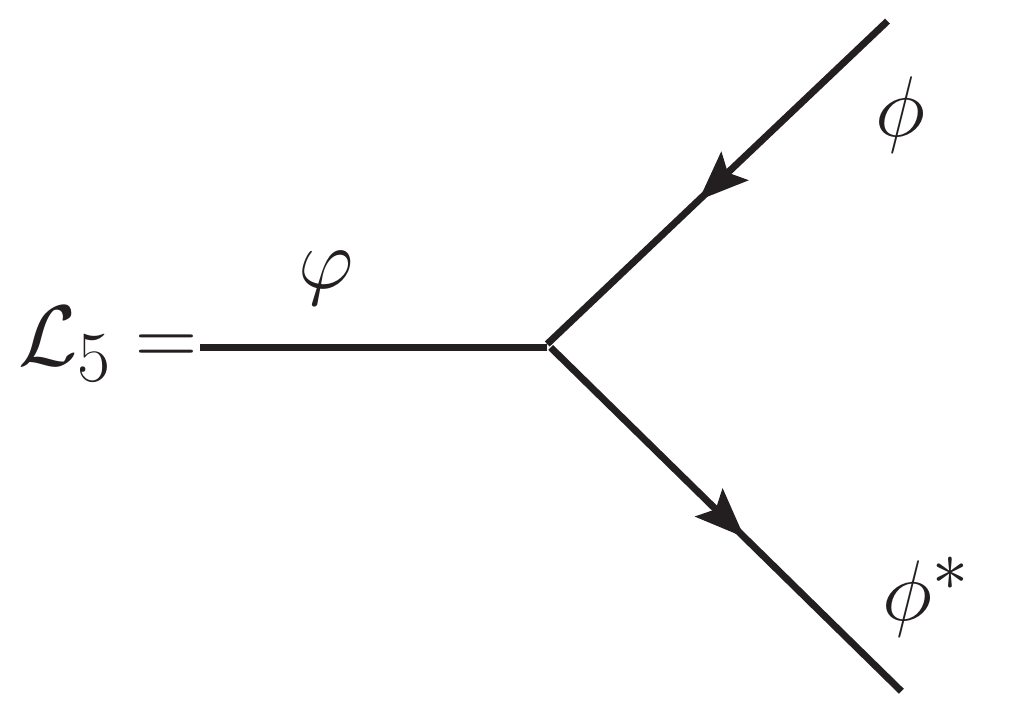}}
}

\newcommand{\dgcincop}[1]{
\raisebox{-.5\height}{\includegraphics[height=#1cm]{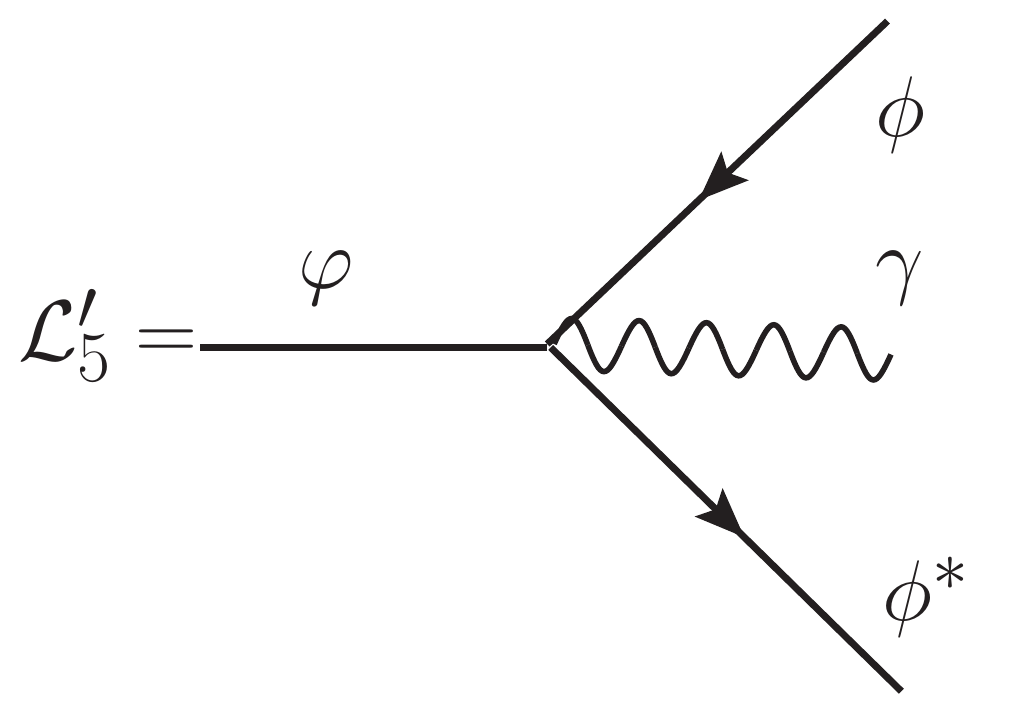}}
}

\begin{document}

\preprint{ULB-TH/16-20}

\author{Camilo Garcia-Cely}
\email{Camilo.Alfredo.Garcia.Cely@ulb.ac.be}
\affiliation{Service de Physique Th\'eorique, Universit\'e Libre de Bruxelles, Boulevard du Triomphe, CP225, 1050 Brussels, Belgium}

\author{Andres \surname{Rivera}}
\email{afelipe.rivera@udea.edu.co}
\affiliation{Instituto de F\'isica, Universidad de Antioquia,
Calle~70~No.~52-21, Medell\'in, Colombia}
\affiliation{Service de Physique Th\'eorique, Universit\'e Libre de Bruxelles, Boulevard du Triomphe, CP225, 1050 Brussels, Belgium}

\title{
General calculation of the cross section for dark matter annihilations into two photons}

\begin{abstract}
Assuming that the underlying  model satisfies some general requirements such as renormalizability and CP conservation, we calculate the non-relativistic one-loop cross section for any self-conjugate dark matter particle annihilating into two photons.  We accomplish this  by carefully classifying all possible one-loop diagrams and, from them,  reading off the  dark matter interactions with the particles running in the loop. Our approach is general and leads to the same results found in the literature for popular dark matter candidates such as the neutralinos of the MSSM, minimal dark matter, inert Higgs and Kaluza-Klein dark matter. 
\end{abstract}

\maketitle
\tableofcontents

\section{Introduction}

Little is known about the nature of Dark Matter (DM), even though its existence has been firmly established by multiple astrophysical
and cosmological observations.
We know its abundance ( $\Omega_\text{DM} h^2 = 0.1197\pm 0.0022$~\cite{Ade:2015xua}), the fact that interacts very weakly
with normal matter and that was cold during the time when  the first structures formed in the early universe.
These properties naturally arise  in scenarios where DM is a Weakly Interacting Massive Particle (WIMP) and make the latter very compelling DM models (For reviews see Refs.~\cite{Bergstrom:2000pn,Bertone:2004pz}).  One of the chief predictions of such models is the possibility that  DM can annihilate into SM particles. 
Among these, gamma rays are particularly important because, in contrast to charged particles, they are not deflected when they propagate through astrophysical environments and thus point towards the region where they were produced. 

Even more important are gamma-ray lines: since no astrophysical  process is known to produce them, the observation of one of them would strongly suggest the existence of WIMP DM, specially if they come from a region where the concentration of DM is known to be high (For a review, see e.g. Ref~\cite{Bringmann:2012ez}). In fact,  the non-observation of statistically significant gamma-ray lines\footnote{The DM interpretation of the 130\,GeV line found in the Fermi-LAT data coming from the Galactic Center~\cite{Bringmann:2012vr, Weniger:2012tx,Su:2012ft}  has been disfavored because no evidence of the line was found coming from  DM-dominated objects like dwarf galaxies~\cite{GeringerSameth:2012sr} or coming from the Galactic Center by  other gamma-ray telescopes~\cite{Abdalla:2016olq}.  Also, because the line was  hinted in places where it could not be due to DM annihilation such as   the Earth's Limb~\cite{Finkbeiner:2012ez, Hektor:2012ev} and the vicinity of the Sun~\cite{Whiteson:2013cs}. In fact, the origin of the line was very likely due to instrumental reasons as implied by the fact that a later analysis of the telescope data showed no evidence of the line~\cite{Ackermann:2015lka}.
} by  Fermi-LAT~\cite{Ackermann:2013uma,Ackermann:2015lka, Liang:2016bxu, Anderson:2015dpc,Liang:2016pvm,Profumo:2016idl} or H.E.S.S.~\cite{Abramowski:2013ax,Abdalla:2016olq} allows to set stringent limits on the DM annihilation cross section into monochromatic photons. Similar limits have been derived using the CMB anisotropies measured by the Planck satellite~\cite{Slatyer:2015jla} (and references therein). 

Consequently, in a given DM model, it is very important to calculate the cross sections of processes leading to gamma-ray lines. Nevertheless, in contrast to other annihilation channels, this task is not straightforward. Such processes only arise at one-loop level because, typically, DM does not couple to photons. Moreover, the number of Feynman diagrams increases dramatically with the number of charged particles that couple to DM and run in the loop, which leads to  annihilation cross sections that are highly dependent on the DM model. 

In view of this situation, most of the studies that calculate the annihilation cross section into gamma-ray lines have been based  on specific DM models~\cite{Jackson:2009kg, Giacchino:2014moa,Bergstrom:2004nr,Ibarra:2014vya,Bertone:2009cb, Birkedal:2006fz,Bertone:2010fn, Arina:2014fna, Cerdeno:2015jca, Weiner:2012cb, Tulin:2012uq,Choi:2012ap,Chalons:2011ia,Chalons:2012xf,Choi:2012ap}. Early studies focused on certain supersymmetric DM candidates~\cite{Bergstrom:1988fp, Rudaz:1989ij,Giudice:1989kc,Bergstrom:1989jr,Bergstrom:1988jt,Bergstrom:1994mg,Baek:2012ub}, culminating with the works of Refs.~\cite{Bergstrom:1997fh, Bern:1997ng,Ullio:1997ke}, which reported the full one-loop calculation for any neutralino of the MSSM annihilating into one or two photons.  Another common approach to gamma-ray lines  is based  on  effective theories where the annihilation into photons arises from high-dimensional operators in such way that microscopic details of the model are integrated out~\cite{Goodman:2010qn,Abazajian:2011tk,Rajaraman:2012db,Dudas:2014ixa,Aisati:2014nda,Coogan:2015xla,Duerr:2015vna,Arina:2009uq,Chu:2012qy,Wang:2012ts, Bai:2012qy, Fichet:2016clq, Rajaraman:2012fu}. 

In this article, we show that in spite of the complexity of the problem, the cross section for DM annihilations into two photons can be calculated in a general way for \emph{any} DM model meeting a basic set of requirements. Similar attempts in this direction were done in Refs.~\cite{Jackson:2013pjq,Duerr:2015wfa} for DM candidates with s-channel mediators, and more generally in Ref.~\cite{Asano:2012zv} by means of the optical theorem when the DM particles are heavier than the particles running in the loop.

This paper is organized as follows. We start  Sec.~\ref{sec:II} by listing the properties that we assume for the DM, which allows us to determine  the corresponding Feynman diagrams. From them, we read off the interactions between DM and the particles in the loop. In Sec.~\ref{sec:calculation}, we then calculate the annihilation amplitudes and the corresponding cross sections. In Sec.~\ref{sec:discussion}, we summarize our findings and illustrate them with examples from popular DM models. In Sec.~\ref{sec:conclusions}, we present our conclusions. Appendices~\ref{sec:Gauge} discusses  the gauge choice of the vector particles in our work. Appendix~\ref{sec:reduction} gives technical details concerning and the Passarino--Veltman functions for box diagrams in the non-relativistic limit. Finally,  in Appendix~\ref{sec:xs}, we report some formulas needed to compute annihilation amplitudes.

\section{General properties of the annihilation process DM DM $\to \gamma \gamma$}
\label{sec:II}

\subsection{Classification of the diagrams}
\label{sec:cond}

In order to systematically study DM annihilations into two photons, we will assume that the following conditions are satisfied: 

\begin{enumerate}
\item[\label{condition:N}(i)] DM is its own antiparticle and its stability is guaranteed by a $Z_2$ symmetry. This implies  that DM is electrically neutral and that it can not emit photons. 
\item[ \label{condition:R}(ii)]  The underlying DM theory is renormalizable. Consequently, any additional neutral particle, including the DM, do not couple to two photons at tree level. 
\item[\label{condition:C}(iii)] In a cubic vertex, photons couple to particles belonging to the same field. For fermions and scalars, this condition follows from electromagnetic gauge invariance. However, for charged gauge bosons, that is not the case because photons could couple to Goldstone and gauge bosons in the same cubic vertex. As discussed below, by choosing an appropriated gauge, we can nonetheless get rid of such vertices and therefore fulfill this condition.  
\item[ \label{condition:S}(iv)] Particles have spin zero, one-half or one. 
\item[ \label{condition:CP}(v)]  CP is conserved. 
\end{enumerate}

This set of conditions allows us to classify all diagrams leading to DM annihilation into photons. Eventually, from this classification, we will write down the interaction Lagrangians that give rise to DM DM $\to \gamma\gamma$ and calculate the amplitude.

\newsavebox{\Top}
\newsavebox{\Lagrangian}
\newsavebox{\schLagrangian}
\newsavebox{\Aform}
\newsavebox{\Lagrangianone}
\newsavebox{\Lagrangiantwothree}

\newcolumntype{C}[1]{>{\centering\arraybackslash}p{#1}}



%
%
\begin{lrbox}{\Top}
\centering
\begin{tabular}{|C{2cm}|C{7.6cm}|C{6.7cm}|}\hline
{\bf Topology} & {\bf Diagrams}
 & {\bf Interactions}\\\hline\hline
\multirow{2}{*}{\raisebox{-0.5\height}{\includegraphics[height=2cm]{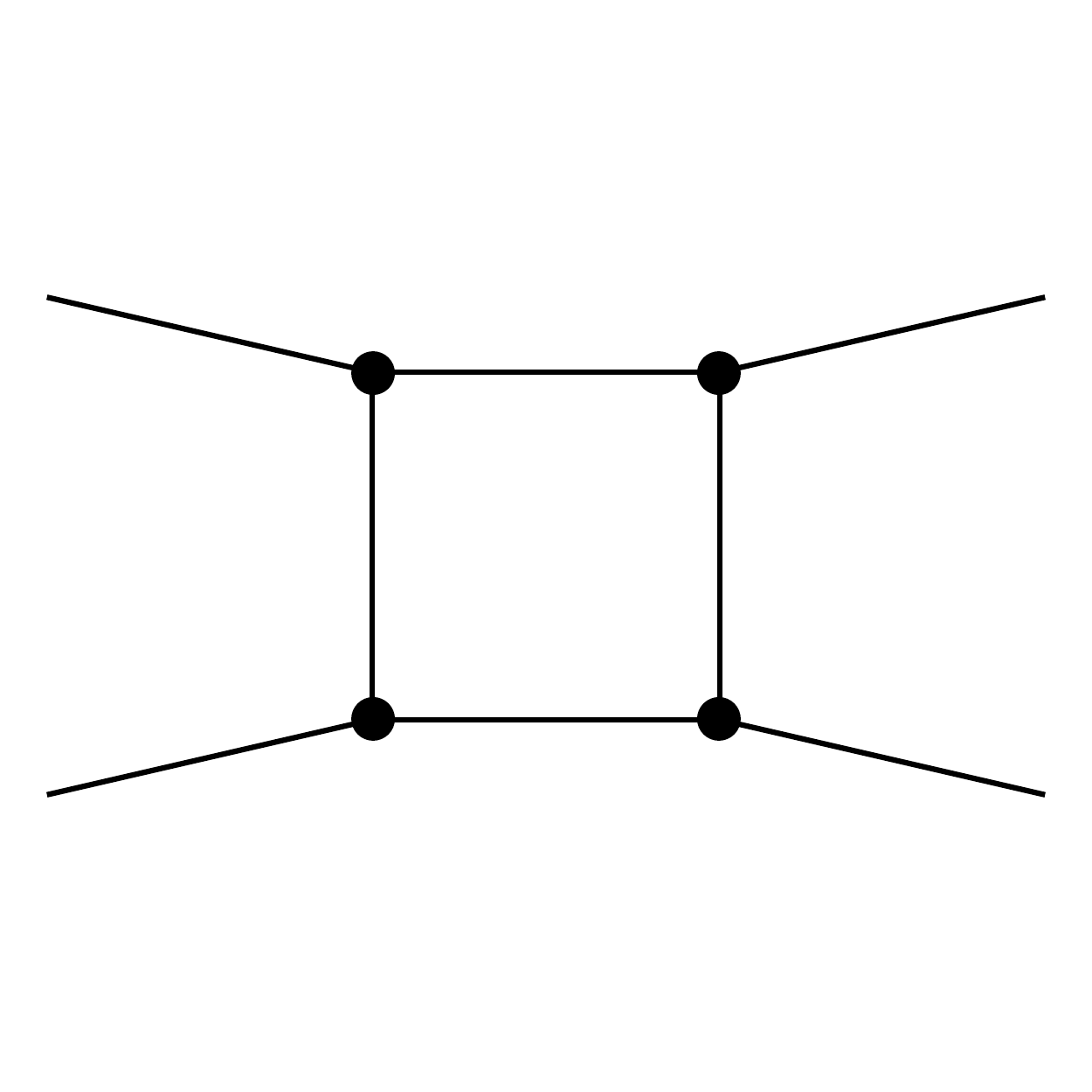}}}
&
\raisebox{-.5\height}{\includegraphics[height=1.3cm]{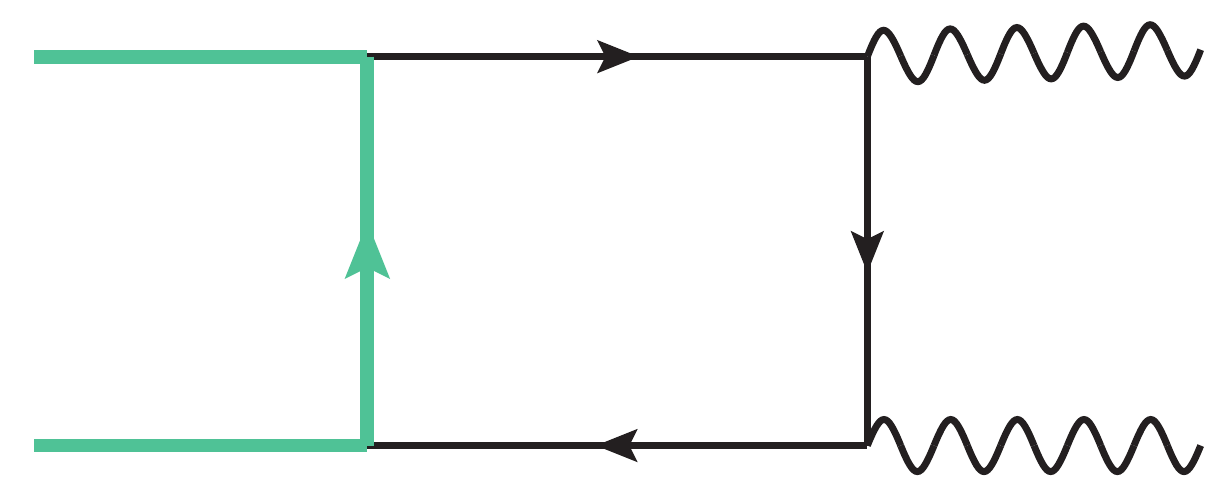}\includegraphics[height=1.3 cm]{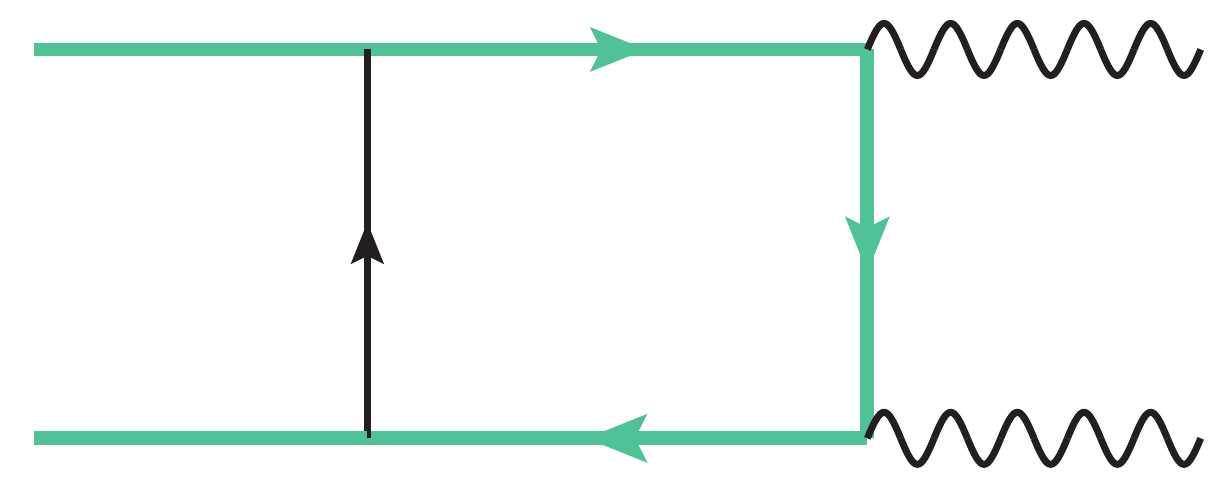}}
&
\multirow{3}{*}{\dguno{1.7}}
\\
\raisebox{-1.5\height}{T1}
&
\raisebox{-.5\height}{\includegraphics[height=1.5cm]{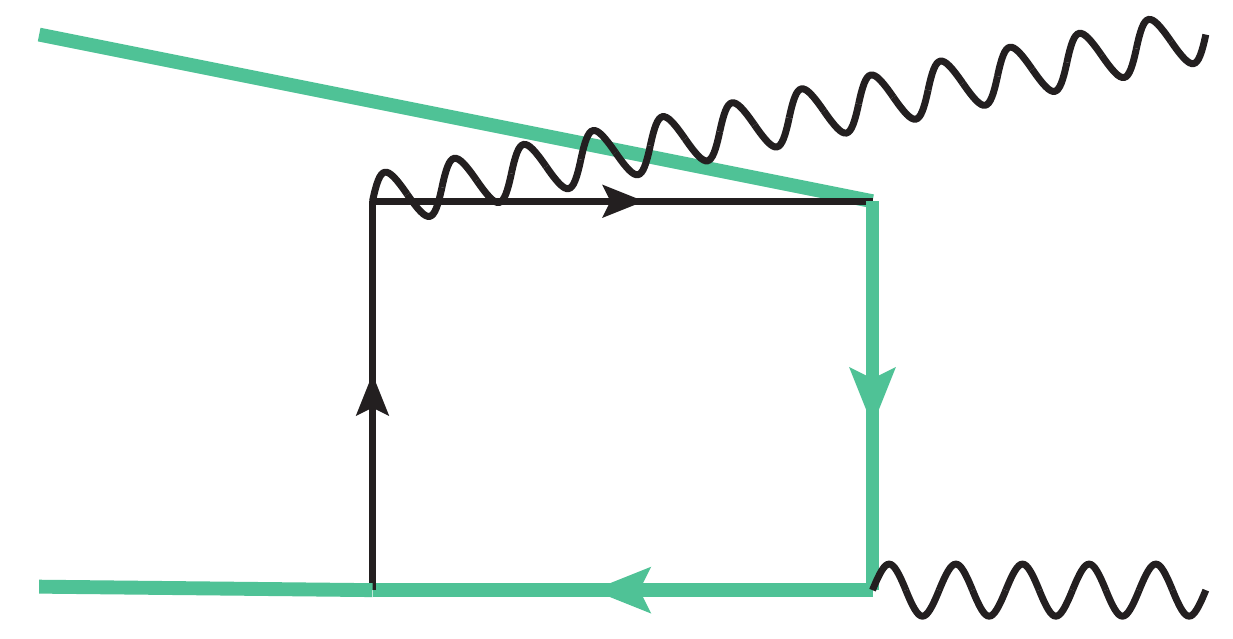}}
&
\\\hline
\multirow{2}{*}{\raisebox{-\height}{\includegraphics[height=2cm]{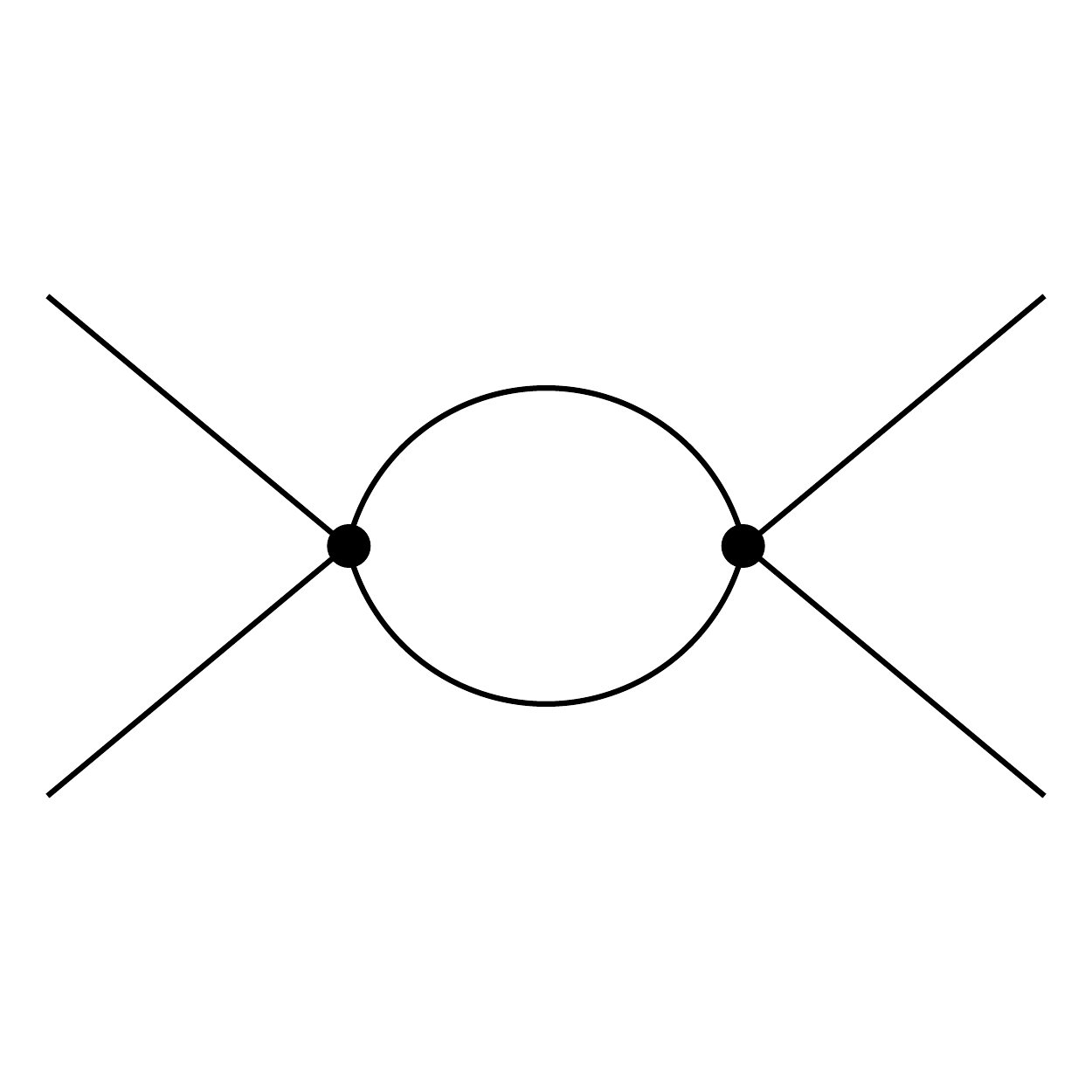}}}
&
\raisebox{-.5\height}{\includegraphics[height=1.4cm]{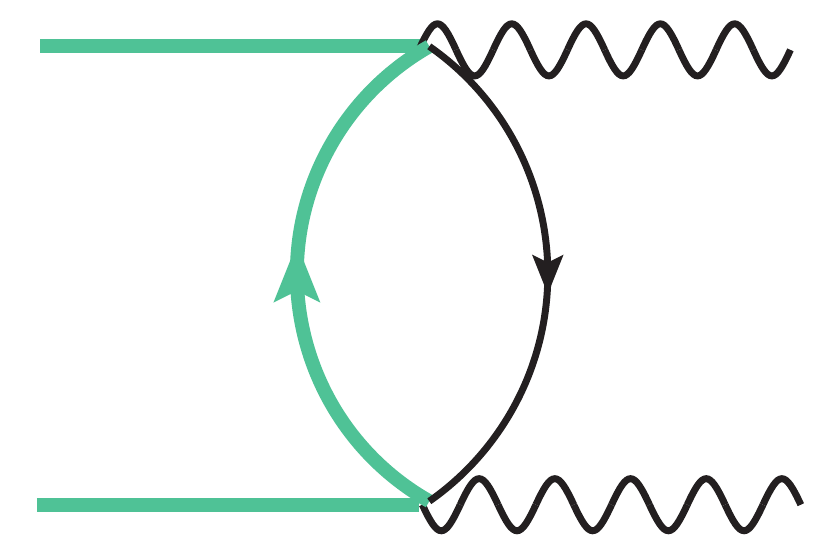}}
&
\dgunop{1.7}
\\\cline{2-3}
\raisebox{-1.8\height}{T2}
&
\raisebox{-0.7\height}{\includegraphics[height=1.5cm]{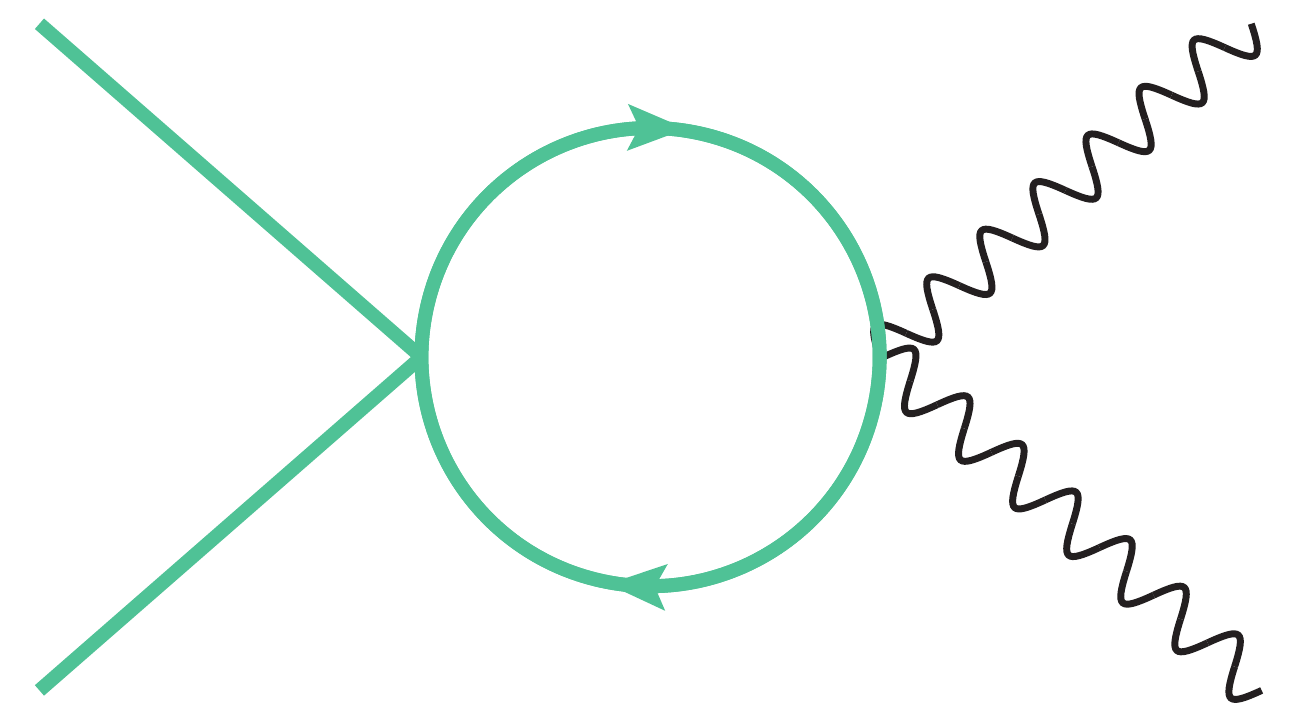}\includegraphics[height=1.5cm]{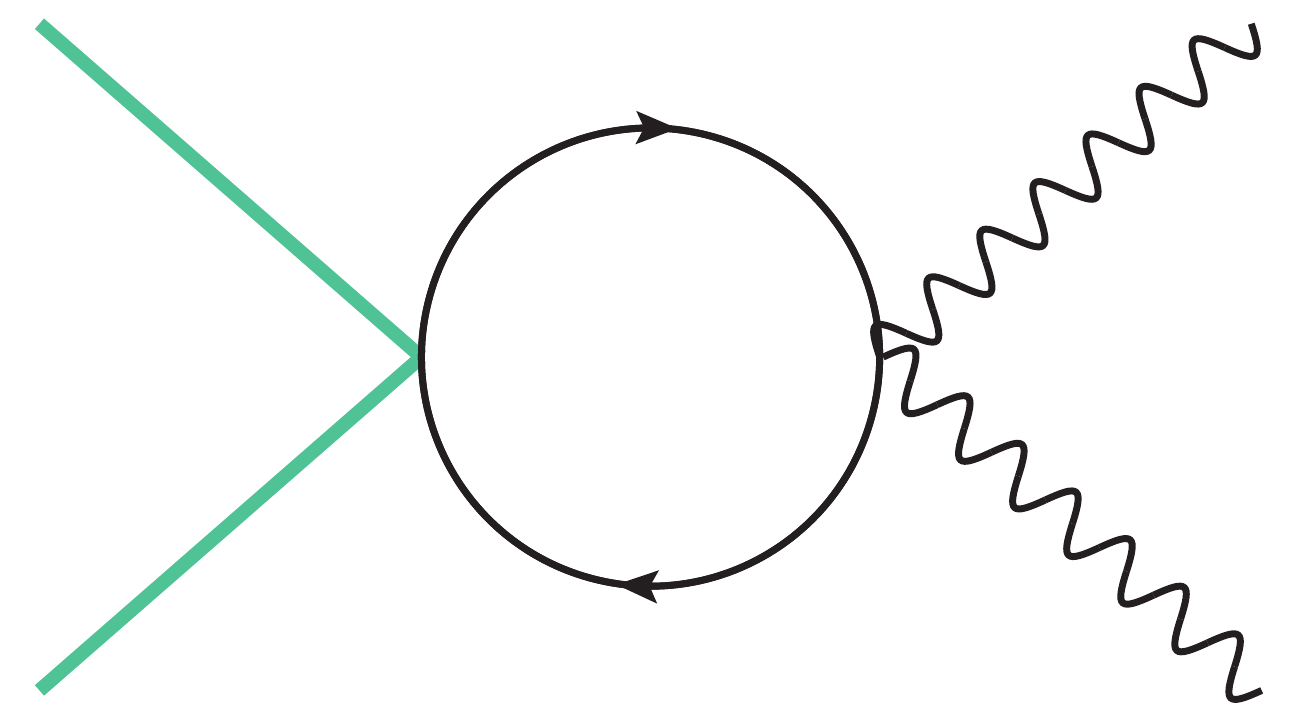}}
&
\multirow{6}{*}{\dgdos{1.7}\dgtres{1.7}}
\\\cline{1-2}
\multirow{3}{*}{\raisebox{-1.5\height}{\includegraphics[height=2cm]{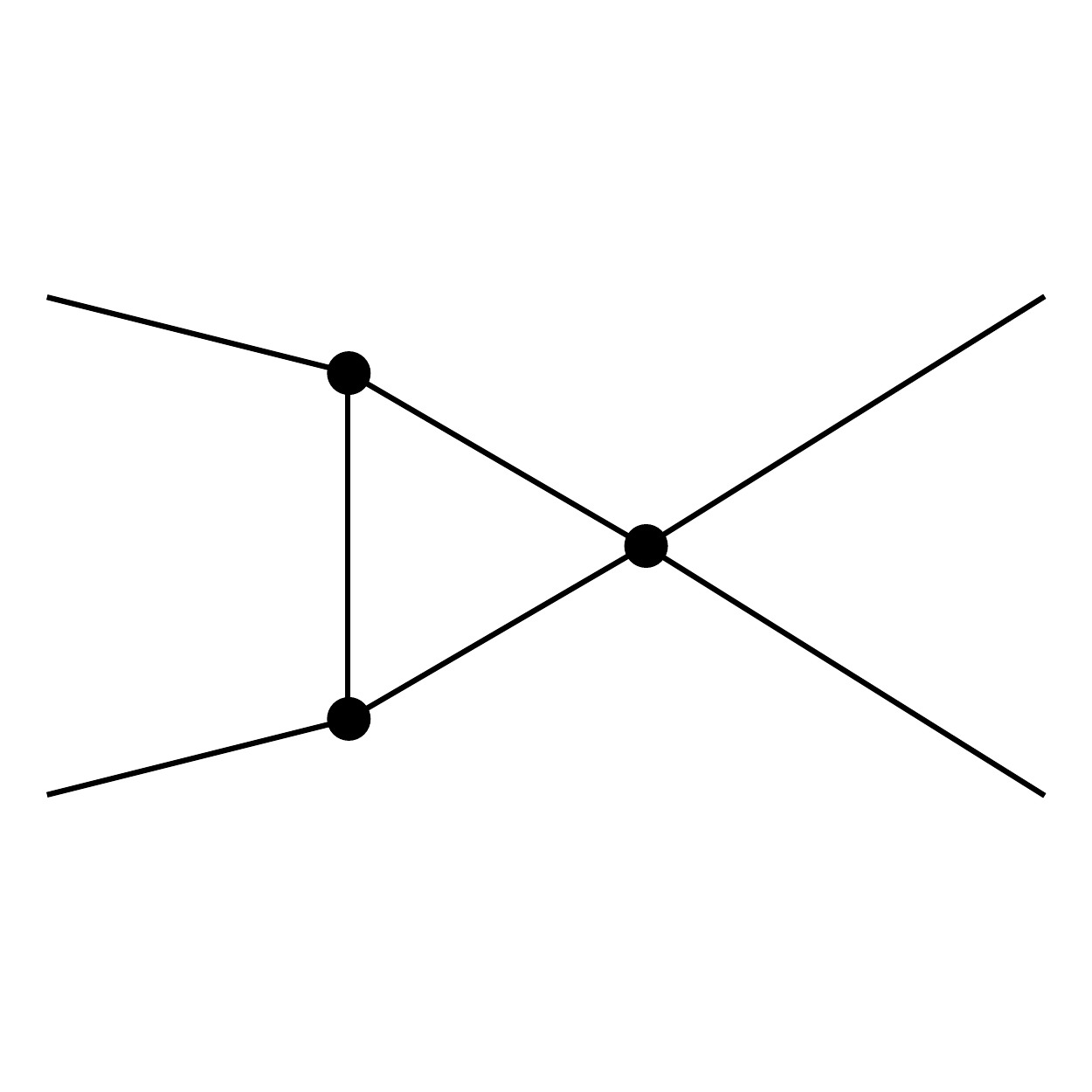}}}
&

\raisebox{-0.8\height}{\includegraphics[height=1.5cm]{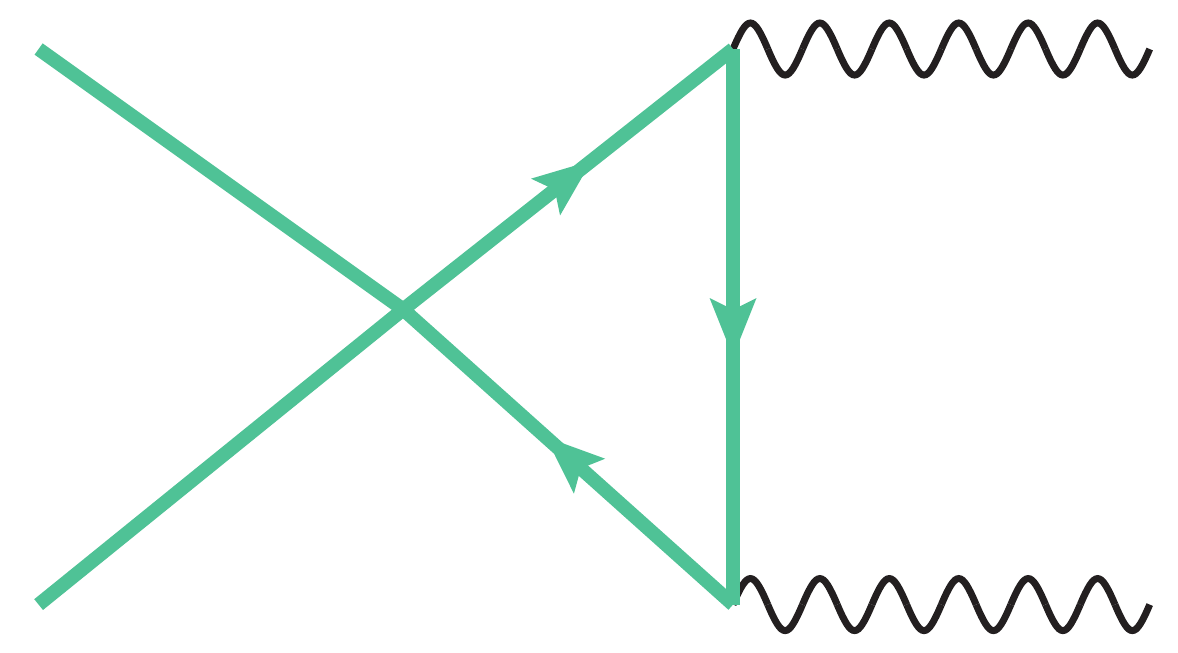}}
\raisebox{-0.7\height}{\includegraphics[height=1.5cm]{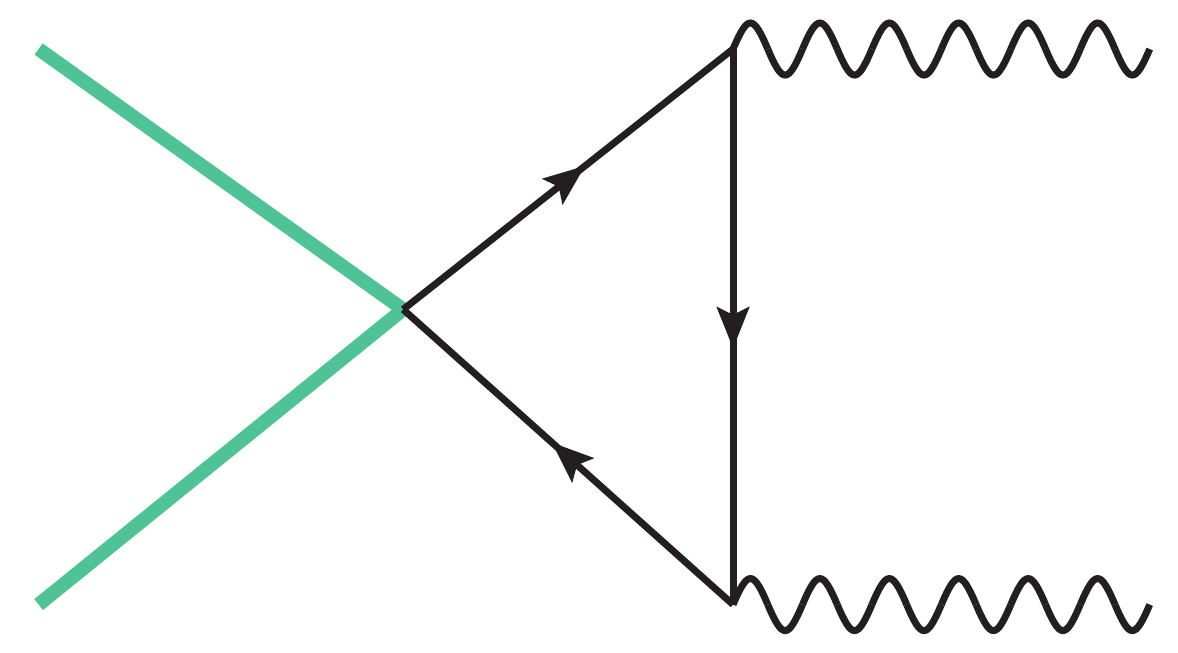}}

&
\\\cline{2-3}
\raisebox{-6.5\height}{T3}
&
\raisebox{-0.7\height}{\includegraphics[height=1.4cm]{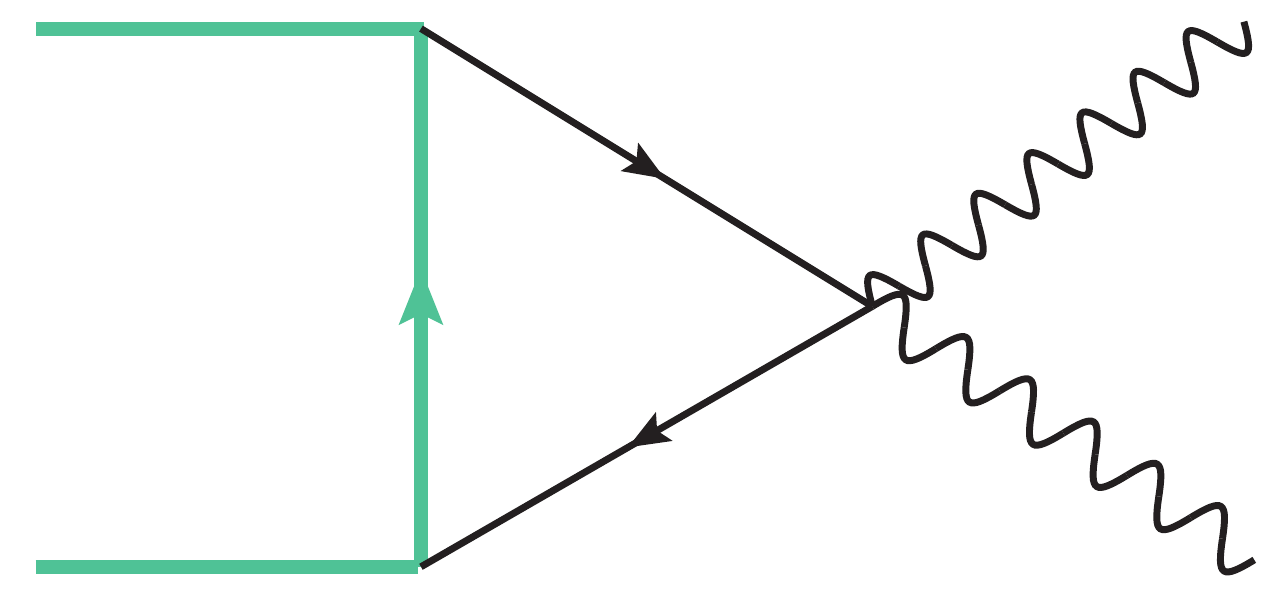}}
\raisebox{-0.7\height}{\includegraphics[height=1.4cm]{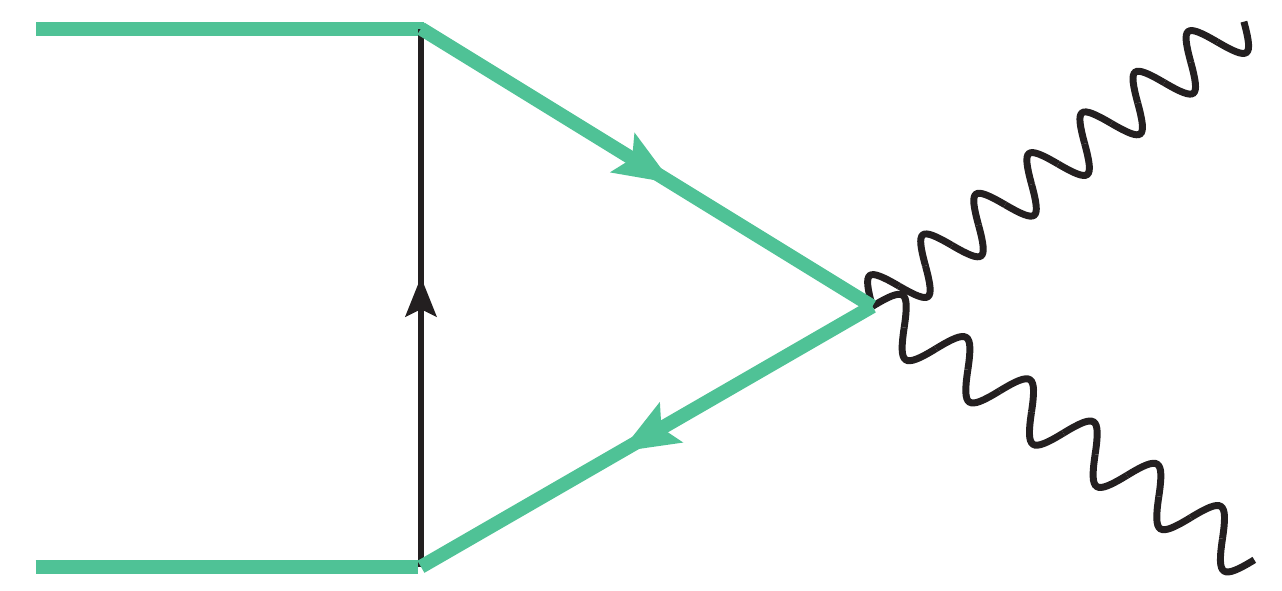}}
&
\multirow{6}{*}{
\dguno{1.7}
\dgunop{1.7}
}
\\
&
\raisebox{-0.4\height}{\includegraphics[height=1.4cm]{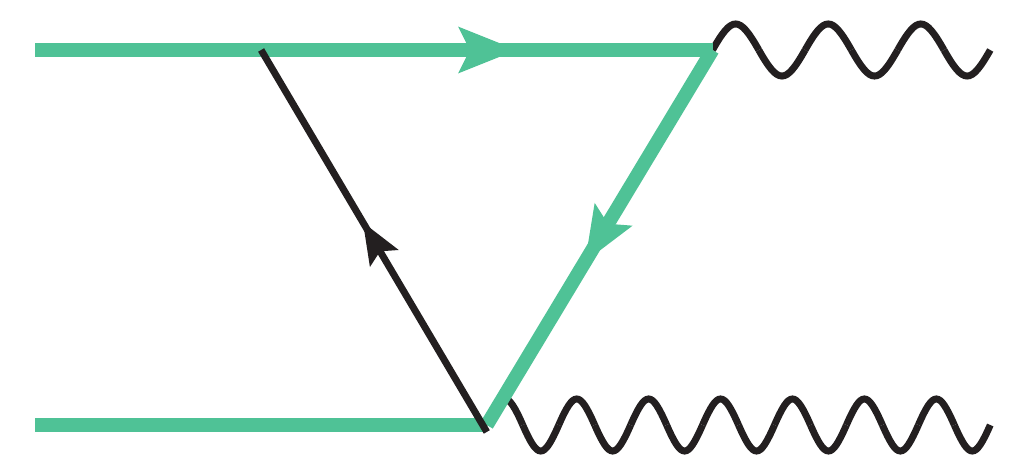}}
\raisebox{-0.4\height}{\includegraphics[height=1.4cm]{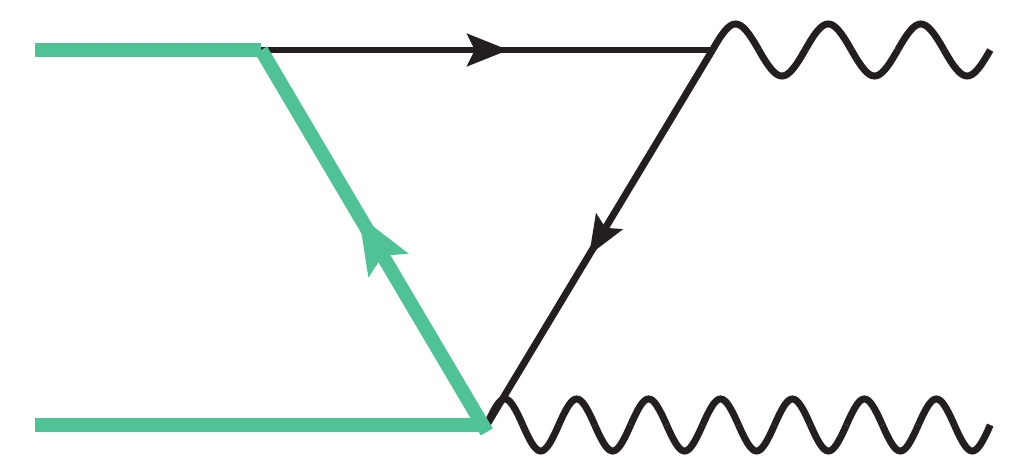}}
&
\\\hline
\raisebox{-.0\height}{\hspace{-20pt}\includegraphics[height=1.8cm]{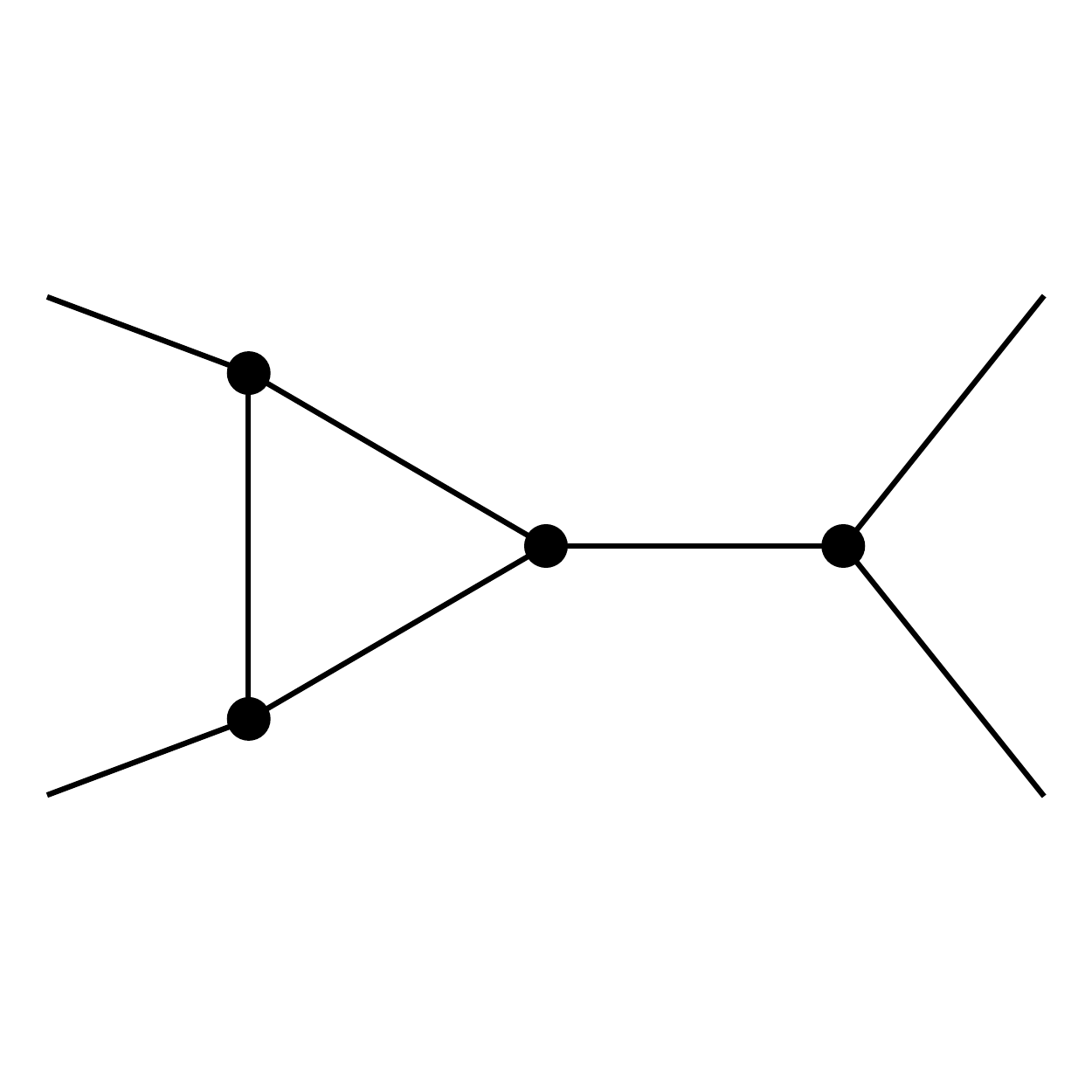}\hspace{-35pt}T4}
&
\raisebox{0.1\height}{\includegraphics[height=1.4cm]{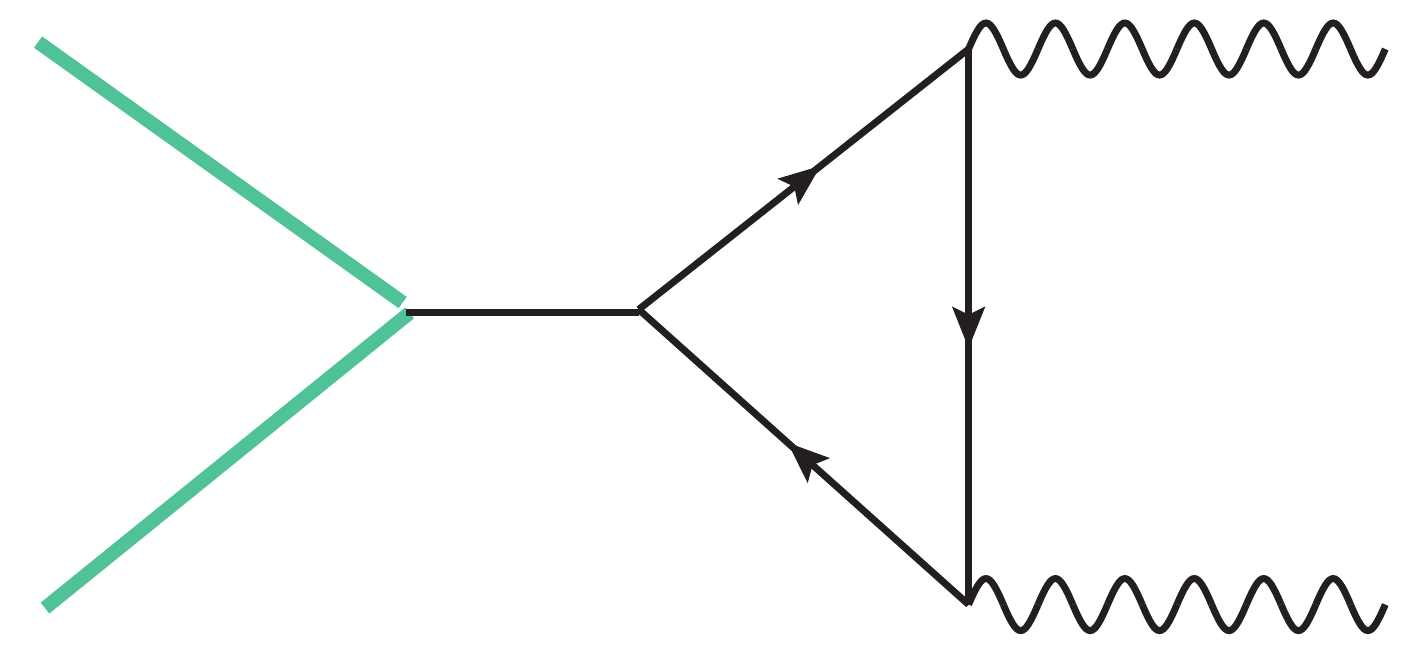}}
&
\multirow{2}{*}{\dgcuatro{1.7}\dgcinco{1.7}}
\\\cline{1-2}
\raisebox{-0.3\height}{\hspace{-20pt}\includegraphics[height=2cm]{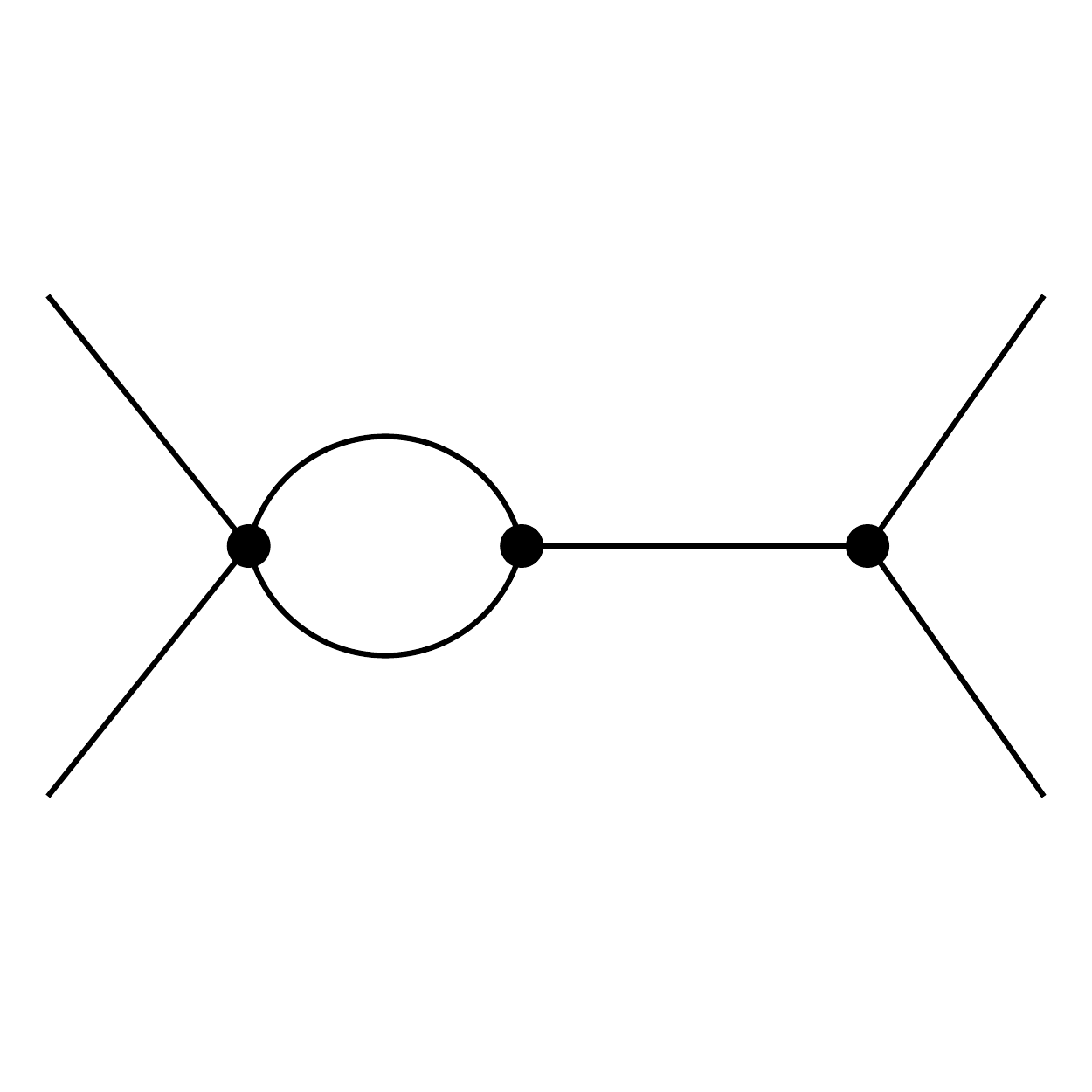}\hspace{-35pt}T5}
&
\raisebox{-.3\height}{\includegraphics[height=1.8cm]{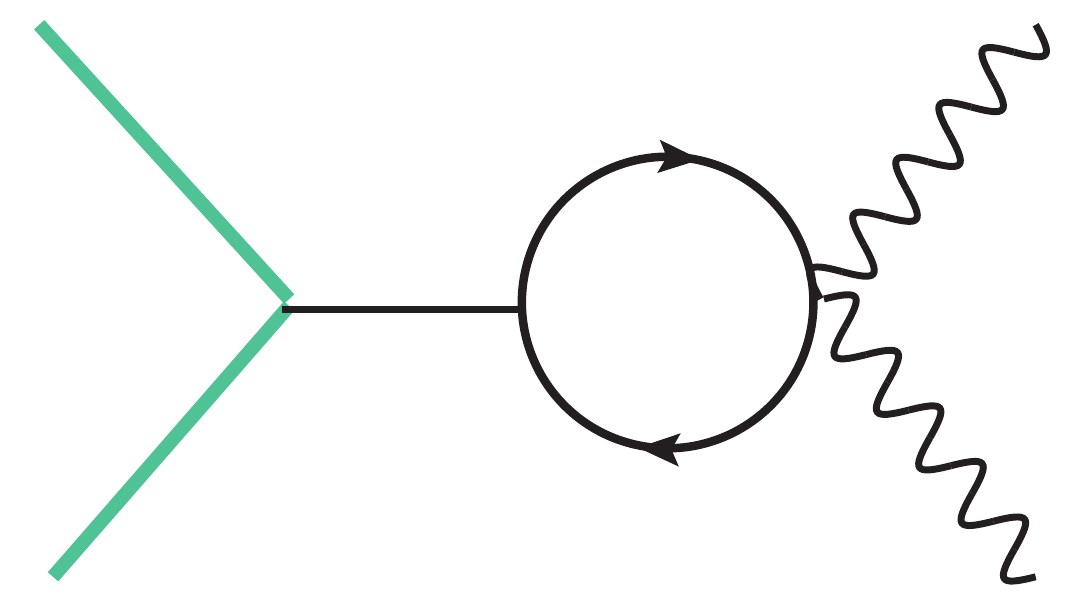}}
&
\\\hline
\raisebox{-0.4\height}{\hspace{-20pt}\includegraphics[height=2cm]{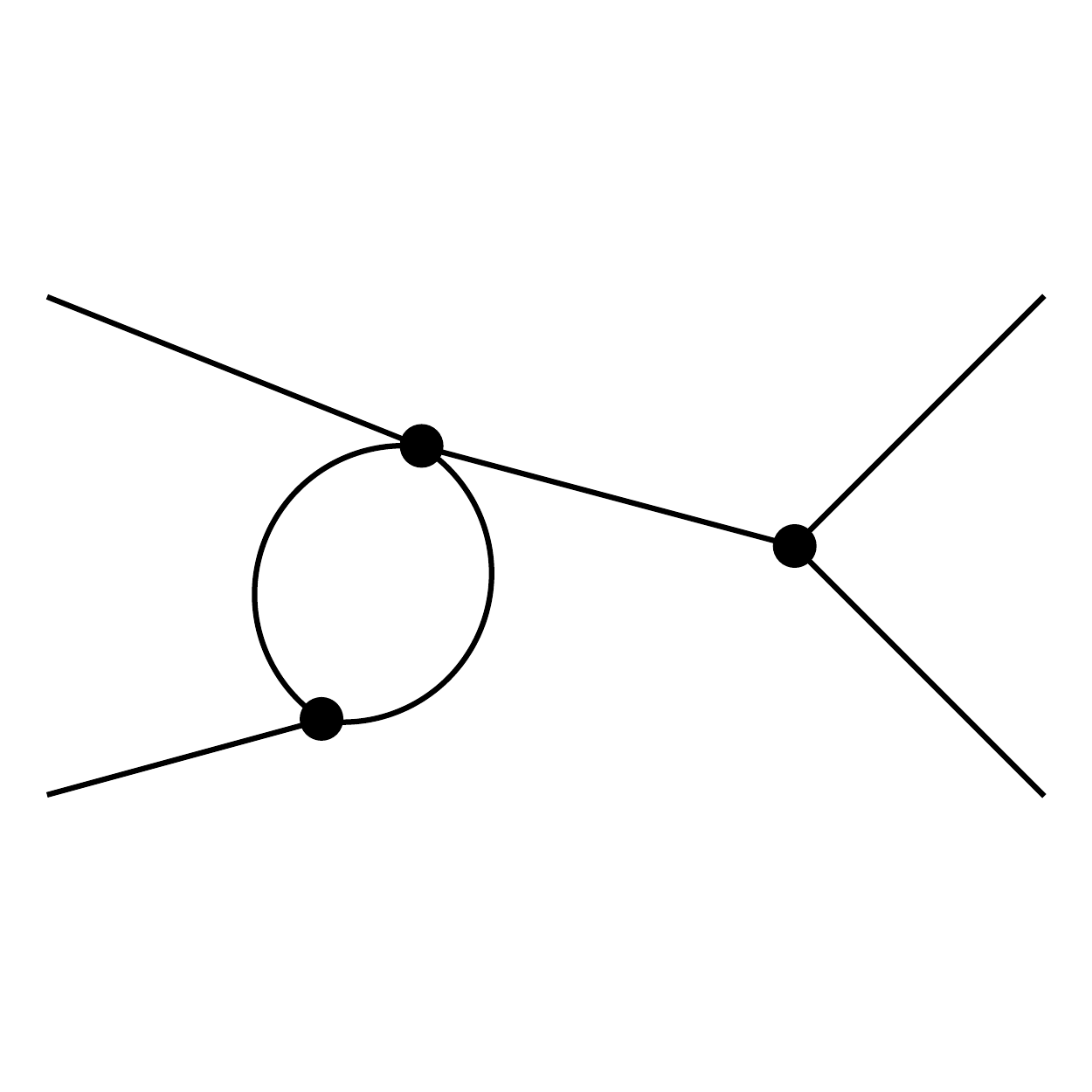}\hspace{-35pt}T6}
&
\raisebox{-0.3\height}{\includegraphics[height=1.6cm]{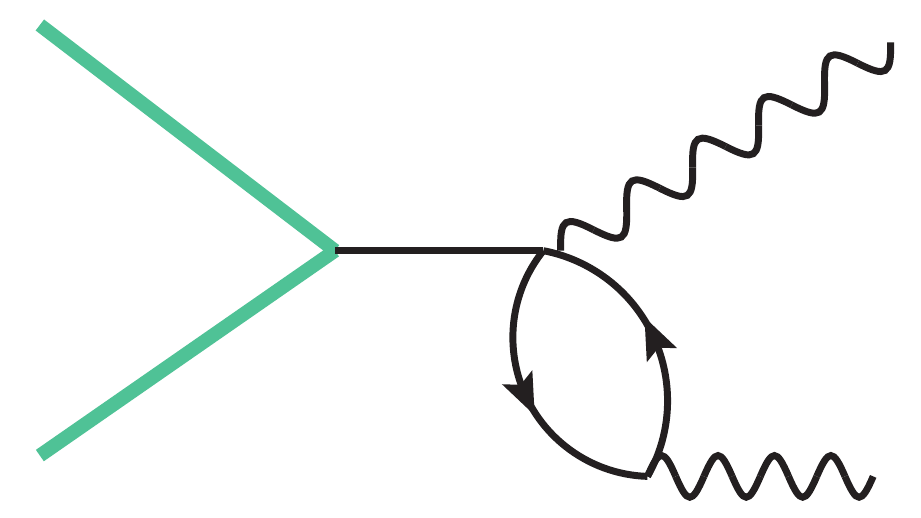}}
&
\raisebox{0.0\height}{\dgcuatro{1.7}\dgcincop{1.7}}
\\\hline
\end{tabular}
%
%
\end{lrbox}

\begin{lrbox}{\Lagrangian}
	\centering
		\begin{tabular}{|c|c|c|c|c|} \hline
\multirow{2}{*}{DM field} & \multicolumn{2}{c|}{\multirow{2}{*}{Mediators}}&  \multirow{3}{*}{\dguno{1.7}}& \multirow{3}{*}{\dgdos{1.7}\dgtres{1.7}}\\
\multirow{2}{*}{ $\DM$}&\multicolumn{2}{c|}{}& &
\\\cline{2-3}
 &\,\,$\medp$\,\,\,&$\DMp$&    & \\\hline\hline
\multirow{3}{*}{Real scalar} & S 
 & S
&$g_1\, \DM \, \DMm  \medp  $
&
$\DM^2 \left(g_2\,  \DMp \DMm + g_3\, \medp \medm\right) $
\\\cline{2-5}
&
F & F
&
$
\, \DM\, \overline{\DMp} \left(g_{1L} P_L 
+g_{1R} P_R\right) \medp
$
&
$0$
\\\cline{2-5}
&
V
& S
&$
i  \, {\medp}^\mu\left(g_1 \DM {\cal D}_\mu  \DMm +g_1'{\cal D}_\mu \DM\DMm \right)
$
&
$
 \DM^2 (g_2\, \DMp \DMm +g_3\, \medp_\mu {\medm}^\mu )
$
\\\hline
\multirow{3}{*}{Majorana}
& S 
& F
& 
$
\medp \overline{\DM} \left(g_{1L} P_L
+g_{1R} \,P_R \right)\DMm 
$
&
\multirow{3}{*}{0}
\\\cline{2-4}
& F
& S
&
$
\DMm\overline{\DM}\left(g_{1L} P_L +g_{1R} P_R \right) \medp 
$
&
\\\cline{2-4}
& V 
& F
&
$
 \overline{\DM}\medp^\mu \gamma_\mu \left( g_{1L} P_L + g_{1R}  P_R \right)\DMm 
$
&
\\\hline
\multirow{2}{*}{Real vector} &
S
 & S
&
$ i g_1  \,\DM^{\mu}  ( {\cal D}_\mu \DMm \medp - \DMm {\cal D}_\mu \medp) $ 
&

$ \DM_\mu \DM^\mu \left(g_2\,  \DMp \DMm+ g_3\, \medm  \medp\right)$
\\\cline{2-5}
&
F & F
&
$
 \overline{\DMp} \DM^{\mu }\gamma_{\mu }  \left( g_{1L} P_L  
+g_{1R}  P_R \right)\medp
$
&
0
\\\hline
\end{tabular}
\end{lrbox}


\begin{lrbox}{\Aform}
	\centering
		\begin{tabular}{|c|c|c|} \hline
\multicolumn{2}{|c|}{\multirow{2}{*}{Mediators}}&  \multirow{3}{*}{\dgcinco{1.70}}\\
\multicolumn{2}{|c|}{}&
\\\cline{1-2}
 \,\, $\medz$\,\,\,&\,\,\,\,\,$\medp$\,\,\,\,\,&     \\\hline\hline
\multirow{3}{*}{CP-even}  
 & S
&$g_5 \medz \, \medm \medp   $
\\\cline{2-3}
& F
&
$g_5 \medz \, \overline{\medp} \medp   $
\\\cline{2-3}
&
V
&
$g_5 \medz {\medm}^\mu{\medp}_\mu$
\\\cline{2-3}
&
Gh
&
$g_5 \medz \, \left(\overline{\med}^-\med^++\overline{\med}^+ \med^- \right)   $
\\\hline
\multirow{3}{*}{CP-odd}  
 & S
&$0  $
\\\cline{2-3}
& F 
&
$i g_5 \medz \, \overline{\medp}\gamma_5 \medp   $
\\\cline{2-3}
&
V
&
0 
\\\cline{2-3}
&
Gh
&
$ig_5 \medz \, \left(\overline{\med}^-\medp^+-\overline{\med}^+ \med^- \right)   $
\\\hline
\end{tabular}
\end{lrbox}

\begin{lrbox}{\schLagrangian}
	\centering
		\begin{tabular}{|c|c|c|} \hline
\multirow{3}{*}{DM field } & \multirow{3}{*}{ Mediator $\medz$}&  \multirow{3}{*}{\dgcuatro{1.70}}\\
& & 
\\ &&\\\hline\hline
\multirow{2}{*}{Real scalar} & CP-even  
&$g_4 \medz \DM^2$
\\\cline{2-3}
& CP-odd & 0
\\\hline
\multirow{2}{*}{Majorana}
& CP-even  
& 
$
g_4\medz \overline{\DM}\DM
$
\\\cline{2-3}
& CP-odd  
& 
$
i g_4\medz \overline{\DM}\gamma_5\DM
$
\\\hline
\multirow{2}{*}{
Real vector
}
&CP-even  
&
$
g_4 \medz \DM_\mu \DM^\mu
$
\\\cline{2-3}
& CP-odd & 0
\\\hline
\end{tabular}
\end{lrbox}

\begin{table}[h]
\usebox{\Top}
\label{table:Top}
\caption{\footnotesize\textit{Left column:} list of one-loop topologies with four external legs. Every one-loop diagram for the process $\DM \DM \to \gamma \gamma$ must have one of these shapes. \textit{Central column:} List of possible Feynman diagrams associated to each topology. Lines in these diagrams follow the conventions of Table~\ref{table:mediators}. \textit{Right column:} Interaction vertices that are necessary for each diagram.  All the interactions of charged mediators with photons are not shown. The hermitian conjugate of each Lagrangian is implicitly assumed. }
\end{table}

Let us start by noticing that conditions \hyperref[condition:R]{(i)} and \hyperref[condition:R]{(ii)}  imply that DM does not annihilate into two photons at tree level. Furthermore,  the corresponding one-loop amplitude  must be finite. The next step is to note that every one-loop diagram must take the form of one of the topologies shown in the left column of Table~\ref{table:Top}, which we enumerate for later convenience. There are no other possible shapes for a one-loop diagram with four external legs. 

Moreover, from condition \hyperref[condition:R]{(i)}, we know that each diagram must have  a $Z_2$ line starting and ending at the DM particles in the initial state. Also, since conditions \hyperref[condition:R]{(i)} and \hyperref[condition:R]{(ii)} forbid the radiation of photons from neutral particles in the diagrams, the fields running in the loop must have electric charge. This means that in addition to the $Z_2$ line, there is a closed line in each  diagram  carrying  electric charge.  

The electric charge loop in a given diagram is associated to only one field according to condition \hyperref[condition:R]{(ii)}, which we generically call $\DMp$ if it is charged under the $Z_2$ symmetry or $\medp$ in the opposite case. 
In addition, some diagrams have neutral particles that are even under the  $Z_2$ symmetry and that we generically call $\medz$.  All these  fields and their quantum numbers are summarized in Table~\ref{table:mediators}, where we also show how we will represent them in Feynman diagrams. In particular, lines associated to the $Z_2$ symmetry are in light blue, whereas, as usual, those associated to the electric charge have an arrow. 
With these assignments, we just proved that every one-loop diagram has a light blue line with its ends in the DM particles of the initial state as well as a loop carrying an arrow.

\begin{table}[t]
\centering
\begin{tabular}{cccc}\hline
Particle &\,\, \,\,\,\,{\bf $Z_2$}\,\,\,\,\,\, & {\bf $U(1)_\text{em}$} & Line\\\hline\hline
$\DM$ & -1 & 0 & \includegraphics[scale=0.4]{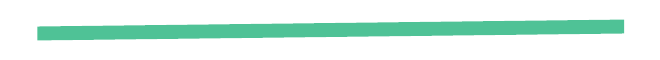} \\
$\DMp$ & -1 & $Q$ & \includegraphics[scale=0.4]{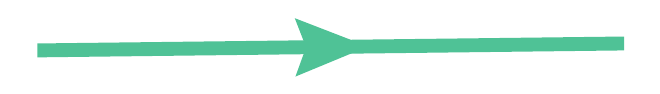} \\
$\medp$ & 1 & $Q$ & \includegraphics[scale=0.4]{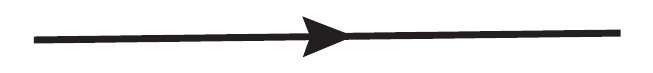}\\
$\medz$ & 1 & 0 & \includegraphics[scale=0.4]{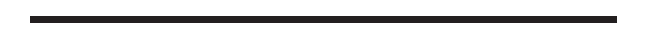}\\
\end{tabular}
\caption{Generic particle content in Feynman diagrams for process $\DM \DM \to \gamma\gamma$. We should stress that, in this work, solid lines represent arbitrary particles and not necessarily  fermions.}
\label{table:mediators}
\end{table}

Using these observations, we can take each topology in the left column of Table~\ref{table:Top} and assign  fields to its lines by following the next procedure\footnote{A similar approach was used in Ref.~\cite{Bonnet:2012kz,Restrepo:2013aga} to systematically study the Weinberg operator at one-loop order.}. First, we consider all the possible permutations of the external legs. Second, we draw the lines carrying the $Z_2$ and the electric charge  
quantum numbers. Finally, we discard the diagrams that violate one of the conditions stated above. In particular, according to the requirements \hyperref[condition:R]{(i)} and \hyperref[condition:R]{(ii)},  we will disregard diagrams whose initial legs radiate photons or have neutral particles directly  coupled to two photons.  Interestingly, this procedure 
determines the vertices  between the DM and the mediators involved in the annihilation process. 

To illustrate the previous procedure, let us first discuss  topologies 1, 2 and 3 of Table~\ref{table:Top}.  None of  
them violates any of our conditions \cond. In fact, they all arise in the one-loop calculation as long as the interaction vertices listed in front exist. These are
\begin{align}&
\dguno{2.5}\,,&
\dgdos{2.5} \,,
\nonumber\\&
\dgtres{2.5}\,,&
\dgunop{2.5}\,.
\label{eq:T123}
\end{align}
We wrote the last interaction as ${\cal L}'_1$ because, in a renormalizable theory, such quartic interaction can only come from a cubic interaction in which the photon is replaced by a covariant derivative. Consequently ${\cal L}'_1 \subset {\cal L}_1$. Notice that the previous vertices involve direct interaction between DM and the charged mediators.

Let us discuss now topologies 4, 5 and 6. Here, the situation is much simpler. A quick look to the left column of Table~\ref{table:Top} reveals that those topologies have an internal line that does not belong to the loop. The external particles attached to such line could not be a DM particle and a photon, because otherwise DM would radiate photons. Similarly, condition \hyperref[condition:R]{(ii)} implies that these external particles can not be two photons. Consequently, all the viable diagrams that can be constructed out of topologies 4, 5 and 6 correspond to s-channel diagrams involving a neutral particle $\medz$, which couples to the DM at tree-level and that subsequently decays into two photons via a loop of charged particles. Hence,  for these particular topologies, our problem is reduced to calculating the off-shell decay of $\medz$. 
For that to be possible,  we need the interactions responsible for the production 
 
\begin{align}
\dgcuatro{2.5}\,,
\label{eq:lscprod}
\end{align}
as well as those associated to the decay
\begin{align}&
\dgcinco{2.5}\,,
&\dgcincop{2.5}\,.
\label{eq:lscdecay}
\end{align}
Notice that  ${\cal L}'_5 \subset{\cal L}_5$.

We arrive to the conclusion that, in any DM model satisfying conditions \cond, the annihilation into two photons has an  amplitude  that  can be split into two pieces. The first one includes diagrams associated to topologies 1,2 and 3, in which DM interacts directly with charged particles  by  means of the vertices in Eq.~\eqref{eq:T123}.  The second piece  is associated to diagrams with topologies 4, 5 and 6, in which DM interacts with charged particles indirectly via the exchange of a neutral particle in the s-channel.

We would like to remark that, even though the total annihilation amplitude is gauge invariant, that is not necessarily  true for the s-channel diagrams separately or the diagrams with topologies 1, 2 and 3. For instance, some parts of the amplitude associated to a s-channel diagram typically cancel with others coming from diagrams with topology 2 or 3. As a result, we have to carefully specify a gauge for our one-loop calculation. 

Conditions \cond  also set restrictions on this matter. For fermions or scalars, condition \hyperref[condition:C]{(iii)} just demands that no FCNC are present. However, for charged gauge bosons, the situation is more involved because photons could couple to Goldstone and gauge bosons in the same cubic vertex. For instance,  vertices such as $\gamma \,G^+ \,W^-$  are present in linear $R_\xi$ gauges of the SM such as the Feynman gauge (Here, $G^+$ is the Goldstone boson associated to the $W^+$ boson). Nevertheless, as pointed out in Refs.~\cite{Fujikawa:1973qs,Bergstrom:1997fh},  such vertices are absent in some non-linear gauges. We refer the reader to appendix~\ref{sec:Gauge} for a detailed discussion. Here, we just mention that, if there are charged vector bosons acting as mediators, we  will always work in the non-linear Feynman gauge in order to satisfy condition \hyperref[condition:C]{(iii)}.    

In addition, for neutral gauge bosons, we will work in their Landau gauge. This because, as also shown in Appendix~\ref{sec:Gauge},   if the particle on the s-channel is a massive gauge boson, its contribution to the annihilation vanishes in that gauge, and only the corresponding  (massless) Goldstone boson must be taken into account.  In particular,  this implies that ${\cal L}'_5$  must vanish because its neutral mediator is a scalar and the Lorentz index of the photon field can not be contracted with the resulting bilinear of charge mediators. Hence, topology 6 is not present. 

We are now ready to translate the interactions vertices into Lagrangian terms.   

\subsection{Interactions of the DM and the mediators}

Let us start by pointing out that, according to condition\hyperref[condition:S]{(iv)}, the DM field  must be a real scalar, a Majorana fermion or a real vector field. With respect to the mediators with electric charge, we consider the following possibilities for their fields
\begin{align}
\begin{array}{l}
Z_2\text{-even}\\
\text{mediator }\medp\\
\end{array}:
\left\{
\begin{array}{l}
\text{scalar (S)}\\
\text{fermion (F)}\\
\text{gauge boson (V)}\\
\text{Faddeev--Popov ghost (Gh)}
\end{array}
\right.
&&
\begin{array}{l}
Z_2\text{-odd}\\
\text{mediator } \DMp\\
\end{array}:
\left\{
\begin{array}{l}
\text{scalar (S)}\\
\text{fermion (F)}\\
\end{array}
\right.\nonumber
\end{align}
We do not consider the possibility of a mediator $\DMp$ as a vector boson or a ghost because it is charged under the $Z_2$ symmetry. In fact, electrically charged spin-1 particles can only be described in a renormalizable way by means of a non-abelian gauge boson, which must not be charged under a $Z_2$ symmetry. To see this, consider the covariant derivative $\partial_\mu - i g V_\mu$. The whole object must transform in the same way under the $Z_2$ symmetry, if this is preserved. Because the first term is even, the second one must be even too.
 
Concerning the neutral mediators $\medz$, Table~\ref{table:Top}  clearly shows that they are $Z_2$-even bosons with no electric charge. In the Landau gauge, neutral gauge bosons do not contribute (only their Goldstone bosons do). Thus,  without loss of generality, we will only consider the possibility of a $\medz$ as a scalar field. Furthermore, because we are assuming that CP is conserved, we will classify neutral mediators according to their CP-parity.

Using this, we can now write down the most general Lagrangians associated to the interaction vertices of  Eq~\eqref{eq:T123}, that are compatible with electromagnetic gauge invariance. This is shown in Table~\ref{table:Lagrangian}. 
There, the letters S, F and V specify the type of charged field, and stand schematically for scalar, fermionic and vector, respectively. Furthermore, in each case, we use generic couplings whose subindex corresponds to the Lagrangian they belong to.  Since we assume that CP is conserved,  $g_2$ and $g_3$ are real, while $g_1$ is either real or purely imaginary depending on whether DM is CP-even or CP-odd, respectively.  

Notice that we did not write down the Lagrangians ${\cal L}'_1$ of Eq.\eqref{eq:T123}, as it is included in ${\cal L}_1$, as explained above.  Note also that Faddeev--Popov ghosts are not present in Table~\ref{table:Lagrangian} because DM can not couple to them.
Ghosts would only couple to DM  if this were present in the gauge fixing-term, but that is not possible because it is charged under $Z_2$.  Furthermore, we do not consider the case of vector DM interacting with other spin-1 particle because for that, on the basis of renormalizability,  we would need a non-abelian gauge structure, which is not possible because the DM is charged under $Z_2$.
 
\begin{table}
\usebox{\Lagrangian}
\caption{Interactions between DM and the charged mediators. }
\label{table:Lagrangian}
\begin{minipage}{0.42\textwidth}

\hspace{-40pt}
\usebox{\schLagrangian}
\caption{Interactions of DM with neutral mediators. }
\label{table:schLagrangian}
\end{minipage}
\hspace{65pt}
\begin{minipage}{0.42\textwidth}
\vspace{40pt}
\usebox{\Aform}
\caption{Interactions among neutral and charged mediators. }
\label{table:Acal}
\end{minipage}

\end{table}

Similarly, we can write down the interactions of the neutral mediator.  
This is shown in Tables~\ref{table:schLagrangian} and \ref{table:Acal}, where we write  ${\cal L}_4$ and ${\cal L}_5$, respectively. 
The couplings $g_4$ and $g_5$ are real.  Note that,  in contrast to case of DM couplings to charged mediators, here we do need to take into account the presence of ghosts because a scalar particle can interact with them if it also couples to the corresponding charged gauge bosons.

It remains to specify the interactions with photons. 
For scalar and fermions, this is fixed by gauge invariance and given by the usual expressions
\begin{align}
{\cal L}\bigg|_{\substack{\text{Scalar}\\\text{Mediators}}} &=  {\cal D}_\mu \medm  {\cal D}^\mu \medp + 
 {\cal D}_\mu \DMm  {\cal D}^\mu \DMp - \mmed^2\, \medm \medp -\mDMp^2\, \DMm \DMp \,,\label{eq:ScalarMediator}
\\
{\cal L}\bigg|_{\substack{\text{Fermionic}\\\text{Mediators}}} &=i\overline{\medp}  \slashed{\cal D}\medp +
 i\overline{\DMp}  \slashed{\cal D}\DMp - \mmed\, \overline{\medp} \medp- \mDMp\, \overline{\DMp} \DMp\,, \label{eq:FermionicMediator}
\end{align}
where ${\cal D}= \partial - ie Q A$ is the electromagnetic covariant derivative for field with charge $Q$ and $A$ is the photon field. 

In contrast, 
the interactions of the gauge field $\medp^\mu$ with photons are not uniquely determined by electromagnetic gauge invariance. For instance, if $F$ is the electromagnetic field strength, the coupling  in front of the renormalizable  interaction $ F^{\mu\nu} \medpVcov{\mu}\medp_{\nu}$ is in principle not fixed by gauge invariance\footnote{When $\med$ is the $W$ boson of the SM, such coupling arises from the $SU(2)$ structure of the electroweak interactions. It has been shown that theories with charged gauge bosons with a different coupling from that of the SM have problems with unitarity~\cite{Ferrara:1992yc}.}.
For concreteness, from now on we will assume that the couplings of the vector mediator with photons resembles those of the SM $W$ boson, with a possibly different charge. This assumption is not so restrictive as it allows to study DM  with electroweak quantum number as well as other  scenarios where DM interacts with other gauge bosons arising from larger gauge symmetries  such as $W'$ bosons in left-right symmetric DM theories (See e.g. Ref.~\cite{Heeck:2015qra,Garcia-Cely:2015quu}) or 3-3-1 scenarios (See e.g. Ref.~\cite{TavaresVelasco:2001vb}). Therefore, the vector boson Lagrangian is given by 
\begin{equation}
{\cal L }\bigg|_{\substack{\text{Vector}\\\text{Mediators}}} =-\frac{1}{2}\left({\cal D}_\mu \medpVcov{\nu} -{\cal D}_\nu \medpVcov{\mu} \right)\left({\cal D}^\mu \medp^{\nu} -{\cal D}^\nu \medp^{\mu} \right)+ \mmed^2\, \medm^{\mu}\medp_\mu-ie\, Q\,F^{\mu\nu} \medpVcov{\mu}\medp_{\nu} +{\cal L}_\text{gf}\, ,
\label{eq:VectorMediator}
\end{equation}
where ${\cal L}_\text{gf}$ is the piece of the interaction obtained by the gauge-fixing  procedure in the  non-linear Feynman gauge(see Appendix~\ref{sec:Gauge} for details or Ref.~\cite{Pasukonis:2007fu}).

This is not the whole story. A massive charged vector field requires also  one complex Goldstone boson and four ghosts. In the non-linear Feynman gauge, the former is just a scalar and is properly described by Eq.~\eqref{eq:ScalarMediator} if its mass is taken equal to that of corresponding gauge boson. The latter, which we denote as $\overline{\phi}^\pm$ and $\phi^\pm$, have the same mass of the gauge bosons and interact with electromagnetic field by means of  
\begin{equation}
{\cal L} =
-ie Q A_\mu\left(\partial_\mu \overline{\phi}^-\phi^+-
\partial_\mu \overline{\phi}^+ \phi^- + \overline{\phi}^+\partial_\mu \phi c^-
- \overline{\phi}^-\partial_\mu \phi^+\right)
- e^2 Q^2 A_\mu A^\mu \left(\overline{\phi}^- \phi^++\overline{\phi}^+ \phi^-\right)\,.
\label{eq:GhostMediator}
\end{equation}

With all these Lagrangians, we will  be able to calculate $\sigma v$ in the next Section.

\section{Calculation of the amplitude}
\label{sec:calculation}

\subsection{Lorentz structure of the annihilation amplitude}

DM moves with non-relativistic speeds in  astrophysical environments. This was also true  during the dark ages, where DM could have potentially alter the CMB if it annihilates producing gamma-ray lines. Therefore, we are only interested in the limit of vanishing DM relative velocity, $v=0$. In that case, we will show that   the amplitude of  $\text{DM}\text{DM}\to\gamma\gamma$ can be specified by one or  few form factors depending on the DM spin.

In the following listing according to DM spin, $q$ and $q'$ are the momenta of the final state photons,  $\sigma$  and $\sigma'$ are their helicities  and $\epsilon$ and $\epsilon'$  are the corresponding polarization vectors. Moreover,   both particles of the initial state have the same four-momentum $p \equiv (\mDM,0,0,0) = (q+q')/2$.

\begin{itemize}
\item  \textbf{Scalar DM:}  In this case, the annihilation amplitude can be cast as
$
{\cal M }_S= {\cal M}^{\mu\nu} \epsilon_{\mu}^* \epsilon_{\nu}'^*\,.
$
The tensor ${\cal M}^{\mu\nu}$ depends only on $q$ and $q'$ and, according to the Ward identities, satisfies $q_{\mu} {\cal M}^{\mu\nu} = q'_{\nu} {\cal M}^{\mu\nu} =0$. This, the property $\epsilon \cdot q =0$ and the fact that two scalar particles at rest form a CP-even state imply that  
\begin{equation}
  {\cal M}_S ={\cal B} \left( -g^{\mu\nu} + \frac{q'^{\mu} q^{\nu}}{2\,\mDM^2} \right) \epsilon_{\mu}^* \epsilon_{\nu}'^*={\cal B}\,\delta_{\sigma\,\sigma'}
\,,
\label{eq:Mscalar}
\end{equation}
where $\cal B$ is a  scalar function. In terms of this, the cross section reads
\begin{equation}
\sigma v \left(\DM\DM \to \gamma\gamma\right) = \frac{c|{\cal B}|^2}{32 \pi\mDM^2}\,,
\label{eq:cross}
\end{equation}
with the spin-average factor $c=1$. Thus, our goal for spin-zero DM is to calculate $\cal B$.

\item \textbf{Majorana DM:} in this case, we first  write the annihilation amplitude as $\overline{v_1} {\cal M}^{\mu\nu} u_2 \epsilon_{\mu}^* \epsilon_{\nu}'^*$. That is,  $\cal M^{\mu\nu}$ is the amplitude after stripping out the spinors of the DM particles in the initial state. This object has more information than we actually need because we are only interested in initial states with  total spin zero. The state with total spin one is banned for identical particles because it is totally symmetric when the two fermions do not move with respect to each other. Following Refs.~\cite{Bergstrom:1997fh,Kuhn:1979bb}, we can obtain the amplitude corresponding to the spin-zero initial configuration as
\begin{equation}
\begin{split}
{\cal M}_\text{F} =- \frac{1}{\sqrt2}
\text{Tr} \left\{ {\cal M}^{\mu\nu} \left(\slashed{p}+\mDM\right) \gamma^5 \right\} \epsilon_{\mu}^* \epsilon_{\nu}'^*.
\end{split}
\label{eq:P0}
\end{equation}
Similar to the scalar case, gauge invariance and CP conservation restrict the annihilation amplitude. Taking into account that two Majorana particles at rest form a CP-odd state, we must have 
\begin{equation}
{\cal M}_\text{F} 
=\frac{{i\,\cal B}}{2\mDM^2}\,{\epsilon}^{\alpha\beta \mu \nu } {q}_\alpha q'_\beta \epsilon_{\mu}^* \epsilon_{\nu}'^*
={\cal B}\, \sigma \delta_{\sigma\,\sigma'}\,,
\label{eq:MMaj}
\end{equation}

where $\cal B$ is a  scalar function. This can be used to calculate the cross section by means of Eq~\eqref{eq:cross} with the spin-average factor $c=1/4$.

Eqs.~\eqref{eq:Mscalar} and \eqref{eq:MMaj} show that the helicities of the photons must equal. This can be understood from the fact that the total angular momentum is zero when the DM relative velocity is zero. For scalar particles, this is because there is no spin. For Majorana particles, that follows from the fact that the spin-one state is not possible.

\item \textbf{Vector DM.} In this case, both the initial and final state particles are vector bosons and we can write the amplitude as $
\mathcal{M}_\text{V}=\mathcal{M}_{\alpha\beta\mu\nu} \epsilon^{\alpha}_1\epsilon^{\beta}_2\epsilon^{\mu*}\epsilon'^{\nu*}\,.
$
Assuming a CP-even initial state, from gauge invariance and Bose statistics, as pointed out in Ref.~\cite{Bergstrom:2004nr}, it follows that this object can be decomposed as 
\begin{align}
\mathcal{M}^{\alpha\beta\mu\nu} =& \,
{\cal B}_2\left[\left(-\frac{p^\mu q^\alpha}{\mDM^2}+g^{\mu\alpha}\right)\left(-\frac{p^\nu q'^\beta}{\mDM^2}+g^{\nu\beta}\right)+\left(\frac{p^\mu q'^\beta}{\mDM^2}+g^{\mu\beta}\right)\left(\frac{p^\nu q^\alpha}{\mDM^2}+g^{\nu\alpha}\right)\right]\nonumber\\
&+ \,\left({\cal B}_1 g^{\alpha \beta}-{2\cal B}_6\frac{q^\alpha q^\beta }{\mDM^2} \right)\left(\frac{p^\mu p^\nu}{\mDM^2}-\frac{g^{\mu\nu}}{2}\right)
\,.\label{eq:Mvec}
\end{align}

Hence, our goal is to calculate the function ${\cal B}_1$, ${\cal B}_2$ and ${\cal B}_6$ (we use this notation to keep the conventions of Ref.~\cite{Bergstrom:2004nr}). In terms of these, the corresponding cross section is given by\footnote{Even though our expression for the amplitude is the same, for the cross section formula, we have a disagreement with Ref.~\cite{Bergstrom:2004nr}}
\begin{equation}
\sigma v= \frac{1}{576 \pi\mDM^2} \left[\frac{3}{2}|{\cal B}_1|^2+12|{\cal B}_2|^2+2|{\cal B}_6|^2-4 \text{Re}\left( {\cal B}_1 ( {\cal B}_2^*+ \frac{{\cal B}_6^*}{2})\right)\right]\,.
\label{eq:crossV}
\end{equation}

\end{itemize}

We are ready to calculate the loop diagrams and the corresponding cross sections. To that end,  by means of \texttt{FeynRules}~\cite{Christensen:2008py,Alloul:2013bka}, we have implement all the Lagrangians quoted Section~\ref{sec:II} in \texttt{FeynArts}~\cite{Hahn:2000kx}.  Then, we calculate the amplitude for the process DMDM$\to\gamma\gamma$ in each case and have \texttt{FormCalc}~\cite{Hahn:1998yk} reduce the tensor loop integrals to scalar Passarino--Veltman functions~\cite{Passarino:1978jh}. 
Since our process of interest has four external legs, our form factors will depend on the two-, three- and four-point functions $B$, $C$ and $D$, respectively. For these functions, we follow the conventions of \texttt{FormCalc}. 
In fact, as we discuss  in Appendix~\ref{sec:reduction}, in the non-relativistic limit the latter must be reduced further to two- and three-point functions. The corresponding form factors ${\cal B}$ thus depends only of Passarino--Veltman functions $B$ and $C$. We now report such form factors for each scenario.

\subsection{Results for topologies 1, 2 and 3: charged mediators interacting directly with DM}
\label{sec:T123}

Let us discuss first the diagrams induced by ${\cal L}_2$
\begin{eqnarray}
\raisebox{-.5\height}{\includegraphics[height=1.7cm]{t3r4}}+
\raisebox{-.5\height}{\includegraphics[height=1.7cm]{t2r2}}.
\end{eqnarray}
Based on general considerations such as Lorentz and gauge invariance, we argued that the corresponding annihilation amplitude can be cast as shown in Eqs.~\eqref{eq:Mscalar} and \eqref{eq:Mvec}. We explicitly  corroborate  that and find the following 
\begin{eqnarray}
\begin{array}{c}
 \text{form factors} \\
\text{induced by }{\cal L}_2
\end{array}:
\left\{
\begin{array}{llll}
{\cal B}{\,\,}=& \frac{Q^2 \alpha g_2}{\pi}\left(1- \rDMp^2 f_\DMp\right)&\text{       for scalar DM and scalar }\DMp \\
{\cal B}_1=& \frac{2 Q^2 \alpha g_2}{\pi}\left(1- \rDMp^2 f_\DMp\right), & \text{      for vector DM and scalar }\DMp\\
{\cal B}=& 0 \text{ (or   } {\cal B}_1 =0) & \text{      otherwise}
\end{array}
\label{eq:BforL2}
\right.
\end{eqnarray}
as well as $ {\cal B}_2 = {\cal B}_6 =0$ for vector DM.  Here, we introduce the notation $r_i \equiv m_i/\mDM$ and
\begin{eqnarray}
\label{eq:f}
f_i\equiv -2 \, C_0\left( 0, 4 , 0, r_i^2,  r_i^2,  r_i^2\right) =  \left\{
\begin{array}{ll}
\arcsin^2\left(\frac{1}{r_i}\right) & \text{if }r_i \geq 1 \\
-\frac{1}{4}\left(\log\left(\frac{1-\sqrt{1-r_i^2}}{1+\sqrt{1-r_i^2}}\right)+i\pi\right)^2 & \text{if } r_i<1 \,.
\end{array}
\right.
\end{eqnarray}

Similarly, the  diagrams associated to ${\cal L}_3$ are
\begin{eqnarray}
\raisebox{-.5\height}{\includegraphics[height=1.7cm]{t3r3}}+
\raisebox{-.5\height}{\includegraphics[height=1.7cm]{t2r1}},
\end{eqnarray}
which give rise to 
\begin{eqnarray}
\begin{array}{c}
 \text{form factors} \\
\text{induced by }{\cal L}_3
\end{array}:
\left\{
\begin{array}{llll}
{\cal B}=& \frac{Q^2 \alpha g_3}{\pi}\left(1- \rmed^2 f_\medp\right)&\text{        for scalar DM and scalar }\medp \\
{\cal B}=& -\frac{4Q^2 \alpha g_3}{\pi}\left(1-(\rmed^2-2)  f_\medp\right)&\text{        for scalar DM and vector } \medp\\
{\cal B}_1=& \frac{2Q^2 \alpha g_3}{\pi}\left(1- \rmed^2 f_\medp\right),& \text{        for vector DM and scalar } \medp\\
{\cal B}=& 0 \text{ (or   } {\cal B}_1 =0) & \text{      otherwise}
\end{array}
\label{eq:BforL3}
\right.
\end{eqnarray}
as well as  $ {\cal B}_2 = {\cal B}_6 =0$ for vector DM. 

For the Feynman diagrams induced by ${\cal L}_1$, the calculation is significantly more difficult because of the presence of box diagrams in the annihilation amplitude such as  

\begin{align}
\raisebox{-.5\height}{\includegraphics[height=1.5cm]{t1r1}},
\raisebox{-.5\height}{\includegraphics[height=1.5cm]{t1r2}},
\raisebox{-.5\height}{\includegraphics[height=1.5cm]{t1r3}}\,.\nonumber
\end{align}

For relative DM velocities approaching zero, i.e. $v\to0$, the algorithm for reducing the  tensor integrals to scalar functions leads to numerical instabilities and even breaks down for $v=0$. This pathological behavior is well-understood and stems from the assumption that the external momenta are linearly independent, which is not true here because both DM particles are assumed to have the same momentum. Following~\cite{Stuart:1987tt, Stuart:1989de, Stuart:1994xf}, we reduce the loop integrals dropping such assumption. For a detailed description of this procedure we refer the reader to Appendix~\ref{sec:reduction}.  Using such method, in the case of scalar or Majorana DM, we find
\begin{align}
\label{eq:BforL1}
{\cal B}\bigg|_{{\cal L}_1}= \dfrac{Q^2 \alpha}{\pi}\bigg[
&x_1 
+x_2 \dfrac{ \, C_0( 0, 1, -1, \rmed ^2, \rmed ^2, \rDMp^2)  }{(\rDMp^2-\rmed^2)(1+\rDMp^2-\rmed^2)}
+x_3 \dfrac{ \, C_0( 0, 1, -1, \rDMp^2, \rDMp^2, \rmed ^2)}{\left(-\rDMp^2+\rmed^2\right)\left(1-\rDMp^2+\rmed^2\right)}\\
&+x_4 \, \frac{C_0 ( 0, 4 \, , 0, \rmed ^2, \rmed ^2, \rmed ^2)}{1+\rDMp^2-\rmed^2}
+ x_5 \,\frac{ C_0 (0, 4 \, , 0, \rDMp^2,\rDMp^2, \rDMp^2 )}{1-\rDMp^2+\rmed^2}
+x_6\, B_0(4 , \rmed^2 ,\rmed^2 )\nonumber\\
&
+x_7\, B_0(4 , \rDMp^2,\rDMp^2)
+x_8\, B_0(-1 , \rDMp^2,\rmed^2 )
-(x_6+x_7+x_8) B_0( 1,\rDMp^2, \rmed^2 )\nonumber
\bigg]\,
\end{align}

where $x_1,..,x_8$ are dimensionless coefficients listed in Appendix~\ref{sec:xs} for the different combinations of mediators, and $Q$ is the charge of the mediators (which must be the same for both). The same expressions hold for  ${\cal B}_1, {\cal B}_2$ and ${\cal B}_6 $ in the case of vector DM. 

Eq.~\eqref{eq:BforL1} can be simplified further as a function of analytical expressions. On the one hand, the Passarino--Veltman  functions $B_0$ can be written in terms of logarithms~\cite{Passarino:1978jh}. On the other hand, the functions $C_0$ in Eq.~\eqref{eq:BforL1} can be cast either in terms of $f_\DMp$ and $f_\med$ by means of Eq.~\eqref{eq:f}, or as real combinations of dilogarithms as shown in, e.g., Ref.~\cite{Bern:1997ng}.

\subsection{Results for topologies 4 and 5: neutral mediators on the s-channel }
\label{sec:schannels}

Combining the information on Tables~\ref{table:schLagrangian} and \ref{table:Acal}, we can calculate the annihilation amplitude for any s-channel process. Concretely, the diagrams associated to ${\cal L}_4\text{ and }{\cal L}_5$ are 
\begin{eqnarray}
\raisebox{-.5\height}{\includegraphics[height=1.7cm]{t4r3}}+
\raisebox{-.5\height}{\includegraphics[height=1.7cm]{t5r1}},
\end{eqnarray}
which give rise to 
{\small
\begin{eqnarray}
\begin{array}{c}
 \text{form} \\ \text{factors} \\
\text{induced}\\\text{by }\\
{\cal L}_4+{\cal L}_5
\end{array}:
\left\{
\begin{array}{llll}
{\cal B}=-{\cal A}\,(1- \rmed^2 f_\medp)&\text{ for scalar DM, CP-even }\medz \text{  and scalar }\medp \\
{\cal B}= 4{\cal A}\,\mmed \left(1-(\rmed^2-1) f_\medp\right)&\text{ for scalar DM, CP-even }\medz \text{ and fermionic }\medp\\
 {\cal B}= 4{\cal A}\,\left(1-(\rmed^2-2) f_\medp\right)&\text{ for scalar DM, CP-even }\medz \text{  and vector }\medp \\
{\cal B}= 2{\cal A}\,\left(1-\rmed^2 f_\medp\right)&\text{ for scalar DM, CP-even }\medz \text{ and ghost }\medp \\
{\cal B}= 4\sqrt{2}{\cal A}\, \mDM^2 r_\medp f_\medp &\text{ for Majorana DM, CP-odd }\medz \text{ and fermionic }\medp \\
{\cal B}_1=-2{\cal A}\,(1- \rmed^2 f_\medp)&\text{ for vector DM, CP-even }\medz \text{  and scalar }\medp \\
{\cal B}_1= 8{\cal A}\,\mmed \left(1-(\rmed^2-1) f_\medp\right)&\text{ for vector DM, CP-even }\medz \text{ and fermionic }\medp\\
 {\cal B}_1= 8{\cal A}\,\left(1-(\rmed^2-2) f_\medp\right)&\text{ for vector DM, CP-even }\medz \text{  and vector }\medp \\
{\cal B}_1= 4{\cal A}\,\left(1-\rmed^2 f_\medp\right)&\text{ for vector DM, CP-even }\medz \text{ and ghost }\medp\\
{\cal B}=0 \text{ (or   } {\cal B}_1 =0) & \text{      otherwise}
\end{array}
\label{eq:BforL45}
\right.
\hspace{-5pt},
\end{eqnarray}
}
with
\begin{equation}
{\cal A} = \dfrac{Q^2 \alpha g_4 g_5 }{\pi\,  \left( 4 \mDM^2-m_\medz^2+ i m_\medz \Gamma_\medz\right)}\,.
\end{equation}
In addition, for vector DM, s-channel diagrams always lead to ${\cal B}_3={\cal B}_6= 0$.  For fermionic mediators, these results fully agree with those of Ref.~\cite{Jackson:2013pjq}.

We would like to discuss, as examples, the case of the Higgs and the $Z$ boson as s-channel mediators. They are very important not only because they arise in many DM models but also because we know their couplings to SM particles and consequently their contribution to the annihilation amplitudes can be calculated precisely.
\begin{itemize}
\item $\medz$ as the Higgs boson. If SM scalar doublet is given by
\begin{equation}
H = \begin{pmatrix} G^+ \\ \frac{v+h +i G^0}{\sqrt{2}}\end{pmatrix}\,,
\end{equation}
the relevant couplings $g_5$ are shown schematically in Fig~\ref{fig:HiggsCouplings}. We will use diagrams like this to represent couplings from now on.

\begin{figure}[t]
\includegraphics[scale=0.5]{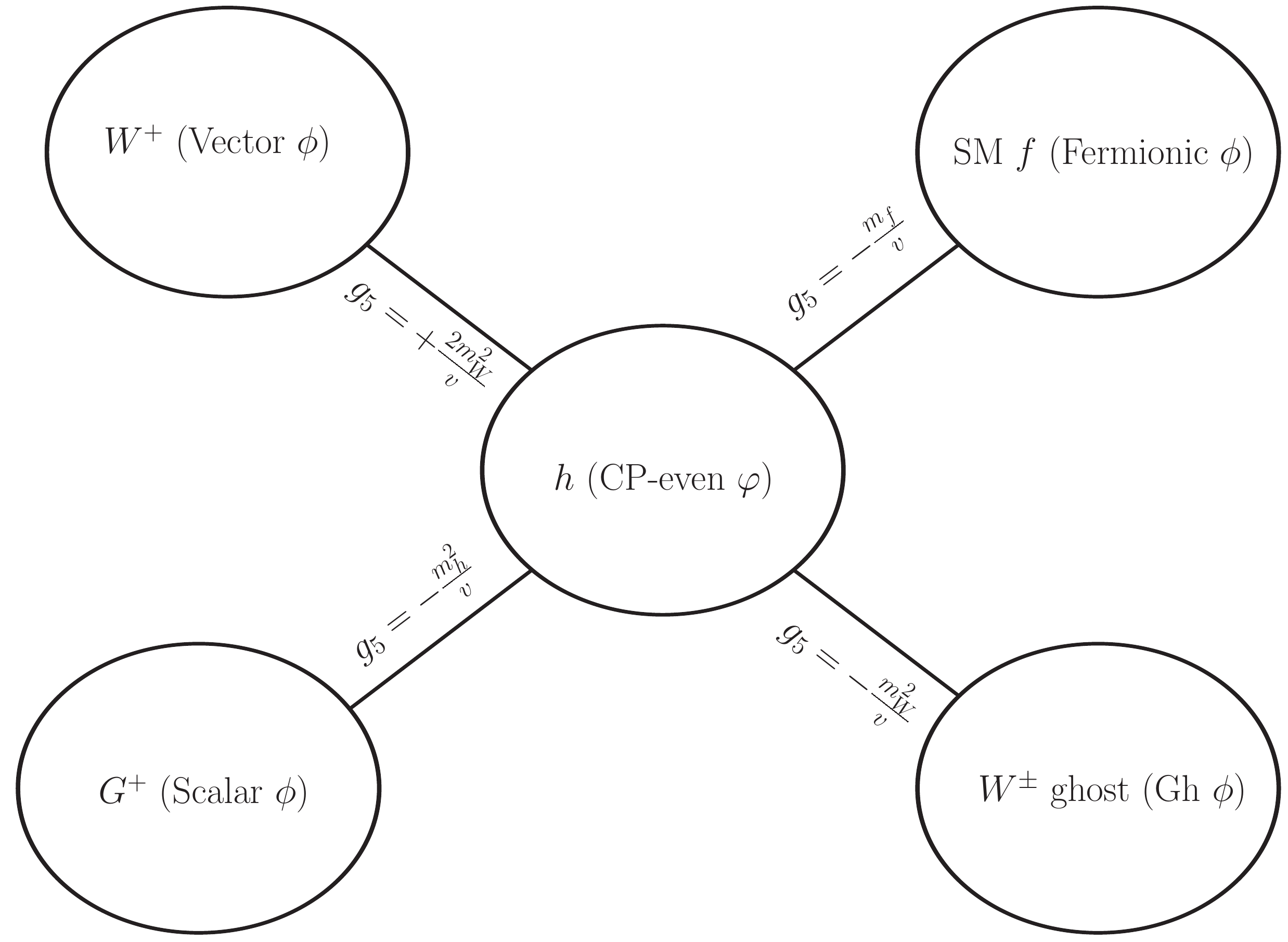}
\caption{Schematic representation of the couplings of the Higgs boson and the charged mediators present in the loop for $h\to \gamma\gamma$ as given in Table~\ref{table:Acal}. }
\label{fig:HiggsCouplings}
\end{figure}

For scalar DM annihilating into photons via the Higgs on the s-channel, we can compute the amplitude by plugging $g_5$ in Eq.~\eqref{eq:BforL45}. We find
\begin{eqnarray}
\label{eq:schHiggslike0}
{\cal B}^h_\text{SM} &=&\frac{g_4 \alpha}{\pi v}  \left[\underbrace{ m_h^2 \left( 1-r_W^2 f_W\right)}_{\text{loop of $G^+$} } \underbrace{-4 \mDM  \sum_f  Q_f^2\, N_f\, m_f r_f \left( 1- (r_f^2-1)f_f\right)}_{\text{loop of SM fermions} } \right.\\
&&\left.
+ \underbrace{8 m_W^2\left(1-(r_W^2-2)f_W \right)}_{\text{loop of $W^+$} } \underbrace{-2  m_W^2\left(1-r_W^2f_W \right)}_{\text{loop of ghosts} }\right]\frac{1}{4\mDM^2-m_h^2+i\Gamma_h m_h}\nonumber\,,
\end{eqnarray}

where we scale the fermions contribution with their electric charge and number of colors. If we define
\begin{align}
A^h_1(r_W) = -(2+3r_W^2)- 3(2-r_W^2)r_W^2 f_W & \,, & A^h_{1/2}(r_f) =  2\,r_f^2\left( 1-(r_f^2-1)f_f\right) \,,
\end{align}
and notice  that $ m_h^2 = -(4\mDM^2-m_h^2+i\Gamma_h m_h)+  4 \mDM^2+ {\cal O}(\alpha)$, Eq.~\eqref{eq:schHiggslike0} can be cast in a more compact form
\begin{eqnarray}
{\cal B}^h_\text{SM} &=&-\frac{2 \mDM^2g_4 \alpha \left[ \sum_f  Q_f^2\, N_f\,  A^h_{1/2}(r_f) + A^h_1(r_W)  \right]}{\pi v(4\mDM^2-m_h^2+i\Gamma_h m_h)}   - \frac{g_4 \alpha}{\pi v} \left( 1-r_W^2 f_W\right)\,.
\label{eq:schHiggslike}
\end{eqnarray}
In the following section, we will see that, in realistic models, the last term typically cancels with another one coming from the amplitude associated to ${\cal L}_3$. 

Regarding Majorana DM, the contribution to ${\cal B}$ involving the Higgs in the s-channel is zero, while for vector DM   ${\cal B}^h_2= {\cal B}^h_6=0$ and ${\cal B}^h_1$ is given by twice  ${\cal B}^h_\text{SM}$ of scalar DM~\cite{Birkedal:2006fz}.

Notice that  if the CP-even scalar $\medz$ is not the Higgs itself  but a neutral particle that mixes with the Higgs and inherits its couplings to the SM particles, we can use the previous expressions for calculating the decay amplitude up to a global factor (obviously, we must also add other possible contributions not present in the SM).

\item $\medz$ as the $Z$ boson. For scalar and vector DM, the amplitude vanishes. For Majorana DM, as explained above, the vector boson $Z$ itself does not contribute to the amplitude in the Landau gauge but we have to account for the contribution of its Goldstone boson, $G^0$, to the annihilation process. In that case, while  ghosts give zero, SM fermions running in the loop give  

\begin{eqnarray}
{\cal B}_\text{SM}^Z &=& \frac{4\sqrt{2}\,g_4\,\alpha\,\mDM^3  \sum_f \pm Q_f^2\, N_f\   r_f^2 f_f  }{\pi v(4\mDM^2-m_{G^0}^2+i\Gamma_{G^0} m_{G^0})} =\sqrt{2}\frac{\,g_4\alpha\mDM}{\pi v}\,  \sum_f \pm Q_f^2  N_f r_f^2 f_f  \,,
\label{eq:schZ}
\end{eqnarray}
\end{itemize}
where we took $g_5= \pm m_f/v$ with a negative sign for the charged leptons, the down, strange and bottom quarks, and a positive sign for the up, charm and top quarks. In the last equation we used the fact that the Goldstone boson is massless in the Landau gauge.

\section{Discussion}
\label{sec:discussion}

\subsection{Summary of the results}

Before discussing concrete examples, we would like to  summarize our findings. In Section~\ref{sec:II}, we proved that every amplitude form factor can be cast as

\begin{eqnarray}
{\cal B}&=&
\sum_{\medp,\DMp}\left(\raisebox{-.5\height}{\includegraphics[height=1.2cm]{t1r1}}+\nonumber
\raisebox{-.5\height}{\includegraphics[height=1.3cm]{t1r2}}+
\raisebox{-.5\height}{\includegraphics[height=1.3cm]{t1r3}}+
\raisebox{-.5\height}{\includegraphics[height=1.3cm]{t3r1}}
\right.\\&&
\underbrace{\left.+
\raisebox{-.5\height}{\includegraphics[height=1.3cm]{t3r2}}+
\raisebox{-.5\height}{\includegraphics[height=1.3cm]{t3r5}}+
\raisebox{-.5\height}{\includegraphics[height=1.3cm]{t3r6}}+
\raisebox{-.5\height}{\includegraphics[height=1.3cm]{t2r3}}\right)\hspace{60pt}}_{\text{Contribution of }{\cal L}_1\text{ given by Eq.~\eqref{eq:BforL1}}}\nonumber\\&&
+\underbrace{\sum_{\DMp}\left(
\raisebox{-.5\height}{\includegraphics[height=1.3cm]{t2r2}}+
\raisebox{-.5\height}{\includegraphics[height=1.4cm]{t3r4}}\right)}_{\text{Contribution of }{\cal L}_2\text{ given by Eq.~\eqref{eq:BforL2}}}+\underbrace{\sum_{\medp}\left(
\raisebox{-.5\height}{\includegraphics[height=1.3cm]{t2r1}}+
\raisebox{-.5\height}{\includegraphics[height=1.3cm]{t3r3}}\right)}_{\text{Contribution of }{\cal L}_3\text{ given by Eq.~\eqref{eq:BforL3}}}\nonumber\\&&
+\underbrace{\sum_{\medz,\medp}\left(
\raisebox{-.5\height}{\includegraphics[height=1.3cm]{t4r3}}+
\raisebox{-.5\height}{\includegraphics[height=1.3cm]{t5r1}}\right)}_{\text{Contribution of }{\cal L}_4+{\cal L}_5\text{ given by Eq.~\eqref{eq:BforL45}}}+\text{\footnotesize Permutations of the external legs.}
\label{eq:master}
\end{eqnarray}

Hence, in order to calculate the amplitude and obtain the cross section for DM annihilations into two photons, we have to add the contribution of each interaction. The general algorithm to do this is the following:
\begin{enumerate}
\item For spin-1 particles carrying electric charge,  use the non-linear Feynman gauge. For the neutral spin-1 bosons,  use the Landau gauge. 
\item Identify the charged particles that couple directly to DM. 
\item If they are charged under $Z_2$ (i.e. their type is $\DMp$), obtain ${\cal L}_2$ and the corresponding coupling $g_2$. This gives a contribution to the form factors equal Eq.~\eqref{eq:BforL2}.
\item If the charged particles are $Z_2$-even (i.e. their type is $\medp$), obtain ${\cal L}_3$ and the corresponding coupling $g_3$.  This gives a contribution to the form factors equal Eq.~\eqref{eq:BforL3}.
\item The interactions ${\cal L}_1$ involve two charged mediators directly coupled to DM, one is $Z_2$-even and the other one is $Z_2$-odd. Extract the couplings $g_1$.  The corresponding contribution to the form factors is given by Eq.~\eqref{eq:BforL1}. 
\item Identify the neutral \emph{scalar} particles $\medz$ that couple to DM. Obtain ${\cal L}_4$ and the corresponding coupling $g_4$. Then, determine the charged mediators to which the neutral particle couples to. This gives ${\cal L}_5$ and,  correspondingly, the coupling $g_5$. The total contribution of these particles to the form factors is given by Eqs.~\eqref{eq:BforL45}.
\item After the form factors have been determined, calculate the cross section by means of Eq.~\eqref{eq:cross} for scalar or Majorana DM, or Eq.~\eqref{eq:crossV} for vector DM.  
\end{enumerate}

We now discuss six different examples in concrete DM models. We will schematically represent  the corresponding mediators with figures like Fig.~\ref{fig:HiggsCouplings}. There, each mediator is  within in a ellipse that further encloses the  couplings associated to vertices where only DM and the mediator are involved (i.e. $g_2$, $g_3$ or $g_4$). In addition, if two different mediators are involved  in the same vertex, we join them with a line and write the corresponding coupling on it (i.e. $g_1$ or $g_5$).

\subsection{Concrete examples}

\subsubsection{Wino and Minimal DM}
Here we consider fermionic DM that belongs to a self-conjugate $SU(2)_L$ multiplet of dimension $N$ with no hypercharge. This sort of scenario includes Wino DM (for $N=3$), or quintuplet Minimal DM ($N=5$). In the first case, a stabilizing symmetry is needed and that is the role of R-parity in the MSSM. In the second case, an accidental symmetry protects the stability of DM at renormalizable level.  

\begin{figure}[t]
\includegraphics[scale=0.5]{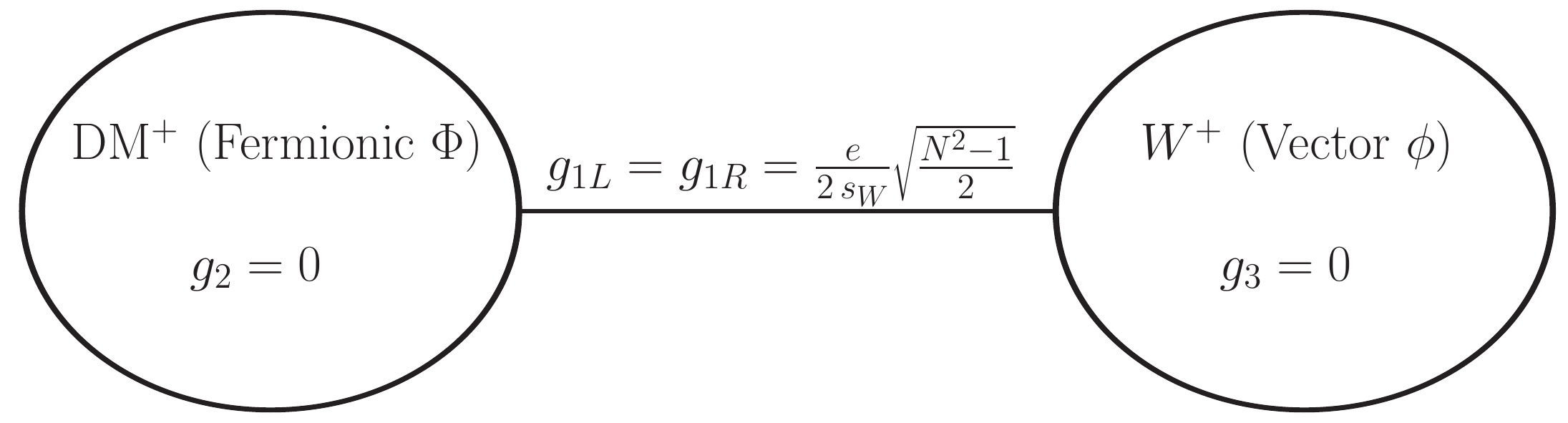}
\caption{Schematic representation of the mediators and the couplings for  Wino and Minimal DM.}
\label{fig:ex1}
\end{figure}

Here, the  only relevant interaction is given by the vertex ${\cal L}_1 \propto$ DM DM$^+ W^-$, where DM$^+$ is the fermion in the multiplet with charge $+e$.  The mediators and the corresponding couplings are shown in Fig.~\ref{fig:ex1}.  

Note that if no radiative correction is taken into account, we have $m_{\text{DM}^+} = \mDM$, i.e. $r_{\text{DM}^+} =1$. Plugging this and the couplings of Fig.~\ref{fig:ex1} in Eqs.~\eqref{eq:xsMFV}, we can calculate the coefficients in front of the Passarino--Veltman functions  for the form factor of Eq.~\eqref{eq:BforL1}. The corresponding  cross section is

\begin{align}
\sigma v =
\bigg|
&\dfrac{ r_W^4+2r_W^2-4 }{(1-r_W^2)(2-r_W^2)} C_0( 0, 1, -1, r_W ^2, r_W ^2, 1) 
+\dfrac{ r_W^2-2}{r_W^2\left(-1+r_W^2\right)}  C_0( 0, 1, -1, 1, 1, r_W ^2)\nonumber\\
&-4\left( \frac{r_W^2-1}{r_W^2-2}\right) \,\, C_0\left( 0, 4  , 0, r_W ^2, r_W ^2, r_W ^2\right)
- \frac{2}{r_W^2} \,\,  C_0\left( 0, 4  , 0, 1,1, 1 \right)%
\bigg|^2  \dfrac{\alpha^4(N^2-1)^2}{16\pi s_W^4 \mDM^2} \,.
\end{align}
For the Wino case, this equation agrees explicitly with Eq.~(22) of Ref.~\cite{Bern:1997ng}. Even though it can be simplified further in terms of dilogarithms, for the sake of illustration, we will only recast the cross section in the limit $\mDM\gg m_W$, that is, when $r_W \to 0$. To that end, notice that in that limit, each of Passarino--Veltman functions diverges at most logarithmically. Notice also that the coefficient in front of $C_0( 0, 1, -1, r_W ^2, r_W ^2, 1)$ and $C_0\left( 0, 4 \, , 0, r_W ^2, r_W ^2, r_W ^2\right)$ are finite in that limit, whereas those in front of $C_0( 0, 1, -1, 1, 1, r_W ^2)$ and $C_0\left( 0, 4 , 0, 1,1, 1 \right)$ diverge like $1/r_W^2$.  Hence, the latter Passarino--Veltman functions dominate the cross section in that limit. They give~\cite{Patel:2015tea}
\begin{align}
C_0\left( 0, 1 , -1 , 1, 1, r_W ^2\right) &\simeq -\frac{\pi^2}{8}+\frac{\pi r_W}{2}+{\cal O}(r_W^2)\,,& C_0\left( 0, 4 , 0, 1, 1, 1\right) &= -\frac{\pi^2}{8}\,.&
\end{align}
Using this, we find
$\sigma v = \pi\alpha^4(N^2-1)^2/16 m_W^2 s_W^4$, in agreement with Refs.~\cite{Bergstrom:1997fh,Cirelli:2005uq}.  The same expression can also be obtained by calculating  the Sommerfeld effect in the limit in which the potential is perturbative~\cite{Hisano:2006nn, Garcia-Cely:2015dda}.

\subsubsection{Scotogenic DM}

In this scenario~\cite{Ma:2006km,Kubo:2006yx}, there two types of fields charged under the $Z_2$ symmetry:  a scalar $H' = (H^+, \frac{1}{\sqrt{2}} (H^0 + iA^0) )^T$ with the same quantum numbers of the SM scalar doublet $H$, and a handful of  right-handed neutrinos $N_j$.  The interactions of the scalar  doublets are described by 
\begin{align}
\mathcal{L}= & 
\left(D_{\mu}H\right)^{\dagger}\left(D^{\mu}H\right)
+\left(D_{\mu}H'\right)^{\dagger}\left(D^{\mu}H'\right)
-\mu_1^2|H|^2-\mu_2^2|H'|^2 \nonumber\\
-&\lambda_1|H|^4-\lambda_2|H'|^4 -\lambda_3|H|^2|H'|^2 - \lambda_4 |H^{\dagger}H'|^2
-\frac{\lambda_5}{2}\left[\left(H^{\dagger}H'\right)^2+\text{h.c.}\right]\,.
\label{eq:LIDM}
\end{align}
For the right-handed neutrinos, the relevant interactions are with the SM lepton doublets $L_\beta$, which are given by
\begin{equation}
{\cal L} = \epsilon_{ab} h_{\beta j} \overline{N_j}P_{L}L^a_{\beta}H'^b\,.
\end{equation}

\begin{figure}[t]
\includegraphics[scale=0.5]{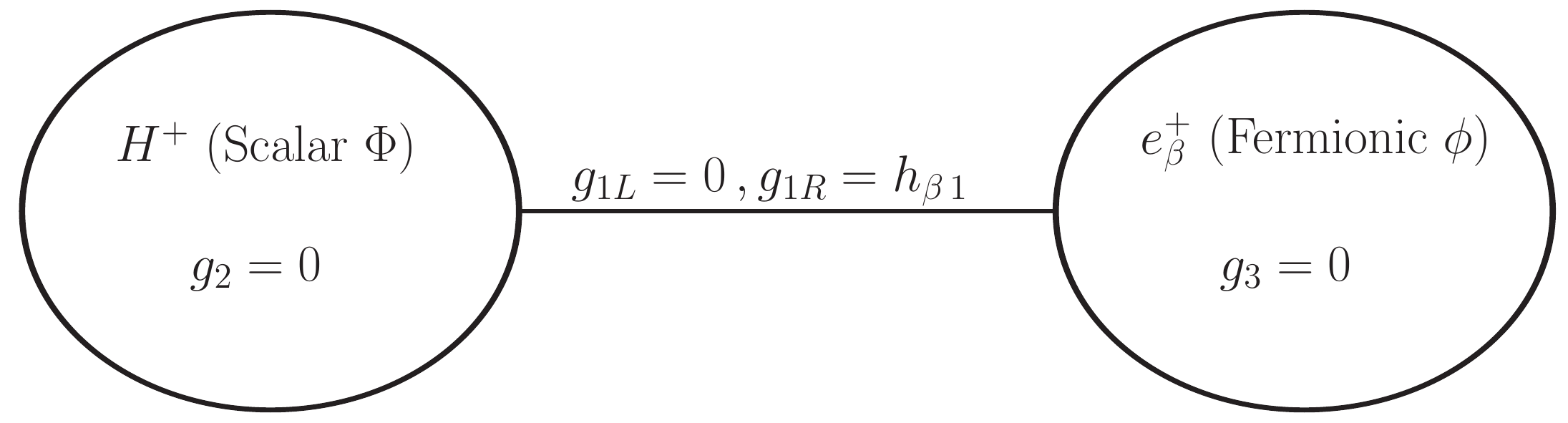}
\caption{Schematic representation of the mediators and the couplings for  Scotogenic DM.}
\label{fig:ex2}
\end{figure}

The DM candidate   is the lightest particle with $Z_2$ charge. 
If such particle is one of the right-handed neutrinos, DM can  annihilate into photons by means of one-loop diagrams containing a SM charged lepton and $H^+$. The corresponding couplings are shown  in Fig.~\ref{fig:ex2}, where DM was taken as $N_1$. Notice that there are no s-channel mediators.  Now, it is straightforward to calculate ${\cal B}$ by means of Eqs.~\eqref{eq:xsMFS} and \eqref{eq:BforL1}. The resulting cross section is
\begin{eqnarray}
\sigma v  &=& \frac{\alpha^2}{64 \pi^3 m_1^2 }\sum_{\beta} \frac{r_H^4 h_{\beta 1}^4}{ \left(r_{e_\beta}^2-r_H^2-1\right)^2 \left(r_{e_\beta}^2-r_H^2\right)^2}
\bigg|
\frac{r_{e_\beta}^2}{r_H^2} \left(1+r_{e_\beta}^2-r_H^2\right)C_0(0,1,-1,r_{e_\beta}^2,r_{e_\beta}^2,r_H^2)\nonumber\\
&&- (1-r_{e_\beta}^2+r_H^2)C_0(0,1,-1,r_H^2,r_H^2,r_{e_\beta}^2)+
\frac{2\,r_{e_\beta}^2(r_H^2-r_{e_\beta}^2)}{r_H^2} C_0(0,4,0,r_{e_\beta^2},r_{e_\beta}^2,r_{e_\beta}^2)\bigg|^2.
\end{eqnarray}

In the limit of $\mDM \gg m_{e_\beta^+}$, i.e. when $r_{e_\beta} \to 0$, this expression gives (e.g. Ref.~\cite{Garny:2015wea})
\begin{align}
\sigma v = & \frac{\alpha^2 h^4_{\beta 1}}{256 \pi^3 m_1^2} |2\, C_0(0,1,-1,r_H^2,r_H^2,0)|^2 
= \frac{\alpha^2 h^4_{\beta 1}}{256 \pi^3 m_1^2} \left|\text{Li}_2\left(\frac{1}{r_H^2}\right)-\text{Li}_2\left(-\frac{1}{r_H^2}\right)\right|^2\,,
\end{align}

\subsubsection{ Singlet scalar DM  }

\begin{figure}[t]
\includegraphics[scale=0.40]{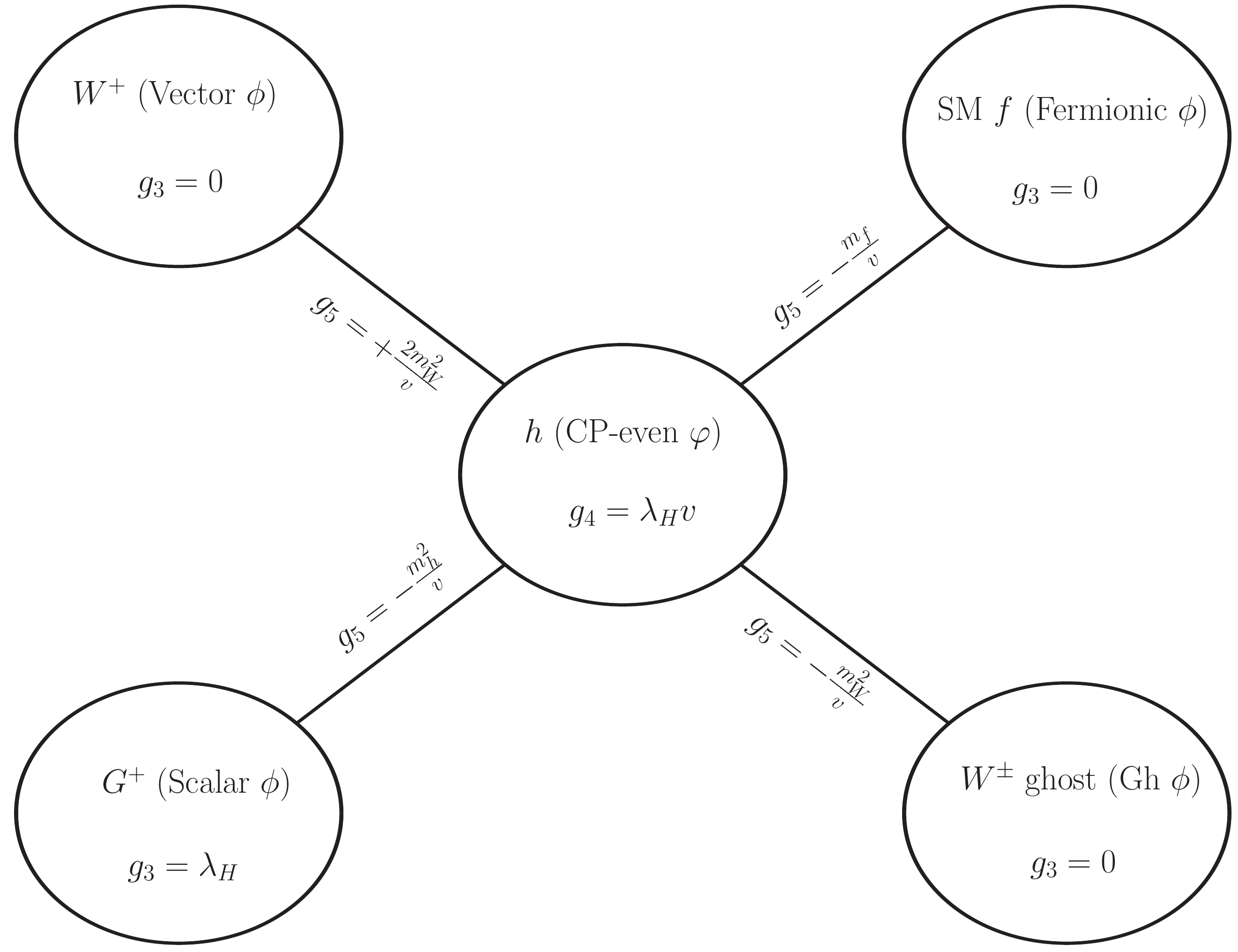}
\caption{Schematic representation of the mediators and the couplings for  singlet scalar DM. }
\label{fig:ex3}
\end{figure}

Suppose that DM is a scalar field $\phi$, which is singlet under $SU(2)_L$~\cite{Silveira:1985rk,McDonald:1993ex}. Then, the only non-trivial interaction of DM with the SM takes place via the so-called Higgs portal ${\cal L}= \lambda_H\, \DM^2 H^\dagger H \supset  \lambda_H \DM^2 ( G^+ G^- + v \,h) $. Hence,  there are five mediators, which are shown in Fig.~\ref{fig:ex3}. First, we have $G^+$, which  is involved in the ${\cal L}_3$ interaction. The corresponding contribution to the form factor can be computed with Eq.~\eqref{eq:BforL3}. Second, we have the Higgs boson, which acts as mediator on the s-channel.  The contribution of the Higgs boson was already calculated and reported in Eq.~\eqref{eq:schHiggslike}.   The total form factor is
\begin{eqnarray}
{\cal B}&=&\frac{\alpha\lambda_H}{\pi} \left(1-r_W^2 f_W\right)+
{\cal B}^h_\text{SM}\nonumber\\
&=& -\dfrac{2 \mDM^2\alpha\,\lambda_H }{\pi \left(4\mDM^2-m_h^2 + i \Gamma_h m_h\right)} 
\left(\sum_f N_f Q_f^2 A^h_{1/2}(r_f)
+A^h_{1}(r_W) \right)\,,
\end{eqnarray}
which, according to Eq.~\eqref{eq:cross}, corresponds to a cross section
\begin{eqnarray}
\sigma v= \dfrac{ \mDM^2\alpha^2\,\lambda_H^2 }{8\pi^3 \left( (4\mDM^2-m_h^2)^2 + m_h^2 \Gamma_h^2 \right)}
\bigg|\sum_f  Q_f^2\, N_f\,  A^h_{1/2}(r_f) + A^h(r_W)\bigg|^2\,.
\end{eqnarray}
This expression is in agreement with the results of the literature (see e.g.~\cite{Duerr:2015aka,Gunion:1989we}).

\subsubsection{Inert Higgs DM}
\begin{figure}[t]
\includegraphics[scale=0.40]{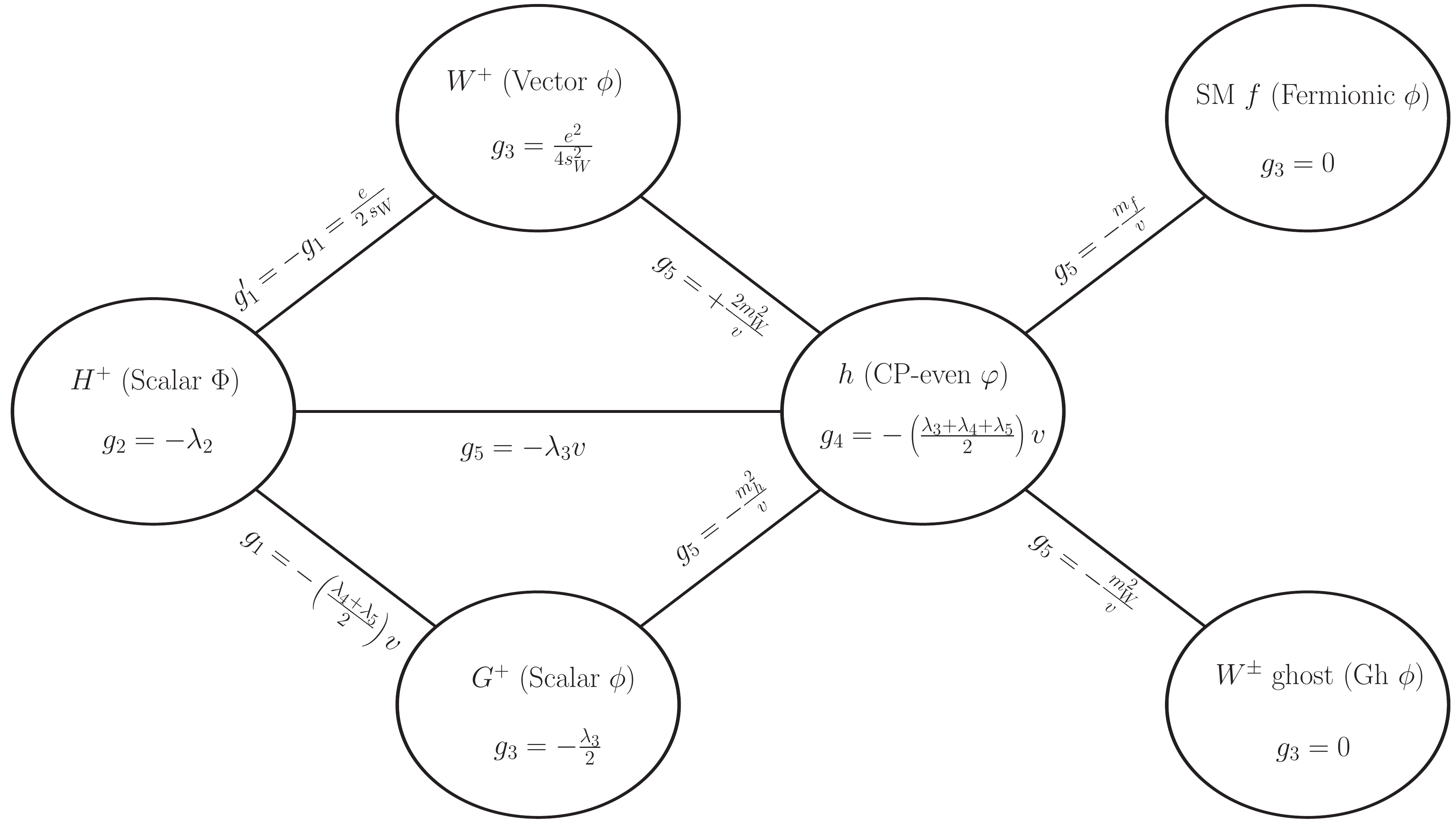}
\caption{Schematic representation of the mediators and the couplings for  inert Higgs DM. }
\label{fig:ex4}
\end{figure}

Suppose that we have an additional scalar doublet $H'$ which is charged under a $Z_2$ symmetry (like the Scotogenic model above, but without right-handed neutrinos).  Hence,  the relevant interactions are described by Eq.~\eqref{eq:LIDM} and the DM candidate is the lightest particle that is charged under $Z_2$. Without loss of generality, we assume this is the $H^0$ boson.

In this case, the calculation of the cross section is significantly more difficult than in the previous examples.  First, we have diagrams with the Higgs on the s-channel, receiving contributions not only from SM particles (computed already in  Eq.~\eqref{eq:schHiggslike0}) but also from the additional scalar $H^+$. Second, and more importantly, the direct interactions of DM with charged mediators 
-which give rise to diagrams with topologies 1, 2 and 3- are of two kinds. One of them is of scalar nature, in which the mediators are scalars $H^+$ and $G^+$; the other one is associated to the gauge interaction, whose charged mediators are $H^+$ again and the $W^+$ boson. 
All this is schematically represented in Fig.~\ref{fig:ex4}. 

The contribution of ${\cal L}_2$ and ${\cal L}_3$ to the form factor is given by Eqs.~\eqref{eq:BforL2} and \eqref{eq:BforL3}
\begin{eqnarray}
B\bigg|_{{\cal L}_2+{\cal L}_3} = \underbrace{-\frac{\alpha \lambda_2}{\pi} \left(1-r_H^2 f_H\right)}_{\text{Loop of } H^+} \underbrace{-\frac{\alpha \lambda_3}{2\pi} \left(1-r_W^2 f_W\right)}_{\text{Loop of } G^+}\underbrace{-\frac{4\pi \alpha^2}{s_W^2} \left(1- (r_W^2-2) f_W\right)}_{\text{Loop of } W^+}\,.
\end{eqnarray}

For the Higgs on the s-channel, we can use Eq.~\eqref{eq:schHiggslike} for the SM piece and add, by means of Eq.~\eqref{eq:BforL45}, the contribution associated to the additional charged particle. This is
\begin{eqnarray}
{\cal B}^h&=&{\cal B}^h_\text{SM }
-\dfrac{\alpha\lambda_3\lambda_h v^2}{\pi \left(4-r_h^2 + i r_h\Gamma_h/m_{H^0} \right)} \left(1-r_H^2 f_H \right) \\
&=& \left(
\dfrac{2 \pi\alpha \left( \sum_f N_f Q_f^2 A^h_{1/2}(r_f)
+A^h_{1}(r_W)\right)-s_W^2 r_W^2 \lambda_3(1-r_H^2 f_H)}{\pi^2\left(4-r_h^2 + i r_h \Gamma_h/m_{H^0}\right)} 
+\frac{\alpha\left(1- r_W^2 f_W\right)}{\pi} \right)\lambda_h\nonumber\,, 
\end{eqnarray}
with 
\begin{equation}
\lambda_h = \frac{1}{2} \left(\lambda_3+\lambda_4+\lambda_5\right)  = \frac{\lambda_3}{2} -\frac{\pi \alpha(r_H^2-1)}{s_W^2 r_W^2}\,.
\end{equation}
In the last line, we use $m_W = e\, v/2s_W$, and the fact that,  after electroweak symmetry breaking, the charged particle in the doublet and the DM candidate are not longer degenerate  in mass. Indeed, $m_{H^+}^2 = m_{H^0}^2-(\lambda_4+\lambda_5)v^2/2$. This shows that even though there are many masses and parameters, the form factor ${\cal B}$ depends only  on four unknown variables: $r_W, r_H, \lambda_2$ and $\lambda_3$. Notice that the first one is just the inverse of the DM mass in units of $m_W$. Using this, we can calculate the contribution of ${\cal L}_1$ by means of Eq.~\eqref{eq:BforL1}, with the corresponding coefficients extracted from Eqs.~\eqref{eq:xsSSS} and \eqref{eq:xsSSV}. 

Finally, putting everything together, we obtain

\begin{align}
\sigma v =&\Bigg|
\frac{r_H^2+3r_W^2-1}{r_W^2}+\frac{\tilde{\lambda}_2 s_W^2}{\pi \alpha}\left(1-r_H^2 f_H\right)-
\frac{2\lambda_h s_W^2}{\pi \alpha}\left(\dfrac{ \sum_f N_f Q_f^2 A^h_{1/2}(r_f)
+A^h_{1}(r_W)}{4-r_h^2 + i r_h \Gamma_h/m_{H^0}}  \right)
\nonumber\\
&+
\left(6-3r_W^2+  \frac{4r_W^2-4}{1+r_H^2-r_W^2}\right)f_W+\left(5-r_H^2+\frac{4r_H^2-4}{1+r_W^2-r_H^2}\right)\frac{r_H^2 f_H}{r_W^2}\nonumber\\
&-2
\left( 1+\frac{5 r_H^2-r_W^2+1}{(r_H^2-r_W^2)(1+r_H^2-r_W^2)}\right)
C_0(0,1,-1,r_W^2, r_W^2, r_H^2)
\nonumber\\
&-
\frac{2r_H^2}{r_W^2}\left(1+\frac{5 r_W^2- r_H^2 +1}{(r_W^2-r_H^2)(1+r_W^2-r_H^2)}\right)
C_0(0,1,-1,r_H^2, r_H^2, r_W^2)
\Bigg|^2 \frac{\alpha^4}{32\pi s_W^4 m_{H^0}^2},
\end{align}
with
\begin{equation}
\tilde{\lambda}_2  = \lambda_2 +\dfrac{s_W^2 r_W^2 \lambda_3\lambda_h}{\pi \alpha\left(4-r_h^2 + i r_h\Gamma_h/m_{H^0} \right)}\,.
\end{equation}
This result agrees with those of Refs.~\cite{Gustafsson:2007pc, GarciaCely:2014jha}, which were found numerically but not analytically. It also agrees with the cross section obtained with the Sommerfeld effect in the limit of perturbative potential~\cite{Garcia-Cely:2015khw}. 
\subsubsection{Singlet-doublet DM and Higgsinos}

\begin{figure}[t]
\includegraphics[scale=0.4]{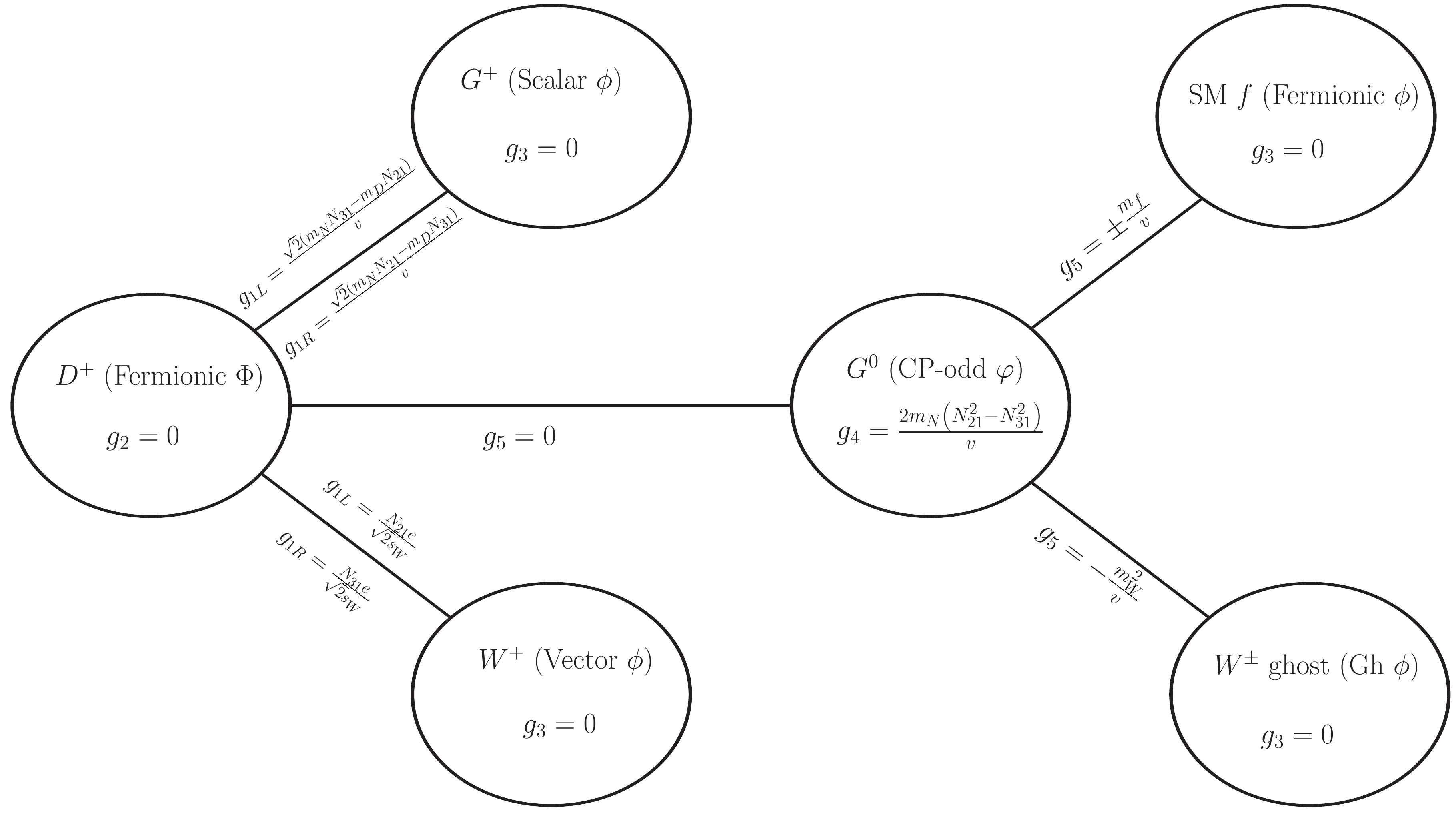}
\caption{Schematic representation of the mediators and the couplings for  singlet-doublet DM. }
\label{fig:ex5}
\end{figure}

In this case, the fields that are  charged under $Z_2$ are all chiral fermions, two of them which are $SU(2)_L$-doublets  with hypercharge $\pm1/2$ and one gauge singlet. 
After electroweak symmetry breaking, one charged fermion $D^+$ is obtained along with three Majorana particles. The lightest of the latter is the DM candidate, which we call $N$. The reader can find more  details of the model and its phenomenology in Refs.\cite{D'Eramo:2007ga,Abe:2014gua,Calibbi:2015nha, Restrepo:2015ura}. Here just mention that this DM candidate interacts with the SM gauge bosons (and its Goldstone bosons) by means of 
\begin{eqnarray}
{\cal L} &=& \frac{2m_N}{v}\left(N_{21}^2-N_{31}^2\right) G^0 \overline{N}\gamma_5 N+ \left(\dfrac{e}{2 s_W}  \overline{N}\slashed{W}^+ (N_{21} P_L + N_{31} P_R) D^-\right.\nonumber\\
&&\left. +\frac{\sqrt2}{v} G^+ \overline{N}\left((m_N N_{31}-m_D N_{21}) P_L +(m_N N_{21}-m_D N_{31}\right) P_R)D^-+ \text{h.c.}\right)\,.
\end{eqnarray}
where $N_{21}$ and $N_{31}$ are mixing parameters. Higgsino DM is a particular realization of this scenario when such parameters take specific values and the $Z_2$ symmetry corresponds to  R-parity. Furthermore, the pure Higgsino limit is the situation for which $N_{21}=N_{31}=1/\sqrt{2}$ and the charged particle and the DM have the same mass, i.e. $m_D = m_N$.  Notice that we omit the coupling to $Z$ boson, as it is not relevant in the Landau gauge. Also, note that the $D^+$ does not interact with the Goldstone boson, $G^0$. Such interaction is not renormalizable because it requires at least two scalar doublets and two fermionic doublets.

The calculation of the annihilation cross section is similar to that of the inert doublet model. First, we have diagrams with the $Z$ on the s-channel, which nevertheless only receive contributions from the SM fields (see Fig~\ref{fig:ex5}). Hence, we can use Eq.~\eqref{eq:schZ} directly. Second,  interactions of DM with charged mediators 
-which give rise interactions type ${\cal L}_1$- are between $D^+$ and $G^+$ or $W^+$. All this is summarized in Fig.~\ref{fig:ex5}.

With those couplings, we can use Eqs.~\eqref{eq:xsMFS} and \eqref{eq:xsMFV} to determine the coefficients in front of Passarino--Veltman functions in Eq.~\eqref{eq:BforL1}. The final result is 

\begin{eqnarray}
\sigma v  &=& \frac{\alpha^4 (N_{21}^2+N_{31}^2)^2}{16 \pi s_W^4 \mDM^2 } \Bigg|
 \frac{ (N_{21}^2-N_{31}^2) \sum_f\left( \pm Q_f^2  N_f r_f^2 f_f\right)}{  r_W^2 (N_{21}^2+N_{31}^2)}\,  
\nonumber\\
&&+\left(\frac{2 r_W^4-r_W^2 \left(r_D^2+1\right)-r_D^4+6 r_D^2-1}{\left(r_W^2-r_D^2-1\right) \left(r_W^2-r_D^2\right)}+\frac{12y  r_D }{r_W^2-r_D^2}\right)C_0(0,1,-1,r_W^2,r_W^2,r_D^2)\nonumber\\
&&-\frac{ r_D^2 \left(-2 r_W^4+r_W^2
   \left(r_D^2-3\right)+\left(r_D^2-1\right)^2\right)+12y r_D  r_W^2}{r_W^2 \left(r_W^2-r_D^2\right) \left(r_W^2-r_D^2+1\right)}C_0(0,1,-1,r_D^2,r_D^2,r_W^2)\nonumber\\
&&
+\frac{4  \left(r_W^2-1\right)}{r_W^2-r_D^2-1}f_W
-\frac{ r_D \left( r_D \left(2 r_W^2-r_D^2+1\right)+ y \left(-8 r_W^2+2 r_D^2-2\right)\right)}{r_W^2 \left(r_W^2-r_D^2+1\right)} f_D
\Bigg|^2\,,
\end{eqnarray}
with $y =N_{21}\,N_{31}/(N_{21}^2+N_{31}^2) $. The pure Higgsino limit, when $m_N \gg m_W$, gives $\sigma v= \pi \alpha^4/4 m_W^2 s_W^4 $ in agreement with Ref.~\cite{Rudaz:1989ij,Bern:1997ng}.

\subsubsection{Vector Kaluza-Klein DM }

\begin{figure}[t]
\includegraphics[scale=0.4]{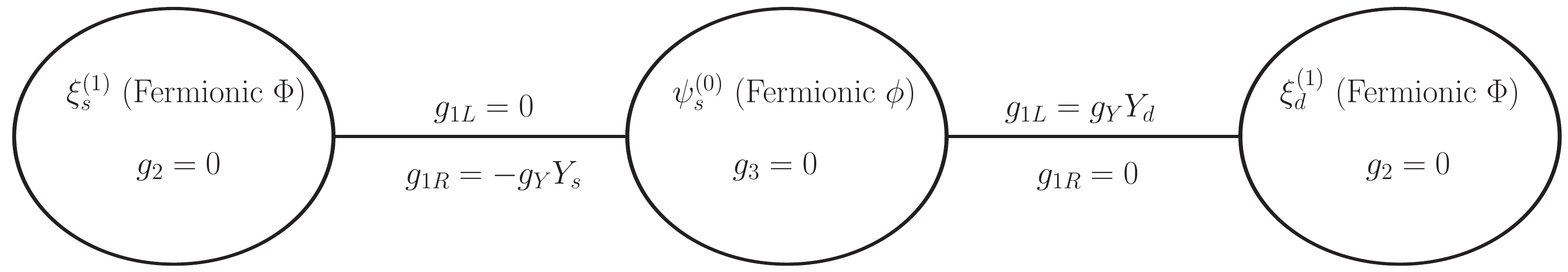}
\caption{Schematic representation of the mediators and the couplings for  Kaluza-Klein DM. }
\label{fig:ex6}
\end{figure}

The annihilation of the Kaluza-Klein (KK) DM particle $B^{(1)}$ into two photons was analyzed in the Ref.~\cite{Bergstrom:2004nr}. This is a nice example of vector DM where  the $Z_2$ symmetry corresponds to the KK parity (indicated as superscript). 
The particle content of this model consists of a zero level KK fermion $\psi^{(0)}$ as well as of  $\xi_s^{(1)}$ and $\xi_d^{(1)}$, which are its first singlet and doublet excitation with hypercharge $Y_s$ and $Y_d$, respectively.  The DM particle couples to the other fermions according to the Lagrangian
\begin{align}
\mathcal{L}^{'}= -g_Y Y_s B_{\mu}^{(1)}\bar{\xi}_s^{(1)}\gamma^{\mu}P_R\psi^{(0)}
+g_Y Y_d B_{\mu}^{(1)}\bar{\xi}_d^{(1)}\gamma^{\mu}P_L\psi^{(0)} + \text{h.c.}  \,,
\end{align}
which gives rise to the mediator classification of Fig.~\ref{fig:ex6}. Only interactions type ${\cal L}_1$ are present. Using Eqs.~\eqref{eq:xsVFF1},\eqref{eq:xsVFF2} and \eqref{eq:xsVFF3}  we can calculate the form factors ${\cal B}_1$, ${\cal B}_2$ and ${\cal B}_6$, respectively. Moreover, we can compute the corresponding cross section by means of Eq.~\eqref{eq:crossV}.
This calculation  was previously  performed in Ref.~\cite{Bergstrom:2004nr} for the case when the zero-level KK excitation $\psi^{(0)}$ is massless. Our results, valid for arbitrary masses, are in agreement with the expressions reported in that paper in the limit $m_\psi\to0$.

\section{Conclusions}
\label{sec:conclusions}

Gamma-ray lines produced in WIMP annihilations play a significant role for indirect DM searches because they stand out of the soft featureless background and no astrophysical process is known to produce them. A model-independent study of gamma-ray lines is nevertheless challenging  because the calculation of the corresponding cross sections crucially depend on multiple details of the underlying DM model. This work is a step towards such study.    

By means of a careful classification of the one-loop diagrams leading to DM annihilation into two photons, we have shown that for any model satisfying conditions \cond, the annihilation amplitude - and consequently, the cross section- can be calculated by just adding different expressions that we report in Eq.~\eqref{eq:master}.  Our results were summarized and exemplified in Sec.~\ref{sec:discussion}. We also provide
a Mathematica notebook where this is done~\footnote{\url{http://www.desy.de/~camilog/gamma-ray-lines.html}}. We find an agreement with previous works done in the context of popular DM models. 

A natural extension of this article is applying the same methods for calculating the annihilation cross sections associated  to the final states  $Z\gamma$ and $h\gamma$. In addition, one can consider going beyond the conditions \cond, for instance by calculating the annihilation cross section into two photons for Dirac or complex scalar DM. We leave this for a future work. 
\vspace{-0.6cm}
\section*{Acknowledgments}
We thank  Julian Heeck, Hiren Patel, Diego Restrepo and Mathias Garny for useful discussions.
CGC is supported by the IISN and the Belgian Federal Science Policy through the Interuniversity Attraction Pole P7/37 ``Fundamental Interactions''. AR is supported by COLCIENCIAS though the PhD fellowship 6172 and the Grant No.~111-565-842691, and by UdeA through the Grant of Sostenibilidad-GFIF.
AR is thankful for the hospitality of Universit\'e Libre de Bruxelles.
We acknowledge the use of \texttt{Package-X}~\cite{Patel:2015tea} and \texttt{JaxoDraw}~\cite{Binosi:2003yf}.

\appendix

\section{Gauge choice for vector boson mediators}
\label{sec:Gauge}

\subsection{Charged gauge bosons }

In order to calculate the annihilation amplitude, we assume that the underlying model meets conditions \cond. The third one in particular is not satisfied by $W^+$ boson in ordinary $R_\xi$ gauges, because of the presence of the interaction
\begin{equation}
\delta {\cal L} =  e M_W\, G^+ W^{-\mu} A_\mu+ \text{h.c.}\,
\label{eq:cubrxi}
\end{equation}
The solution to this problem is to work in a different gauge. The gauge fixing term in the ordinary Feynman gauge is given by ${\cal L}_\text{gf} =- f^*f$ with $f = \partial_\mu W^{+\mu} -  i m_W G^+$ . If we work instead with $f = \partial_\mu W^{+\mu} -  i  m_W G^+  + ie A_\mu W^{+\mu}$, we clearly cancel the interaction term in Eq.~\eqref{eq:cubrxi}. In fact, this procedure replaces such term by the following interactions between the $W$ bosons and the photons
\begin{equation}
\delta {\cal L}= 
 - e^2 A^\mu A^\nu  W^-_\mu W^+_\nu
+i  e A^\mu ( W^+_\mu \partial^\nu W^-_\nu-  W^-_\mu \partial^\nu W^+_\nu)\,.
\label{eq:terms1}
\end{equation}
This is the so-called Feynman non-linear gauge. The new gauge fixing term gives rise to the following interactions between the Faddeev--Popov ghosts associated to the $W^\pm$ boson and photons~\cite{Pasukonis:2007fu}
\begin{equation}
{\cal L} =
-ie A_\mu\left(\partial_\mu \overline{c}^-c^+-
\partial_\mu \overline{c}^+ c^-\right)
\underbrace{-ie A_\mu\left( \overline{c}^+\partial_\mu c^-
- \overline{c}^-\partial_\mu c^+\right)
- e^2 A_\mu A^\mu \left(\overline{c}^- c^++\overline{c}^+ c^-\right)}_\text{only present in the Feynman non-linear gauge}\,.
\label{eq:terms2}
\end{equation}
Even though the expressions reported here are those associated to the $W$ boson, they can be generalized to any charged gauge boson by rescaling the electric charge. Because of that, for arbitrary vector charged mediators $\medp$, we assume that terms like Eq.~\eqref{eq:cubrxi} are not present, and include Eq.~\eqref{eq:terms1} to their interactions with photons. Furthermore, we describe the corresponding ghosts by means of Eq.~\eqref{eq:terms2}.

\subsection{Neutral gauge bosons in the s-channel }
In this appendix, we show that when DM annihilates into two photons via a massive gauge boson in the s-channel, the corresponding
amplitude can be calculated by considering only the associated Goldstone boson in the Landau gauge. This has been used 
in Ref.~\cite{Moretti:2014rka} in order to calculate the contribution of the process $q\overline{q} \to Z^* \to \gamma \gamma$
to the SM background for a diphoton signal. Here we generalize their arguments to an arbitrary neutral gauge boson and apply them to DM annihilations.

Let us start by considering the off-shell decay of a vector particle into two photons $\medz^\rho (k) \to \gamma^\mu (q)\gamma^\nu (q')$. After stripping the polarization vectors, the most general  decay amplitude, compatible with Bose statistics and Lorentz invariance, is given by
\begin{eqnarray} 
{\cal M}^{\rho\mu\nu}=&&C_1\, (q^{\nu} g^{\mu\rho}+q'^{\mu} g^{\nu\rho})+C_2\, k^{\rho} g^{\mu\nu} +  C_3\, k^\rho q^{\nu}q'^{\mu}\nonumber\\
  +&&C_4\, \epsilon^{\rho\mu\nu\alpha}(q-q')_\alpha  +\left(  C_5 k^\rho  {\epsilon}^{ \mu \nu\alpha\beta} + C_6 (q^{\nu} {\epsilon}^{\mu\rho\alpha\beta} -q'^{\mu} {\epsilon}^{\nu\rho\alpha \beta})\,q_{\alpha} q'_{\beta}  \right)\,,
\end{eqnarray}
where $C_i$ are scalar functions.  This expression  can be simplified further in the center-of-mass frame. First, there the photons move with opposite three-momentum  and consequently their polarization vectors not only satisfy  $q\cdot \epsilon=0$ and $q'\cdot \epsilon'=0$ but also  $q'\cdot \epsilon=0$ and $q\cdot \epsilon'=0$. This makes $C_1$, $C_3$ and $C_6$ irrelevant once $M^{\rho\mu\nu}$ is contracted with the polarization vectors. In addition, for the same reason, $\{k,q-q',\epsilon,\epsilon'\}$ is an orthogonal basis in the center-of-mass frame, which can be used to prove that\footnote{ Notice that this  relation might not be true in an arbitrary frame because the photon polarization vectors are not true four-vectors (for instance, their zero component vanishes in any frame).} $\epsilon^{\rho\mu\nu\alpha} \epsilon^*_{\mu}\epsilon'^*_{\nu}(q-q')_\alpha  = - k^\rho  {\epsilon}^{ \mu \nu\alpha\beta} \epsilon^*_{\mu}\epsilon'^*_{\nu}q_{\alpha} q'_{\beta} /q\cdot q' $, and consequently that $C_4$ can be absorbed into $C_5$. We conclude that the amplitude is determined by
\begin{eqnarray} 
{\cal M}^{\rho\mu\nu}=& k^\rho \left( C_2\,  g^{\mu\nu} + C_5 {\epsilon}^{ \mu \nu\alpha\beta} \,q_{\alpha} q'_{\beta}  \right)\,.
\end{eqnarray}
This is just  a restatement of the Landau--Yang theorem~\cite{Landau,Yang:1950rg}. If the gauge boson is on its mass-shell, its polarization vector $\epsilon(k)$ satisfies $k \cdot \epsilon(k) =0$ and, according to the previous equation, the decay amplitude vanishes. Furthermore, on an arbitrary $R_\xi$ gauge (linear or not), the amplitude for the process $\DM\DM\to\medz^*\to\gamma\gamma$ is proportional to
\begin{align}
\left(g_{\sigma\rho}+\frac{ (\xi-1) k_\sigma k_\rho}{k^2-\xi m_\medz^2} \right) M^{\rho\mu\nu} \epsilon^*_{\mu}\epsilon'^*_{\nu}
=\xi \, k_\sigma \frac{\left(k^2- m_\medz^2\right)\left( C_2\,  g^{\mu\nu} + C_5 {\epsilon}^{ \mu \nu\alpha\beta} \,q_{\alpha} q'_{\beta}  \right)\epsilon^*_{\mu}\epsilon'^*_{\nu}}{k^2-\xi m_\medz^2}  \,.
\end{align}
When the vector particle is on-shell, this expression vanishes as expected from the Landau--Yang theorem. Most importantly, in the Landau gauge, $\xi=0$, the expression vanishes even off-shell.  
The decay of the gauge bosons into two photons is thus given only by the Goldstone boson contribution. Since the latter is a massless scalar, we can calculate the annihilation amplitude by applying the results presented in Sec.~\ref{sec:schannels}.

\section{Reduction of tensor integrals in the non-relativistic limit}
\label{sec:reduction}

Box diagrams in the annihilation amplitude such as  

\begin{equation}
\raisebox{-.5\height}{\includegraphics[height=1.5cm]{t1r1}}+
\raisebox{-.5\height}{\includegraphics[height=1.5cm]{t1r2}}+
\raisebox{-.5\height}{\includegraphics[height=1.5cm]{t1r3}}
\end{equation}
lead to the next four-point loop integrals~\cite{Passarino:1978jh}
\begin{align}
\label{eq:D-point-integral}
D_0;&D_{\mu};D_{\mu\nu};D_{\mu\nu\rho};D_{\mu\nu\rho\sigma}\left(k_1^2,(k_2-k_1)^2,(k_3-k_2)^2,(k_4-k_3)^2,k_2^2,(k_3-k_1)^2,m_1^2,m_2^2,m_3^2,m_4^2\right) \nonumber\\
&=\int \frac{d^Dl}{i \pi^2} \dfrac{ 1;l_{\mu};l_{\mu}l_{\nu};l_{\mu}l_{\nu}l_{\rho};l_{\mu}l_{\nu}l_{\rho}l_{\sigma} }{[l^2-m_1^2][(l+k_1)^2-m_2^2][(l+k_2)^2-m_3^2][(l+k_3)^2-m_4^2]} \\
&=D_0;D_{\mu};D_{\mu\nu};D_{\mu\nu\rho};D_{\mu\nu\rho\sigma} \left(1,2,3,4\right)  \,,
\end{align} 
where the $k_N$ are related to the external momenta $p_i$ as $k_N=\sum_{i=1}^N p_i$. In the original Passarino--Veltman schema, it is possible to reduce the four-point tensor integrals to scalar expressions. 
However,  such procedure is based on the assumption of independent external momenta $p_i$, which is not our case because the DM  legs have the same momentum $(\mDM,0,0,0)$ in the non-relativistic limit. Nevertheless, since there are three independent momenta,  we can still reduce the tensor four-point integrals to a linear combination of tensor three-point integrals, which can be reduced to scalar functions. For this, we closely follow the  algebraic reduction 
of Refs.~\cite{Stuart:1987tt, Stuart:1989de, Stuart:1994xf}, but expanding Eq.~\eqref{eq:D-point-integral} in terms of the momenta $k_i$ instead of the external momenta $p_i$.

As an example, we consider  the scalar reduction of the tensor $D_{\mu}$. Very schematically, we have
{\small
\begin{align}
\label{eq:du}
D_{\mu}& = \int \dfrac{d^Dl}{i \pi^2} \dfrac{ l_{\mu} }{[l^2-m_1^2][(l+k_1)^2-m_2^2][(l+k_2)^2-m_3^2][(l+k_3)^2-m_4^2]}
=\int \frac{d^Dl}{i \pi^2} \dfrac{ l_{\mu} }{[1][2][3][4]} \nonumber \\
=&\alpha_{123}\int \frac{d^Dl}{i \pi^2} \dfrac{ l_{\mu} }{[1][2][3]} 
+\alpha_{124}\int \frac{d^Dl}{i \pi^2} \dfrac{ l_{\mu} }{[1][2][4]} 
+\alpha_{134}\int \frac{d^Dl}{i \pi^2} \dfrac{ l_{\mu} }{[1][3][4]} 
+\alpha_{234}\int \frac{d^Dl}{i \pi^2} \dfrac{ l_{\mu} }{[2][3][4]}\nonumber \\
=& \alpha_{123}C_{\mu}(1,2,3)+\alpha_{124}C_{\mu}(1,2,4)+\alpha_{134}C_{\mu}(1,3,4) +\alpha_{234}\left(C_{\mu}(2,3,4)-k_{1\mu}C_{0}(2,3,4)\right)\,,
\end{align}
}
where we did the substitution $l+k_1\rightarrow l^{'}$ in the last integral to cast it in the canonical form of a three-point function. Also, the coefficients $\alpha_{ijk}$ can be obtained by solving the system~\cite{Stuart:1987tt}
\begin{align}
\begin{pmatrix}
1 & 1 & 1 & 1 \\
0 & p_1^2 & (p_1^2-p_2^2+p_5^2)/2  & (p_1^2+p_4^2-p_6^2)/2\\
0 & (-p_1^2-p_2^2+p_5^2)/2 & (-p_1^2+p_2^2+p_5^2)/2  & (-p_1^2-p_3^3+p_5^2+p_6^2)/2\\
 -m_1^2 & p_1^2-m_2^2 & p_5^2-m_3^2 & p_4^2-m_4^2
\end{pmatrix}
\begin{pmatrix}
\alpha_{234} \\ \alpha_{134} \\ \alpha_{124} \\ \alpha_{123}
\end{pmatrix}=
\begin{pmatrix}
0 \\ 0 \\ 0 \\ 1
\end{pmatrix}\,,
\end{align}
with $p_5=p_1+p_2$ and $p_6=p_2+p_3$~.   The three-point tensor integrals can now be reduced to scalar integrals
\begin{align}
D_{\mu} =&\alpha_{123}\left(k_{1\mu}C_1(1,2,3)+k_{2\mu}C_2(1,2,3)\right)
+\alpha_{124}\left(k_{1\mu}C_1(1,2,4)+k_{3\mu}C_2(1,2,4)\right)\nonumber \\
&+\alpha_{134}\left(k_{2\mu}C_1(1,3,4)+k_{3\mu}C_2(1,3,4)\right)\nonumber \\
&+\alpha_{234}\left((k_{2\mu}-k_{1\mu})C_1(2,3,4)+(k_{3\mu}-k_{1\mu})C_2(2,3,4)-k_{1\mu}C_0(2,3,4)\right)\,.
\end{align} 
This expression must be compared against the defining expression for the scalar functions $D_{\mu}=k_{1\mu}D_1+k_{2\mu}D_2+k_{3\mu}D_3$, which leads to
\begin{align}
\label{eq:Di-reduction}
D_1 =&\, \alpha_{123}C_1  + \alpha_{124}C_1                    - \alpha_{234}\left(C_0+C_1+C_2\right)\nonumber \\
D_2 =&\, \alpha_{123}C_2                     + \alpha_{134}C_1 + \alpha_{234}C_1\nonumber \\
D_3 =&\,                    \alpha_{124}C_2  + \alpha_{134}C_2 + \alpha_{234}C_2\,,
\end{align}
with $\alpha_{ijk}C_l\equiv \alpha_{ijk}C_l(i,j,k)$. A similar reduction can be applied to the scalar integral $D_0$, which gives rise to
\begin{align}
D_0 =&\,  \alpha_{123}C_0 + \alpha_{124}C_0  + \alpha_{134}C_0 + \alpha_{234}C_0\,. 
\end{align}

\section{Coefficients in the Passarino--Veltman reduction of  the amplitudes}
\label{sec:xs}

This appendix reports the coefficients in Eq.~\eqref{eq:BforL1} according to the spin of the  particles involved in the one-loop diagram. 

\paragraph*{Scalar  $\DM$\\}

For the scalar case we always find $x_6= x_7 =x_8 =0$. The  non-zero coefficients for each mediator combination are:
\begin{itemize}[leftmargin=*]
\item Scalar $\DMp$ and scalar $\medp$
\begin{align}
x_1&=0 \,,&
x_2 &= x_4 = \frac{2 \rmed^2 g_1^2}{\mDM^2}  \,,&
x_3 =x_5 &= x_2\Big|_{\rmed\leftrightarrow \rDMp}\,.
\label{eq:xsSSS}
\end{align}
\item Fermionic $\DMp$ and fermionic $\medp$
\begin{align}
x_1=& 2 \left(g_{1L}^2+g_{1R}^2\right)\,,\nonumber\\
x_2=&2  \rmed \left( \rmed
   \left(1-\rDMp^2-\rmed^2\right)\left(g_{1L}^2+g_{1R}^2\right)+4  \rDMp \left(1-\rmed^2\right)g_{1L} g_{1R}\right)\,, & 
x_3 &= x_2\Big|_{\rmed\leftrightarrow \rDMp}\,,
\nonumber\\   
x_4=&4  \left(1-\rmed^2\right) \rmed \left( \rmed \left(g_{1L}^2+g_{1R}^2\right)+2 \rDMp\, g_{1L} g_{1R} \right)\,,&
x_5&=x_4\Big|_{\rmed \leftrightarrow\rDMp}\,.
\label{eq:SFF-xi}
\end{align}
\item Scalar $\DMp$ and vector $\medp$
\begin{align}
x_1=&  g_1 ^2 \nonumber\\
x_2=&  2 \left(-\rmed^2+2 \rDMp^2\right) g_1'^2+2 g_1 \left(\left(1-\rmed^2\right) \rmed^2+\left(-2+\rmed^2\right) r_{\DMp }^2-2 \rDMp^2\right) g_1'+2 g_1^2 \left(2-\rmed^2\right) \rDMp^2  \nonumber\\
x_3=&  2 g_1'^2 \rDMp^2-2 g_1 g_1' \left(\rmed^2-\rDMp^2+3\right) \rDMp^2+2 g_1^2
   \left(2-\rDMp^2\right) \rDMp^2 \nonumber\\
x_4=&-2 \left(\rmed^4-3 \rmed^2+2\right) g_1^2-8 g_1' \left(\rmed^2-1\right) g_1-2 g_1'^2 \left(2-\rmed^2\right) \nonumber\\
x_5=& 2 g_1'^2 \rDMp^2-4 g_1 g_1' \rDMp^2-2 g_1^2 \left(\rDMp^2-2\right) \rDMp^2\,.
\label{eq:xsSSV}
\end{align}
\end{itemize}

\paragraph*{Majorana $\DM$\\}

In this case,  we always obtain $x_1=x_6= x_7 =x_8 =0$. As for the non-zero coefficients, they are listed in the following according the mediators in each diagram. 
\begin{itemize}[leftmargin=*]
\item Fermionic $\DMp$ and scalar $\medp$
\begin{align}
\label{eq:xsMFS}
x_2=& \sqrt{2}  \rmed^2 (-1+\rmed^2-\rDMp^2)(g_{1L}^2+g_{1R}^2)\,,\nonumber
\hspace{15pt}
x_5= -2\sqrt{2}\rDMp \left(\rDMp(g_{1L}^2+g_{1R}^2)+2g_{1L}g_{1R}\right) \,,&
\nonumber\\
 x_3=&  \sqrt{2} \rDMp^2 (-1+\rmed^2-\rDMp^2) (g_{1L}^2+g_{1R}^2)+4\sqrt{2}  \rDMp (\rmed^2-\rDMp^2) g_{1L}g_{1R}\,,\hspace{20pt}   
x_4=0\,.
\end{align}

\item Scalar $\DMp$ and fermionic $\medp$. We find that the expressions are the same as the ones of previous case after doing $x_2 \leftrightarrow x_3,  x_4 \leftrightarrow x_5$ with $\rmed \leftrightarrow \rDMp$. Such behavior can be directly inferred from the Lagrangians in Table~\ref{table:Lagrangian} and Eq.~\eqref{eq:BforL1}.
\item Fermionic $\DMp$ and vector $\medp$
\begin{align}
x_2=& 2\sqrt{2}  \left( (\rmed^4+4\,\rDMp^2-\rmed^2(1+\rDMp^2)) (g_{1L}^2+g_{1R}^2)-8 \rDMp(1-\rmed^2+\rDMp^2 ) g_{1L}g_{1R}  \right)\,,\nonumber\\   
 x_3=& -2\sqrt{2} \left(\rDMp^2(-3-\rmed^2+\rDMp^2) (g_{1L}^2+g_{1R}^2)+8\rDMp g_{1L}g_{1R}\right)\,,\label{eq:xsMFV}\\  
x_4=& 8\sqrt{2} (-1+\rmed^2)(g_{1L}^2+g_{1R}^2)\,, \hspace{20pt}
x_5=  4\sqrt{2} \rDMp \left(\rDMp(g_{1L}^2+g_{1R}^2)-4 g_{1L}g_{1R}\right) \nonumber\,.
\end{align}
\end{itemize}

\paragraph*{Vector  $\DM$\\} 
 
In this case, we find $x_3 = x_2\Big|_{\rmed\leftrightarrow \rDMp}$,  
$x_5=x_4\Big|_{\rmed \leftrightarrow\rDMp}$ and $x_7= x_6\Big|_{\rmed\leftrightarrow \rDMp}$ in all cases. In addition 
\begin{itemize}[leftmargin=*]
\item Scalar $\DMp$ and scalar $\medp$  
\begin{eqnarray}
{\cal B}_1&:&
\left\{
\begin{array}{ll}
x_1= -\frac{2 \left(r_{\DMp }^2 \left(\log \left(\frac{r_{\DMp
   }^2}{r_{\medp }^2}\right)-4\right)+2 \left(\log \left(\frac{r_{\DMp }^2}{r_{\medp
   }^2}\right)+2\right)+4 r_{\medp }^2\right)}{3 \left(r_{\medp }^2-r_{\DMp
   }^2+1\right)} g_1^2\\
x_2=   4 \rmed^4 g_1^2\\
x_4=  4 \rmed^4  g_1^2 \\
x_6=  \frac{2 \left(r_{\medp }^2+2\right)}{3(1-r_{\medp}^2+r_{\DMp}^2)}g_1^2 \\
x_8= -g_1^2 
\end{array}
\right.\\
{\cal B}_2&:&
\left\{
\begin{array}{ll}
x_1=&  \frac{ \left(r_{\DMp }^2 \left(5 \log \left(\frac{r_{\DMp
   }^2}{r_{\medp }^2}\right)+4\right)-2 \left(\log \left(\frac{r_{\DMp }^2}{r_{\medp
   }^2}\right)+2 r_{\medp }^2+2\right)\right)}{3 \left(r_{\medp }^2-r_{\DMp
   }^2+1\right)} g_1^2\\
x_2=&  -2  r_{\medp }^2 \left(-r_{\medp }^2 \left(4 r_{\DMp }^2+1\right)+2 r_{\medp }^4+2 \left(r_{\DMp
   }^4+r_{\DMp }^2\right)\right)g_1^2\\
x_4=& - 2 \rmed^4 g_1^2 \\
x_6=&  \frac{\left(2-5 r_{\medp }^2\right)}{3(1-r_{\medp}^2+r_{\DMp}^2)} g_1^2\\
x_8=& \frac{g_1^2}{2}
\end{array}
\right.\\
{\cal B}_6&:&
\left\{
\begin{array}{ll}
x_1=&  \frac{ \left(-r_{\DMp }^2 \left(13 \log \left(\frac{r_{\DMp
   }^2}{r_{\medp }^2}\right)+2\right)-2 \log \left(\frac{r_{\DMp }^2}{r_{\medp
   }^2}\right)+2 r_{\medp }^2+2\right)}{3 \left(r_{\medp }^2-r_{\DMp }^2+1\right)}  g_1^2\\
x_2=& 2 \rmed^2 (3 \rmed^4+3\rmed^2-6 \rmed^2\rDMp^2-2\rDMp^2+3\rDMp^4-1) g_1^2\\
x_4=& 2 \rmed^2(4+\rmed^2) g_1^2 \\
x_6=& \frac{\left(13 r_{\medp }^2+2\right)}{3(1-r_{\medp}^2+r_{\DMp}^2)} g_1^2\\
x_8=& -\frac{5}{2} g_1^2
\end{array}
\right.
\end{eqnarray}
\item Fermionic $\DMp$ and fermionic $\medp$
\begin{eqnarray}
\label{eq:xsVFF1}
{\cal B}_1&:&
\left\{
\begin{array}{ll}
x_1=& \frac{2 \left(g_{1L}^2+g_{1R}^2\right) \left(r_{\DMp }^2
   \left(\log \left(\frac{r_{\DMp }^2}{r_{\medp }^2}\right)+2\right)-4 \log
   \left(\frac{r_{\DMp }^2}{r_{\medp }^2}\right)-2 r_{\medp }^2-2\right)}{3
   \left(r_{\medp }^2-r_{\DMp }^2+1\right)} \\
x_2=& 4 r_{\medp } \left(\left(g_{1L}^2+g_{1R}^2\right) r_{\medp } \left(r_{\DMp }^2+1\right)-4 g_{1L} g_{1R} \left(r_{\medp }^2-1\right) r_{\DMp }\right) \\
x_4=&4 \rmed \left( \rmed^3(g_{1L}^2+g_{1R}^2)+ 4 \rDMp (1-\rmed^2) g_{1L}g_{1R}\right) \\
x_6=&  - \frac{2\left(r_{\medp }^2-4\right)}{3(1-r_{\medp}^2+r_{\DMp}^2)}  \left(g_{1L}^2+g_{1R}^2\right)  \\
x_8=& g_{1L}^2+g_{1R}^2
\end{array}
\right.\\
\label{eq:xsVFF2}
{\cal B}_2&:&
\left\{
\begin{array}{ll}
x_1=&  \frac{\left(g_{1L}^2+g_{1R}^2\right) \left(-r_{\DMp }^2 \left(5
   \log \left(\frac{r_{\DMp }^2}{r_{\medp }^2}\right)+4\right)-4 \log
   \left(\frac{r_{\DMp }^2}{r_{\medp }^2}\right)+4 r_{\medp }^2+4\right)}{3
   \left(r_{\medp }^2-r_{\DMp }^2+1\right)} \\
x_2=& 2 r_{\medp } \left(g_{1L}^2+g_{1R}^2\right) \left(r_{\medp }^3 \left(1-4 r_{\DMp }^2\right)+2 r_{\medp } r_{\DMp }^4+2 r_{\medp }^5\right)\\
&+ 8 g_{1L} g_{1R} r_{\DMp }r_{\medp } \left(-r_{\medp }^2+r_{\DMp }^2+1\right)\\
x_4=&  2 \rmed^2(2+\rmed^2) (g_{1L}^2+g_{1R}^2) \\
x_6=& -\frac{\left(5 r_{\medp }^2+4\right) }{3(1-r_{\medp}^2+r_{\DMp}^2)} \left(g_{1L}^2+g_{1R}^2\right) \\
x_8=& \frac{1}{2} \left(-g_{1L}^2-g_{1R}^2\right)
\end{array}
\right.\\
\label{eq:xsVFF3}
{\cal B}_6&:&
\left\{
\begin{array}{ll}
x_1=&  \frac{\left(g_{1L}^2+g_{1R}^2\right) \left(r_{\DMp }^2 \left(13
   \log \left(\frac{r_{\DMp }^2}{r_{\medp }^2}\right)+2\right)-4 \log
   \left(\frac{r_{\DMp }^2}{r_{\medp }^2}\right)-2 r_{\medp }^2-2\right)}{3
   \left(r_{\medp }^2-r_{\DMp }^2+1\right)} \\
x_2=& -2\left(g_{1L}^2+g_{1R}^2\right) r_{\medp }^2 \left(r_{\medp }^2 \left(1-6
   r_{\DMp }^2\right)+3 r_{\medp }^4+3 r_{\DMp }^4-1\right)\\
&+ 8 r_{\medp }  g_{1L} g_{1R} r_{\DMp } \left(-r_{\medp }^2+r_{\DMp }^2+1\right)\\
x_4=& -2 \rmed^2(2+\rmed^2) (g_{1L}^2+g_{1R}^2) \\
x_6=& \frac{\left(4-13 r_{\medp }^2\right)}{3(1-r_{\medp}^2+r_{\DMp}^2)} \left(g_{1L}^2+g_{1R}^2\right)  \\
x_8=& \frac{5}{2} \left(g_{1L}^2+g_{1R}^2\right)
\end{array}
\right.
\end{eqnarray}

\end{itemize}

\bibliographystyle{apsrev}
\bibliography{text}

\begin{thebibliography}{99}
\expandafter\ifx\csname natexlab\endcsname\relax\def\natexlab#1{#1}\fi
\expandafter\ifx\csname bibnamefont\endcsname\relax
  \def\bibnamefont#1{#1}\fi
\expandafter\ifx\csname bibfnamefont\endcsname\relax
  \def\bibfnamefont#1{#1}\fi
\expandafter\ifx\csname citenamefont\endcsname\relax
  \def\citenamefont#1{#1}\fi
\expandafter\ifx\csname url\endcsname\relax
  \def\url#1{\texttt{#1}}\fi
\expandafter\ifx\csname urlprefix\endcsname\relax\def\urlprefix{URL }\fi
\providecommand{\bibinfo}[2]{#2}
\providecommand{\eprint}[2][]{\url{#2}}

\bibitem[{\citenamefont{Ade et~al.}(2016)}]{Ade:2015xua}
\bibinfo{author}{\bibfnamefont{P.~A.~R.} \bibnamefont{Ade}}
  \bibnamefont{et~al.} (\bibinfo{collaboration}{Planck}),
  \bibinfo{journal}{Astron. Astrophys.} \textbf{\bibinfo{volume}{594}},
  \bibinfo{pages}{A13} (\bibinfo{year}{2016}), \eprint{1502.01589}.

\bibitem[{\citenamefont{Bergström}(2000)}]{Bergstrom:2000pn}
\bibinfo{author}{\bibfnamefont{L.}~\bibnamefont{Bergström}},
  \bibinfo{journal}{Rept. Prog. Phys.} \textbf{\bibinfo{volume}{63}},
  \bibinfo{pages}{793} (\bibinfo{year}{2000}), \eprint{hep-ph/0002126}.

\bibitem[{\citenamefont{Bertone et~al.}(2005)\citenamefont{Bertone, Hooper, and
  Silk}}]{Bertone:2004pz}
\bibinfo{author}{\bibfnamefont{G.}~\bibnamefont{Bertone}},
  \bibinfo{author}{\bibfnamefont{D.}~\bibnamefont{Hooper}}, \bibnamefont{and}
  \bibinfo{author}{\bibfnamefont{J.}~\bibnamefont{Silk}},
  \bibinfo{journal}{Phys. Rept.} \textbf{\bibinfo{volume}{405}},
  \bibinfo{pages}{279} (\bibinfo{year}{2005}), \eprint{hep-ph/0404175}.

\bibitem[{\citenamefont{Bringmann and Weniger}(2012)}]{Bringmann:2012ez}
\bibinfo{author}{\bibfnamefont{T.}~\bibnamefont{Bringmann}} \bibnamefont{and}
  \bibinfo{author}{\bibfnamefont{C.}~\bibnamefont{Weniger}},
  \bibinfo{journal}{Phys. Dark Univ.} \textbf{\bibinfo{volume}{1}},
  \bibinfo{pages}{194} (\bibinfo{year}{2012}), \eprint{1208.5481}.

\bibitem[{\citenamefont{Bringmann et~al.}(2012)\citenamefont{Bringmann, Huang,
  Ibarra, Vogl, and Weniger}}]{Bringmann:2012vr}
\bibinfo{author}{\bibfnamefont{T.}~\bibnamefont{Bringmann}},
  \bibinfo{author}{\bibfnamefont{X.}~\bibnamefont{Huang}},
  \bibinfo{author}{\bibfnamefont{A.}~\bibnamefont{Ibarra}},
  \bibinfo{author}{\bibfnamefont{S.}~\bibnamefont{Vogl}}, \bibnamefont{and}
  \bibinfo{author}{\bibfnamefont{C.}~\bibnamefont{Weniger}},
  \bibinfo{journal}{JCAP} \textbf{\bibinfo{volume}{1207}}, \bibinfo{pages}{054}
  (\bibinfo{year}{2012}), \eprint{1203.1312}.

\bibitem[{\citenamefont{Weniger}(2012)}]{Weniger:2012tx}
\bibinfo{author}{\bibfnamefont{C.}~\bibnamefont{Weniger}},
  \bibinfo{journal}{JCAP} \textbf{\bibinfo{volume}{1208}}, \bibinfo{pages}{007}
  (\bibinfo{year}{2012}), \eprint{1204.2797}.

\bibitem[{\citenamefont{Su and Finkbeiner}(2012)}]{Su:2012ft}
\bibinfo{author}{\bibfnamefont{M.}~\bibnamefont{Su}} \bibnamefont{and}
  \bibinfo{author}{\bibfnamefont{D.~P.} \bibnamefont{Finkbeiner}}
  (\bibinfo{year}{2012}), \eprint{1206.1616}.

\bibitem[{\citenamefont{Geringer-Sameth and
  Koushiappas}(2012)}]{GeringerSameth:2012sr}
\bibinfo{author}{\bibfnamefont{A.}~\bibnamefont{Geringer-Sameth}}
  \bibnamefont{and} \bibinfo{author}{\bibfnamefont{S.~M.}
  \bibnamefont{Koushiappas}}, \bibinfo{journal}{Phys. Rev.}
  \textbf{\bibinfo{volume}{D86}}, \bibinfo{pages}{021302}
  (\bibinfo{year}{2012}), \eprint{1206.0796}.

\bibitem[{\citenamefont{Abdalla et~al.}(2016)}]{Abdalla:2016olq}
\bibinfo{author}{\bibfnamefont{H.}~\bibnamefont{Abdalla}} \bibnamefont{et~al.}
  (\bibinfo{collaboration}{HESS}), \bibinfo{journal}{Phys. Rev. Lett.}
  \textbf{\bibinfo{volume}{117}}, \bibinfo{pages}{151302}
  (\bibinfo{year}{2016}), \eprint{1609.08091}.

\bibitem[{\citenamefont{Finkbeiner et~al.}(2013)\citenamefont{Finkbeiner, Su,
  and Weniger}}]{Finkbeiner:2012ez}
\bibinfo{author}{\bibfnamefont{D.~P.} \bibnamefont{Finkbeiner}},
  \bibinfo{author}{\bibfnamefont{M.}~\bibnamefont{Su}}, \bibnamefont{and}
  \bibinfo{author}{\bibfnamefont{C.}~\bibnamefont{Weniger}},
  \bibinfo{journal}{JCAP} \textbf{\bibinfo{volume}{1301}}, \bibinfo{pages}{029}
  (\bibinfo{year}{2013}), \eprint{1209.4562}.

\bibitem[{\citenamefont{Hektor et~al.}(2013)\citenamefont{Hektor, Raidal, and
  Tempel}}]{Hektor:2012ev}
\bibinfo{author}{\bibfnamefont{A.}~\bibnamefont{Hektor}},
  \bibinfo{author}{\bibfnamefont{M.}~\bibnamefont{Raidal}}, \bibnamefont{and}
  \bibinfo{author}{\bibfnamefont{E.}~\bibnamefont{Tempel}},
  \bibinfo{journal}{Eur. Phys. J.} \textbf{\bibinfo{volume}{C73}},
  \bibinfo{pages}{2578} (\bibinfo{year}{2013}), \eprint{1209.4548}.

\bibitem[{\citenamefont{Whiteson}(2013)}]{Whiteson:2013cs}
\bibinfo{author}{\bibfnamefont{D.}~\bibnamefont{Whiteson}},
  \bibinfo{journal}{Phys. Rev.} \textbf{\bibinfo{volume}{D88}},
  \bibinfo{pages}{023530} (\bibinfo{year}{2013}), \eprint{1302.0427}.

\bibitem[{\citenamefont{Ackermann et~al.}(2015)}]{Ackermann:2015lka}
\bibinfo{author}{\bibfnamefont{M.}~\bibnamefont{Ackermann}}
  \bibnamefont{et~al.} (\bibinfo{collaboration}{Fermi-LAT}),
  \bibinfo{journal}{Phys. Rev.} \textbf{\bibinfo{volume}{D91}},
  \bibinfo{pages}{122002} (\bibinfo{year}{2015}), \eprint{1506.00013}.

\bibitem[{\citenamefont{Ackermann et~al.}(2013)}]{Ackermann:2013uma}
\bibinfo{author}{\bibfnamefont{M.}~\bibnamefont{Ackermann}}
  \bibnamefont{et~al.} (\bibinfo{collaboration}{Fermi-LAT}),
  \bibinfo{journal}{Phys. Rev.} \textbf{\bibinfo{volume}{D88}},
  \bibinfo{pages}{082002} (\bibinfo{year}{2013}), \eprint{1305.5597}.

\bibitem[{\citenamefont{Liang et~al.}(2016{\natexlab{a}})\citenamefont{Liang,
  Xia, Shen, Li, Jiang, Yuan, Fan, Feng, Liang, and Chang}}]{Liang:2016bxu}
\bibinfo{author}{\bibfnamefont{Y.-F.} \bibnamefont{Liang}},
  \bibinfo{author}{\bibfnamefont{Z.-Q.} \bibnamefont{Xia}},
  \bibinfo{author}{\bibfnamefont{Z.-Q.} \bibnamefont{Shen}},
  \bibinfo{author}{\bibfnamefont{X.}~\bibnamefont{Li}},
  \bibinfo{author}{\bibfnamefont{W.}~\bibnamefont{Jiang}},
  \bibinfo{author}{\bibfnamefont{Q.}~\bibnamefont{Yuan}},
  \bibinfo{author}{\bibfnamefont{Y.-Z.} \bibnamefont{Fan}},
  \bibinfo{author}{\bibfnamefont{L.}~\bibnamefont{Feng}},
  \bibinfo{author}{\bibfnamefont{E.-W.} \bibnamefont{Liang}}, \bibnamefont{and}
  \bibinfo{author}{\bibfnamefont{J.}~\bibnamefont{Chang}},
  \bibinfo{journal}{Phys. Rev.} \textbf{\bibinfo{volume}{D94}},
  \bibinfo{pages}{103502} (\bibinfo{year}{2016}{\natexlab{a}}),
  \eprint{1608.07184}.

\bibitem[{\citenamefont{Anderson et~al.}(2016)\citenamefont{Anderson, Zimmer,
  Conrad, Gustafsson, Sánchez-Conde, and Caputo}}]{Anderson:2015dpc}
\bibinfo{author}{\bibfnamefont{B.}~\bibnamefont{Anderson}},
  \bibinfo{author}{\bibfnamefont{S.}~\bibnamefont{Zimmer}},
  \bibinfo{author}{\bibfnamefont{J.}~\bibnamefont{Conrad}},
  \bibinfo{author}{\bibfnamefont{M.}~\bibnamefont{Gustafsson}},
  \bibinfo{author}{\bibfnamefont{M.}~\bibnamefont{Sánchez-Conde}},
  \bibnamefont{and} \bibinfo{author}{\bibfnamefont{R.}~\bibnamefont{Caputo}},
  \bibinfo{journal}{JCAP} \textbf{\bibinfo{volume}{1602}}, \bibinfo{pages}{026}
  (\bibinfo{year}{2016}), \eprint{1511.00014}.

\bibitem[{\citenamefont{Liang et~al.}(2016{\natexlab{b}})\citenamefont{Liang,
  Shen, Li, Fan, Huang, Lei, Feng, Liang, and Chang}}]{Liang:2016pvm}
\bibinfo{author}{\bibfnamefont{Y.-F.} \bibnamefont{Liang}},
  \bibinfo{author}{\bibfnamefont{Z.-Q.} \bibnamefont{Shen}},
  \bibinfo{author}{\bibfnamefont{X.}~\bibnamefont{Li}},
  \bibinfo{author}{\bibfnamefont{Y.-Z.} \bibnamefont{Fan}},
  \bibinfo{author}{\bibfnamefont{X.}~\bibnamefont{Huang}},
  \bibinfo{author}{\bibfnamefont{S.-J.} \bibnamefont{Lei}},
  \bibinfo{author}{\bibfnamefont{L.}~\bibnamefont{Feng}},
  \bibinfo{author}{\bibfnamefont{E.-W.} \bibnamefont{Liang}}, \bibnamefont{and}
  \bibinfo{author}{\bibfnamefont{J.}~\bibnamefont{Chang}},
  \bibinfo{journal}{Phys. Rev.} \textbf{\bibinfo{volume}{D93}},
  \bibinfo{pages}{103525} (\bibinfo{year}{2016}{\natexlab{b}}),
  \eprint{1602.06527}.

\bibitem[{\citenamefont{Profumo et~al.}(2016)\citenamefont{Profumo, Queiroz,
  and Yaguna}}]{Profumo:2016idl}
\bibinfo{author}{\bibfnamefont{S.}~\bibnamefont{Profumo}},
  \bibinfo{author}{\bibfnamefont{F.~S.} \bibnamefont{Queiroz}},
  \bibnamefont{and} \bibinfo{author}{\bibfnamefont{C.~E.} \bibnamefont{Yaguna}}
  (\bibinfo{year}{2016}), \eprint{1602.08501}.

\bibitem[{\citenamefont{Abramowski et~al.}(2013)}]{Abramowski:2013ax}
\bibinfo{author}{\bibfnamefont{A.}~\bibnamefont{Abramowski}}
  \bibnamefont{et~al.} (\bibinfo{collaboration}{H.E.S.S.}),
  \bibinfo{journal}{Phys. Rev. Lett.} \textbf{\bibinfo{volume}{110}},
  \bibinfo{pages}{041301} (\bibinfo{year}{2013}), \eprint{1301.1173}.

\bibitem[{\citenamefont{Slatyer}(2016)}]{Slatyer:2015jla}
\bibinfo{author}{\bibfnamefont{T.~R.} \bibnamefont{Slatyer}},
  \bibinfo{journal}{Phys. Rev.} \textbf{\bibinfo{volume}{D93}},
  \bibinfo{pages}{023527} (\bibinfo{year}{2016}), \eprint{1506.03811}.

\bibitem[{\citenamefont{Jackson et~al.}(2010)\citenamefont{Jackson, Servant,
  Shaughnessy, Tait, and Taoso}}]{Jackson:2009kg}
\bibinfo{author}{\bibfnamefont{C.~B.} \bibnamefont{Jackson}},
  \bibinfo{author}{\bibfnamefont{G.}~\bibnamefont{Servant}},
  \bibinfo{author}{\bibfnamefont{G.}~\bibnamefont{Shaughnessy}},
  \bibinfo{author}{\bibfnamefont{T.~M.~P.} \bibnamefont{Tait}},
  \bibnamefont{and} \bibinfo{author}{\bibfnamefont{M.}~\bibnamefont{Taoso}},
  \bibinfo{journal}{JCAP} \textbf{\bibinfo{volume}{1004}}, \bibinfo{pages}{004}
  (\bibinfo{year}{2010}), \eprint{0912.0004}.

\bibitem[{\citenamefont{Giacchino et~al.}(2014)\citenamefont{Giacchino,
  Lopez-Honorez, and Tytgat}}]{Giacchino:2014moa}
\bibinfo{author}{\bibfnamefont{F.}~\bibnamefont{Giacchino}},
  \bibinfo{author}{\bibfnamefont{L.}~\bibnamefont{Lopez-Honorez}},
  \bibnamefont{and} \bibinfo{author}{\bibfnamefont{M.~H.~G.}
  \bibnamefont{Tytgat}}, \bibinfo{journal}{JCAP}
  \textbf{\bibinfo{volume}{1408}}, \bibinfo{pages}{046} (\bibinfo{year}{2014}),
  \eprint{1405.6921}.

\bibitem[{\citenamefont{Bergstrom et~al.}(2005)\citenamefont{Bergstrom,
  Bringmann, Eriksson, and Gustafsson}}]{Bergstrom:2004nr}
\bibinfo{author}{\bibfnamefont{L.}~\bibnamefont{Bergstrom}},
  \bibinfo{author}{\bibfnamefont{T.}~\bibnamefont{Bringmann}},
  \bibinfo{author}{\bibfnamefont{M.}~\bibnamefont{Eriksson}}, \bibnamefont{and}
  \bibinfo{author}{\bibfnamefont{M.}~\bibnamefont{Gustafsson}},
  \bibinfo{journal}{JCAP} \textbf{\bibinfo{volume}{0504}}, \bibinfo{pages}{004}
  (\bibinfo{year}{2005}), \eprint{hep-ph/0412001}.

\bibitem[{\citenamefont{Ibarra et~al.}(2014)\citenamefont{Ibarra, Totzauer, and
  Wild}}]{Ibarra:2014vya}
\bibinfo{author}{\bibfnamefont{A.}~\bibnamefont{Ibarra}},
  \bibinfo{author}{\bibfnamefont{M.}~\bibnamefont{Totzauer}}, \bibnamefont{and}
  \bibinfo{author}{\bibfnamefont{S.}~\bibnamefont{Wild}},
  \bibinfo{journal}{JCAP} \textbf{\bibinfo{volume}{1404}}, \bibinfo{pages}{012}
  (\bibinfo{year}{2014}), \eprint{1402.4375}.

\bibitem[{\citenamefont{Bertone et~al.}(2009)\citenamefont{Bertone, Jackson,
  Shaughnessy, Tait, and Vallinotto}}]{Bertone:2009cb}
\bibinfo{author}{\bibfnamefont{G.}~\bibnamefont{Bertone}},
  \bibinfo{author}{\bibfnamefont{C.~B.} \bibnamefont{Jackson}},
  \bibinfo{author}{\bibfnamefont{G.}~\bibnamefont{Shaughnessy}},
  \bibinfo{author}{\bibfnamefont{T.~M.~P.} \bibnamefont{Tait}},
  \bibnamefont{and}
  \bibinfo{author}{\bibfnamefont{A.}~\bibnamefont{Vallinotto}},
  \bibinfo{journal}{Phys. Rev.} \textbf{\bibinfo{volume}{D80}},
  \bibinfo{pages}{023512} (\bibinfo{year}{2009}), \eprint{0904.1442}.

\bibitem[{\citenamefont{Birkedal et~al.}(2006)\citenamefont{Birkedal, Noble,
  Perelstein, and Spray}}]{Birkedal:2006fz}
\bibinfo{author}{\bibfnamefont{A.}~\bibnamefont{Birkedal}},
  \bibinfo{author}{\bibfnamefont{A.}~\bibnamefont{Noble}},
  \bibinfo{author}{\bibfnamefont{M.}~\bibnamefont{Perelstein}},
  \bibnamefont{and} \bibinfo{author}{\bibfnamefont{A.}~\bibnamefont{Spray}},
  \bibinfo{journal}{Phys. Rev.} \textbf{\bibinfo{volume}{D74}},
  \bibinfo{pages}{035002} (\bibinfo{year}{2006}), \eprint{hep-ph/0603077}.

\bibitem[{\citenamefont{Bertone et~al.}(2012)\citenamefont{Bertone, Jackson,
  Shaughnessy, Tait, and Vallinotto}}]{Bertone:2010fn}
\bibinfo{author}{\bibfnamefont{G.}~\bibnamefont{Bertone}},
  \bibinfo{author}{\bibfnamefont{C.~B.} \bibnamefont{Jackson}},
  \bibinfo{author}{\bibfnamefont{G.}~\bibnamefont{Shaughnessy}},
  \bibinfo{author}{\bibfnamefont{T.~M.~P.} \bibnamefont{Tait}},
  \bibnamefont{and}
  \bibinfo{author}{\bibfnamefont{A.}~\bibnamefont{Vallinotto}},
  \bibinfo{journal}{JCAP} \textbf{\bibinfo{volume}{1203}}, \bibinfo{pages}{020}
  (\bibinfo{year}{2012}), \eprint{1009.5107}.

\bibitem[{\citenamefont{Arina et~al.}(2014)\citenamefont{Arina, Bringmann,
  Silk, and Vollmann}}]{Arina:2014fna}
\bibinfo{author}{\bibfnamefont{C.}~\bibnamefont{Arina}},
  \bibinfo{author}{\bibfnamefont{T.}~\bibnamefont{Bringmann}},
  \bibinfo{author}{\bibfnamefont{J.}~\bibnamefont{Silk}}, \bibnamefont{and}
  \bibinfo{author}{\bibfnamefont{M.}~\bibnamefont{Vollmann}},
  \bibinfo{journal}{Phys. Rev.} \textbf{\bibinfo{volume}{D90}},
  \bibinfo{pages}{083506} (\bibinfo{year}{2014}), \eprint{1409.0007}.

\bibitem[{\citenamefont{Cerdeno et~al.}(2016)\citenamefont{Cerdeno, Peiro, and
  Robles}}]{Cerdeno:2015jca}
\bibinfo{author}{\bibfnamefont{D.~G.} \bibnamefont{Cerdeno}},
  \bibinfo{author}{\bibfnamefont{M.}~\bibnamefont{Peiro}}, \bibnamefont{and}
  \bibinfo{author}{\bibfnamefont{S.}~\bibnamefont{Robles}},
  \bibinfo{journal}{JCAP} \textbf{\bibinfo{volume}{1604}}, \bibinfo{pages}{011}
  (\bibinfo{year}{2016}), \eprint{1507.08974}.

\bibitem[{\citenamefont{Weiner and Yavin}(2012)}]{Weiner:2012cb}
\bibinfo{author}{\bibfnamefont{N.}~\bibnamefont{Weiner}} \bibnamefont{and}
  \bibinfo{author}{\bibfnamefont{I.}~\bibnamefont{Yavin}},
  \bibinfo{journal}{Phys. Rev.} \textbf{\bibinfo{volume}{D86}},
  \bibinfo{pages}{075021} (\bibinfo{year}{2012}), \eprint{1206.2910}.

\bibitem[{\citenamefont{Tulin et~al.}(2013)\citenamefont{Tulin, Yu, and
  Zurek}}]{Tulin:2012uq}
\bibinfo{author}{\bibfnamefont{S.}~\bibnamefont{Tulin}},
  \bibinfo{author}{\bibfnamefont{H.-B.} \bibnamefont{Yu}}, \bibnamefont{and}
  \bibinfo{author}{\bibfnamefont{K.~M.} \bibnamefont{Zurek}},
  \bibinfo{journal}{Phys. Rev.} \textbf{\bibinfo{volume}{D87}},
  \bibinfo{pages}{036011} (\bibinfo{year}{2013}), \eprint{1208.0009}.

\bibitem[{\citenamefont{Choi and Seto}(2012)}]{Choi:2012ap}
\bibinfo{author}{\bibfnamefont{K.-Y.} \bibnamefont{Choi}} \bibnamefont{and}
  \bibinfo{author}{\bibfnamefont{O.}~\bibnamefont{Seto}},
  \bibinfo{journal}{Phys. Rev.} \textbf{\bibinfo{volume}{D86}},
  \bibinfo{pages}{043515} (\bibinfo{year}{2012}), \bibinfo{note}{[Erratum:
  Phys. Rev.D86,089904(2012)]}, \eprint{1205.3276}.

\bibitem[{\citenamefont{Chalons and Semenov}(2011)}]{Chalons:2011ia}
\bibinfo{author}{\bibfnamefont{G.}~\bibnamefont{Chalons}} \bibnamefont{and}
  \bibinfo{author}{\bibfnamefont{A.}~\bibnamefont{Semenov}},
  \bibinfo{journal}{JHEP} \textbf{\bibinfo{volume}{12}}, \bibinfo{pages}{055}
  (\bibinfo{year}{2011}), \eprint{1110.2064}.

\bibitem[{\citenamefont{Chalons et~al.}(2013)\citenamefont{Chalons, Dolan, and
  McCabe}}]{Chalons:2012xf}
\bibinfo{author}{\bibfnamefont{G.}~\bibnamefont{Chalons}},
  \bibinfo{author}{\bibfnamefont{M.~J.} \bibnamefont{Dolan}}, \bibnamefont{and}
  \bibinfo{author}{\bibfnamefont{C.}~\bibnamefont{McCabe}},
  \bibinfo{journal}{JCAP} \textbf{\bibinfo{volume}{1302}}, \bibinfo{pages}{016}
  (\bibinfo{year}{2013}), \eprint{1211.5154}.

\bibitem[{\citenamefont{Bergstrom and Snellman}(1988)}]{Bergstrom:1988fp}
\bibinfo{author}{\bibfnamefont{L.}~\bibnamefont{Bergstrom}} \bibnamefont{and}
  \bibinfo{author}{\bibfnamefont{H.}~\bibnamefont{Snellman}},
  \bibinfo{journal}{Phys. Rev.} \textbf{\bibinfo{volume}{D37}},
  \bibinfo{pages}{3737} (\bibinfo{year}{1988}).

\bibitem[{\citenamefont{Rudaz}(1989)}]{Rudaz:1989ij}
\bibinfo{author}{\bibfnamefont{S.}~\bibnamefont{Rudaz}},
  \bibinfo{journal}{Phys. Rev.} \textbf{\bibinfo{volume}{D39}},
  \bibinfo{pages}{3549} (\bibinfo{year}{1989}).

\bibitem[{\citenamefont{Giudice and Griest}(1989)}]{Giudice:1989kc}
\bibinfo{author}{\bibfnamefont{G.~F.} \bibnamefont{Giudice}} \bibnamefont{and}
  \bibinfo{author}{\bibfnamefont{K.}~\bibnamefont{Griest}},
  \bibinfo{journal}{Phys. Rev.} \textbf{\bibinfo{volume}{D40}},
  \bibinfo{pages}{2549} (\bibinfo{year}{1989}).

\bibitem[{\citenamefont{Bergstrom}(1989{\natexlab{a}})}]{Bergstrom:1989jr}
\bibinfo{author}{\bibfnamefont{L.}~\bibnamefont{Bergstrom}},
  \bibinfo{journal}{Phys. Lett.} \textbf{\bibinfo{volume}{B225}},
  \bibinfo{pages}{372} (\bibinfo{year}{1989}{\natexlab{a}}).

\bibitem[{\citenamefont{Bergstrom}(1989{\natexlab{b}})}]{Bergstrom:1988jt}
\bibinfo{author}{\bibfnamefont{L.}~\bibnamefont{Bergstrom}},
  \bibinfo{journal}{Nucl. Phys.} \textbf{\bibinfo{volume}{B325}},
  \bibinfo{pages}{647} (\bibinfo{year}{1989}{\natexlab{b}}).

\bibitem[{\citenamefont{Bergstrom and Kaplan}(1994)}]{Bergstrom:1994mg}
\bibinfo{author}{\bibfnamefont{L.}~\bibnamefont{Bergstrom}} \bibnamefont{and}
  \bibinfo{author}{\bibfnamefont{J.}~\bibnamefont{Kaplan}},
  \bibinfo{journal}{Astropart. Phys.} \textbf{\bibinfo{volume}{2}},
  \bibinfo{pages}{261} (\bibinfo{year}{1994}), \eprint{hep-ph/9403239}.

\bibitem[{\citenamefont{Baek et~al.}(2014)\citenamefont{Baek, Ko, Okada, and
  Senaha}}]{Baek:2012ub}
\bibinfo{author}{\bibfnamefont{S.}~\bibnamefont{Baek}},
  \bibinfo{author}{\bibfnamefont{P.}~\bibnamefont{Ko}},
  \bibinfo{author}{\bibfnamefont{H.}~\bibnamefont{Okada}}, \bibnamefont{and}
  \bibinfo{author}{\bibfnamefont{E.}~\bibnamefont{Senaha}},
  \bibinfo{journal}{JHEP} \textbf{\bibinfo{volume}{09}}, \bibinfo{pages}{153}
  (\bibinfo{year}{2014}), \eprint{1209.1685}.

\bibitem[{\citenamefont{Bergstrom and Ullio}(1997)}]{Bergstrom:1997fh}
\bibinfo{author}{\bibfnamefont{L.}~\bibnamefont{Bergstrom}} \bibnamefont{and}
  \bibinfo{author}{\bibfnamefont{P.}~\bibnamefont{Ullio}},
  \bibinfo{journal}{Nucl. Phys.} \textbf{\bibinfo{volume}{B504}},
  \bibinfo{pages}{27} (\bibinfo{year}{1997}), \eprint{hep-ph/9706232}.

\bibitem[{\citenamefont{Bern et~al.}(1997)\citenamefont{Bern, Gondolo, and
  Perelstein}}]{Bern:1997ng}
\bibinfo{author}{\bibfnamefont{Z.}~\bibnamefont{Bern}},
  \bibinfo{author}{\bibfnamefont{P.}~\bibnamefont{Gondolo}}, \bibnamefont{and}
  \bibinfo{author}{\bibfnamefont{M.}~\bibnamefont{Perelstein}},
  \bibinfo{journal}{Phys. Lett.} \textbf{\bibinfo{volume}{B411}},
  \bibinfo{pages}{86} (\bibinfo{year}{1997}), \eprint{hep-ph/9706538}.

\bibitem[{\citenamefont{Ullio and Bergstrom}(1998)}]{Ullio:1997ke}
\bibinfo{author}{\bibfnamefont{P.}~\bibnamefont{Ullio}} \bibnamefont{and}
  \bibinfo{author}{\bibfnamefont{L.}~\bibnamefont{Bergstrom}},
  \bibinfo{journal}{Phys. Rev.} \textbf{\bibinfo{volume}{D57}},
  \bibinfo{pages}{1962} (\bibinfo{year}{1998}), \eprint{hep-ph/9707333}.

\bibitem[{\citenamefont{Goodman et~al.}(2011)\citenamefont{Goodman, Ibe,
  Rajaraman, Shepherd, Tait, and Yu}}]{Goodman:2010qn}
\bibinfo{author}{\bibfnamefont{J.}~\bibnamefont{Goodman}},
  \bibinfo{author}{\bibfnamefont{M.}~\bibnamefont{Ibe}},
  \bibinfo{author}{\bibfnamefont{A.}~\bibnamefont{Rajaraman}},
  \bibinfo{author}{\bibfnamefont{W.}~\bibnamefont{Shepherd}},
  \bibinfo{author}{\bibfnamefont{T.~M.~P.} \bibnamefont{Tait}},
  \bibnamefont{and} \bibinfo{author}{\bibfnamefont{H.-B.} \bibnamefont{Yu}},
  \bibinfo{journal}{Nucl. Phys.} \textbf{\bibinfo{volume}{B844}},
  \bibinfo{pages}{55} (\bibinfo{year}{2011}), \eprint{1009.0008}.

\bibitem[{\citenamefont{Abazajian et~al.}(2012)\citenamefont{Abazajian,
  Agrawal, Chacko, and Kilic}}]{Abazajian:2011tk}
\bibinfo{author}{\bibfnamefont{K.~N.} \bibnamefont{Abazajian}},
  \bibinfo{author}{\bibfnamefont{P.}~\bibnamefont{Agrawal}},
  \bibinfo{author}{\bibfnamefont{Z.}~\bibnamefont{Chacko}}, \bibnamefont{and}
  \bibinfo{author}{\bibfnamefont{C.}~\bibnamefont{Kilic}},
  \bibinfo{journal}{Phys. Rev.} \textbf{\bibinfo{volume}{D85}},
  \bibinfo{pages}{123543} (\bibinfo{year}{2012}), \eprint{1111.2835}.

\bibitem[{\citenamefont{Rajaraman et~al.}(2012)\citenamefont{Rajaraman, Tait,
  and Whiteson}}]{Rajaraman:2012db}
\bibinfo{author}{\bibfnamefont{A.}~\bibnamefont{Rajaraman}},
  \bibinfo{author}{\bibfnamefont{T.~M.~P.} \bibnamefont{Tait}},
  \bibnamefont{and} \bibinfo{author}{\bibfnamefont{D.}~\bibnamefont{Whiteson}},
  \bibinfo{journal}{JCAP} \textbf{\bibinfo{volume}{1209}}, \bibinfo{pages}{003}
  (\bibinfo{year}{2012}), \eprint{1205.4723}.

\bibitem[{\citenamefont{Dudas et~al.}(2014)\citenamefont{Dudas, Heurtier, and
  Mambrini}}]{Dudas:2014ixa}
\bibinfo{author}{\bibfnamefont{E.}~\bibnamefont{Dudas}},
  \bibinfo{author}{\bibfnamefont{L.}~\bibnamefont{Heurtier}}, \bibnamefont{and}
  \bibinfo{author}{\bibfnamefont{Y.}~\bibnamefont{Mambrini}},
  \bibinfo{journal}{Phys. Rev.} \textbf{\bibinfo{volume}{D90}},
  \bibinfo{pages}{035002} (\bibinfo{year}{2014}), \eprint{1404.1927}.

\bibitem[{\citenamefont{El~Aisati et~al.}(2014)\citenamefont{El~Aisati, Hambye,
  and Scarnà}}]{Aisati:2014nda}
\bibinfo{author}{\bibfnamefont{C.}~\bibnamefont{El~Aisati}},
  \bibinfo{author}{\bibfnamefont{T.}~\bibnamefont{Hambye}}, \bibnamefont{and}
  \bibinfo{author}{\bibfnamefont{T.}~\bibnamefont{Scarnà}},
  \bibinfo{journal}{JHEP} \textbf{\bibinfo{volume}{08}}, \bibinfo{pages}{133}
  (\bibinfo{year}{2014}), \eprint{1403.1280}.

\bibitem[{\citenamefont{Coogan et~al.}(2015)\citenamefont{Coogan, Profumo, and
  Shepherd}}]{Coogan:2015xla}
\bibinfo{author}{\bibfnamefont{A.}~\bibnamefont{Coogan}},
  \bibinfo{author}{\bibfnamefont{S.}~\bibnamefont{Profumo}}, \bibnamefont{and}
  \bibinfo{author}{\bibfnamefont{W.}~\bibnamefont{Shepherd}},
  \bibinfo{journal}{JHEP} \textbf{\bibinfo{volume}{08}}, \bibinfo{pages}{074}
  (\bibinfo{year}{2015}), \eprint{1504.05187}.

\bibitem[{\citenamefont{Duerr et~al.}(2016{\natexlab{a}})\citenamefont{Duerr,
  Fileviez~Perez, and Smirnov}}]{Duerr:2015vna}
\bibinfo{author}{\bibfnamefont{M.}~\bibnamefont{Duerr}},
  \bibinfo{author}{\bibfnamefont{P.}~\bibnamefont{Fileviez~Perez}},
  \bibnamefont{and} \bibinfo{author}{\bibfnamefont{J.}~\bibnamefont{Smirnov}},
  \bibinfo{journal}{Phys. Rev.} \textbf{\bibinfo{volume}{D93}},
  \bibinfo{pages}{023509} (\bibinfo{year}{2016}{\natexlab{a}}),
  \eprint{1508.01425}.

\bibitem[{\citenamefont{Arina et~al.}(2010)\citenamefont{Arina, Hambye, Ibarra,
  and Weniger}}]{Arina:2009uq}
\bibinfo{author}{\bibfnamefont{C.}~\bibnamefont{Arina}},
  \bibinfo{author}{\bibfnamefont{T.}~\bibnamefont{Hambye}},
  \bibinfo{author}{\bibfnamefont{A.}~\bibnamefont{Ibarra}}, \bibnamefont{and}
  \bibinfo{author}{\bibfnamefont{C.}~\bibnamefont{Weniger}},
  \bibinfo{journal}{JCAP} \textbf{\bibinfo{volume}{1003}}, \bibinfo{pages}{024}
  (\bibinfo{year}{2010}), \eprint{0912.4496}.

\bibitem[{\citenamefont{Chu et~al.}(2012)\citenamefont{Chu, Hambye, Scarna, and
  Tytgat}}]{Chu:2012qy}
\bibinfo{author}{\bibfnamefont{X.}~\bibnamefont{Chu}},
  \bibinfo{author}{\bibfnamefont{T.}~\bibnamefont{Hambye}},
  \bibinfo{author}{\bibfnamefont{T.}~\bibnamefont{Scarna}}, \bibnamefont{and}
  \bibinfo{author}{\bibfnamefont{M.~H.~G.} \bibnamefont{Tytgat}},
  \bibinfo{journal}{Phys. Rev.} \textbf{\bibinfo{volume}{D86}},
  \bibinfo{pages}{083521} (\bibinfo{year}{2012}), \eprint{1206.2279}.

\bibitem[{\citenamefont{Wang and Han}(2013)}]{Wang:2012ts}
\bibinfo{author}{\bibfnamefont{L.}~\bibnamefont{Wang}} \bibnamefont{and}
  \bibinfo{author}{\bibfnamefont{X.-F.} \bibnamefont{Han}},
  \bibinfo{journal}{Phys. Rev.} \textbf{\bibinfo{volume}{D87}},
  \bibinfo{pages}{015015} (\bibinfo{year}{2013}), \eprint{1209.0376}.

\bibitem[{\citenamefont{Bai and Shelton}(2012)}]{Bai:2012qy}
\bibinfo{author}{\bibfnamefont{Y.}~\bibnamefont{Bai}} \bibnamefont{and}
  \bibinfo{author}{\bibfnamefont{J.}~\bibnamefont{Shelton}},
  \bibinfo{journal}{JHEP} \textbf{\bibinfo{volume}{12}}, \bibinfo{pages}{056}
  (\bibinfo{year}{2012}), \eprint{1208.4100}.

\bibitem[{\citenamefont{Fichet}(2016)}]{Fichet:2016clq}
\bibinfo{author}{\bibfnamefont{S.}~\bibnamefont{Fichet}}
  (\bibinfo{year}{2016}), \eprint{1609.01762}.

\bibitem[{\citenamefont{Rajaraman et~al.}(2013)\citenamefont{Rajaraman, Tait,
  and Wijangco}}]{Rajaraman:2012fu}
\bibinfo{author}{\bibfnamefont{A.}~\bibnamefont{Rajaraman}},
  \bibinfo{author}{\bibfnamefont{T.~M.~P.} \bibnamefont{Tait}},
  \bibnamefont{and} \bibinfo{author}{\bibfnamefont{A.~M.}
  \bibnamefont{Wijangco}}, \bibinfo{journal}{Phys. Dark Univ.}
  \textbf{\bibinfo{volume}{2}}, \bibinfo{pages}{17} (\bibinfo{year}{2013}),
  \eprint{1211.7061}.

\bibitem[{\citenamefont{Jackson et~al.}(2013)\citenamefont{Jackson, Servant,
  Shaughnessy, Tait, and Taoso}}]{Jackson:2013pjq}
\bibinfo{author}{\bibfnamefont{C.~B.} \bibnamefont{Jackson}},
  \bibinfo{author}{\bibfnamefont{G.}~\bibnamefont{Servant}},
  \bibinfo{author}{\bibfnamefont{G.}~\bibnamefont{Shaughnessy}},
  \bibinfo{author}{\bibfnamefont{T.~M.~P.} \bibnamefont{Tait}},
  \bibnamefont{and} \bibinfo{author}{\bibfnamefont{M.}~\bibnamefont{Taoso}},
  \bibinfo{journal}{JCAP} \textbf{\bibinfo{volume}{1307}}, \bibinfo{pages}{021}
  (\bibinfo{year}{2013}), \eprint{1302.1802}.

\bibitem[{\citenamefont{Duerr et~al.}(2015)\citenamefont{Duerr, Fileviez~Perez,
  and Smirnov}}]{Duerr:2015wfa}
\bibinfo{author}{\bibfnamefont{M.}~\bibnamefont{Duerr}},
  \bibinfo{author}{\bibfnamefont{P.}~\bibnamefont{Fileviez~Perez}},
  \bibnamefont{and} \bibinfo{author}{\bibfnamefont{J.}~\bibnamefont{Smirnov}},
  \bibinfo{journal}{Phys. Rev.} \textbf{\bibinfo{volume}{D92}},
  \bibinfo{pages}{083521} (\bibinfo{year}{2015}), \eprint{1506.05107}.

\bibitem[{\citenamefont{Asano et~al.}(2013)\citenamefont{Asano, Bringmann,
  Sigl, and Vollmann}}]{Asano:2012zv}
\bibinfo{author}{\bibfnamefont{M.}~\bibnamefont{Asano}},
  \bibinfo{author}{\bibfnamefont{T.}~\bibnamefont{Bringmann}},
  \bibinfo{author}{\bibfnamefont{G.}~\bibnamefont{Sigl}}, \bibnamefont{and}
  \bibinfo{author}{\bibfnamefont{M.}~\bibnamefont{Vollmann}},
  \bibinfo{journal}{Phys. Rev.} \textbf{\bibinfo{volume}{D87}},
  \bibinfo{pages}{103509} (\bibinfo{year}{2013}), \eprint{1211.6739}.

\bibitem[{\citenamefont{Bonnet et~al.}(2012)\citenamefont{Bonnet, Hirsch, Ota,
  and Winter}}]{Bonnet:2012kz}
\bibinfo{author}{\bibfnamefont{F.}~\bibnamefont{Bonnet}},
  \bibinfo{author}{\bibfnamefont{M.}~\bibnamefont{Hirsch}},
  \bibinfo{author}{\bibfnamefont{T.}~\bibnamefont{Ota}}, \bibnamefont{and}
  \bibinfo{author}{\bibfnamefont{W.}~\bibnamefont{Winter}},
  \bibinfo{journal}{JHEP} \textbf{\bibinfo{volume}{07}}, \bibinfo{pages}{153}
  (\bibinfo{year}{2012}), \eprint{1204.5862}.

\bibitem[{\citenamefont{Restrepo et~al.}(2013)\citenamefont{Restrepo, Zapata,
  and Yaguna}}]{Restrepo:2013aga}
\bibinfo{author}{\bibfnamefont{D.}~\bibnamefont{Restrepo}},
  \bibinfo{author}{\bibfnamefont{O.}~\bibnamefont{Zapata}}, \bibnamefont{and}
  \bibinfo{author}{\bibfnamefont{C.~E.} \bibnamefont{Yaguna}},
  \bibinfo{journal}{JHEP} \textbf{\bibinfo{volume}{11}}, \bibinfo{pages}{011}
  (\bibinfo{year}{2013}), \eprint{1308.3655}.

\bibitem[{\citenamefont{Fujikawa}(1973)}]{Fujikawa:1973qs}
\bibinfo{author}{\bibfnamefont{K.}~\bibnamefont{Fujikawa}},
  \bibinfo{journal}{Phys. Rev.} \textbf{\bibinfo{volume}{D7}},
  \bibinfo{pages}{393} (\bibinfo{year}{1973}).

\bibitem[{\citenamefont{Ferrara et~al.}(1992)\citenamefont{Ferrara, Porrati,
  and Telegdi}}]{Ferrara:1992yc}
\bibinfo{author}{\bibfnamefont{S.}~\bibnamefont{Ferrara}},
  \bibinfo{author}{\bibfnamefont{M.}~\bibnamefont{Porrati}}, \bibnamefont{and}
  \bibinfo{author}{\bibfnamefont{V.~L.} \bibnamefont{Telegdi}},
  \bibinfo{journal}{Phys. Rev.} \textbf{\bibinfo{volume}{D46}},
  \bibinfo{pages}{3529} (\bibinfo{year}{1992}).

\bibitem[{\citenamefont{Heeck and Patra}(2015)}]{Heeck:2015qra}
\bibinfo{author}{\bibfnamefont{J.}~\bibnamefont{Heeck}} \bibnamefont{and}
  \bibinfo{author}{\bibfnamefont{S.}~\bibnamefont{Patra}},
  \bibinfo{journal}{Phys. Rev. Lett.} \textbf{\bibinfo{volume}{115}},
  \bibinfo{pages}{121804} (\bibinfo{year}{2015}), \eprint{1507.01584}.

\bibitem[{\citenamefont{Garcia-Cely and Heeck}(2015)}]{Garcia-Cely:2015quu}
\bibinfo{author}{\bibfnamefont{C.}~\bibnamefont{Garcia-Cely}} \bibnamefont{and}
  \bibinfo{author}{\bibfnamefont{J.}~\bibnamefont{Heeck}}
  (\bibinfo{year}{2015}), \bibinfo{note}{[JCAP1603,021(2016)]},
  \eprint{1512.03332}.

\bibitem[{\citenamefont{Tavares-Velasco and
  Toscano}(2002)}]{TavaresVelasco:2001vb}
\bibinfo{author}{\bibfnamefont{G.}~\bibnamefont{Tavares-Velasco}}
  \bibnamefont{and} \bibinfo{author}{\bibfnamefont{J.~J.}
  \bibnamefont{Toscano}}, \bibinfo{journal}{Phys. Rev.}
  \textbf{\bibinfo{volume}{D65}}, \bibinfo{pages}{013005}
  (\bibinfo{year}{2002}), \eprint{hep-ph/0108114}.

\bibitem[{\citenamefont{Pasukonis}(2007)}]{Pasukonis:2007fu}
\bibinfo{author}{\bibfnamefont{J.}~\bibnamefont{Pasukonis}}, Ph.D. thesis,
  \bibinfo{school}{Vilnius U.} (\bibinfo{year}{2007}), \eprint{0710.0159},
  \urlprefix\url{http://inspirehep.net/record/762568/files/arXiv:0710.0159.pdf}.

\bibitem[{\citenamefont{Kuhn et~al.}(1979)\citenamefont{Kuhn, Kaplan, and
  Safiani}}]{Kuhn:1979bb}
\bibinfo{author}{\bibfnamefont{J.~H.} \bibnamefont{Kuhn}},
  \bibinfo{author}{\bibfnamefont{J.}~\bibnamefont{Kaplan}}, \bibnamefont{and}
  \bibinfo{author}{\bibfnamefont{E.~G.~O.} \bibnamefont{Safiani}},
  \bibinfo{journal}{Nucl. Phys.} \textbf{\bibinfo{volume}{B157}},
  \bibinfo{pages}{125} (\bibinfo{year}{1979}).

\bibitem[{\citenamefont{Christensen and Duhr}(2009)}]{Christensen:2008py}
\bibinfo{author}{\bibfnamefont{N.~D.} \bibnamefont{Christensen}}
  \bibnamefont{and} \bibinfo{author}{\bibfnamefont{C.}~\bibnamefont{Duhr}},
  \bibinfo{journal}{Comput. Phys. Commun.} \textbf{\bibinfo{volume}{180}},
  \bibinfo{pages}{1614} (\bibinfo{year}{2009}), \eprint{0806.4194}.

\bibitem[{\citenamefont{Alloul et~al.}(2014)\citenamefont{Alloul, Christensen,
  Degrande, Duhr, and Fuks}}]{Alloul:2013bka}
\bibinfo{author}{\bibfnamefont{A.}~\bibnamefont{Alloul}},
  \bibinfo{author}{\bibfnamefont{N.~D.} \bibnamefont{Christensen}},
  \bibinfo{author}{\bibfnamefont{C.}~\bibnamefont{Degrande}},
  \bibinfo{author}{\bibfnamefont{C.}~\bibnamefont{Duhr}}, \bibnamefont{and}
  \bibinfo{author}{\bibfnamefont{B.}~\bibnamefont{Fuks}},
  \bibinfo{journal}{Comput. Phys. Commun.} \textbf{\bibinfo{volume}{185}},
  \bibinfo{pages}{2250} (\bibinfo{year}{2014}), \eprint{1310.1921}.

\bibitem[{\citenamefont{Hahn}(2001)}]{Hahn:2000kx}
\bibinfo{author}{\bibfnamefont{T.}~\bibnamefont{Hahn}},
  \bibinfo{journal}{Comput. Phys. Commun.} \textbf{\bibinfo{volume}{140}},
  \bibinfo{pages}{418} (\bibinfo{year}{2001}), \eprint{hep-ph/0012260}.

\bibitem[{\citenamefont{Hahn and Perez-Victoria}(1999)}]{Hahn:1998yk}
\bibinfo{author}{\bibfnamefont{T.}~\bibnamefont{Hahn}} \bibnamefont{and}
  \bibinfo{author}{\bibfnamefont{M.}~\bibnamefont{Perez-Victoria}},
  \bibinfo{journal}{Comput. Phys. Commun.} \textbf{\bibinfo{volume}{118}},
  \bibinfo{pages}{153} (\bibinfo{year}{1999}), \eprint{hep-ph/9807565}.

\bibitem[{\citenamefont{Passarino and Veltman}(1979)}]{Passarino:1978jh}
\bibinfo{author}{\bibfnamefont{G.}~\bibnamefont{Passarino}} \bibnamefont{and}
  \bibinfo{author}{\bibfnamefont{M.~J.~G.} \bibnamefont{Veltman}},
  \bibinfo{journal}{Nucl. Phys.} \textbf{\bibinfo{volume}{B160}},
  \bibinfo{pages}{151} (\bibinfo{year}{1979}).

\bibitem[{\citenamefont{Stuart}(1988)}]{Stuart:1987tt}
\bibinfo{author}{\bibfnamefont{R.~G.} \bibnamefont{Stuart}},
  \bibinfo{journal}{Comput. Phys. Commun.} \textbf{\bibinfo{volume}{48}},
  \bibinfo{pages}{367} (\bibinfo{year}{1988}).

\bibitem[{\citenamefont{Stuart and Gongora}(1990)}]{Stuart:1989de}
\bibinfo{author}{\bibfnamefont{R.~G.} \bibnamefont{Stuart}} \bibnamefont{and}
  \bibinfo{author}{\bibfnamefont{A.}~\bibnamefont{Gongora}},
  \bibinfo{journal}{Comput. Phys. Commun.} \textbf{\bibinfo{volume}{56}},
  \bibinfo{pages}{337} (\bibinfo{year}{1990}).

\bibitem[{\citenamefont{Stuart}(1995)}]{Stuart:1994xf}
\bibinfo{author}{\bibfnamefont{R.~G.} \bibnamefont{Stuart}},
  \bibinfo{journal}{Comput. Phys. Commun.} \textbf{\bibinfo{volume}{85}},
  \bibinfo{pages}{267} (\bibinfo{year}{1995}), \eprint{hep-ph/9409273}.

\bibitem[{\citenamefont{Patel}(2015)}]{Patel:2015tea}
\bibinfo{author}{\bibfnamefont{H.~H.} \bibnamefont{Patel}},
  \bibinfo{journal}{Comput. Phys. Commun.} \textbf{\bibinfo{volume}{197}},
  \bibinfo{pages}{276} (\bibinfo{year}{2015}), \eprint{1503.01469}.

\bibitem[{\citenamefont{Cirelli et~al.}(2006)\citenamefont{Cirelli, Fornengo,
  and Strumia}}]{Cirelli:2005uq}
\bibinfo{author}{\bibfnamefont{M.}~\bibnamefont{Cirelli}},
  \bibinfo{author}{\bibfnamefont{N.}~\bibnamefont{Fornengo}}, \bibnamefont{and}
  \bibinfo{author}{\bibfnamefont{A.}~\bibnamefont{Strumia}},
  \bibinfo{journal}{Nucl. Phys.} \textbf{\bibinfo{volume}{B753}},
  \bibinfo{pages}{178} (\bibinfo{year}{2006}), \eprint{hep-ph/0512090}.

\bibitem[{\citenamefont{Hisano et~al.}(2007)\citenamefont{Hisano, Matsumoto,
  Nagai, Saito, and Senami}}]{Hisano:2006nn}
\bibinfo{author}{\bibfnamefont{J.}~\bibnamefont{Hisano}},
  \bibinfo{author}{\bibfnamefont{S.}~\bibnamefont{Matsumoto}},
  \bibinfo{author}{\bibfnamefont{M.}~\bibnamefont{Nagai}},
  \bibinfo{author}{\bibfnamefont{O.}~\bibnamefont{Saito}}, \bibnamefont{and}
  \bibinfo{author}{\bibfnamefont{M.}~\bibnamefont{Senami}},
  \bibinfo{journal}{Phys. Lett.} \textbf{\bibinfo{volume}{B646}},
  \bibinfo{pages}{34} (\bibinfo{year}{2007}), \eprint{hep-ph/0610249}.

\bibitem[{\citenamefont{Garcia-Cely et~al.}(2015)\citenamefont{Garcia-Cely,
  Ibarra, Lamperstorfer, and Tytgat}}]{Garcia-Cely:2015dda}
\bibinfo{author}{\bibfnamefont{C.}~\bibnamefont{Garcia-Cely}},
  \bibinfo{author}{\bibfnamefont{A.}~\bibnamefont{Ibarra}},
  \bibinfo{author}{\bibfnamefont{A.~S.} \bibnamefont{Lamperstorfer}},
  \bibnamefont{and} \bibinfo{author}{\bibfnamefont{M.~H.~G.}
  \bibnamefont{Tytgat}}, \bibinfo{journal}{JCAP}
  \textbf{\bibinfo{volume}{1510}}, \bibinfo{pages}{058} (\bibinfo{year}{2015}),
  \eprint{1507.05536}.

\bibitem[{\citenamefont{Ma}(2006)}]{Ma:2006km}
\bibinfo{author}{\bibfnamefont{E.}~\bibnamefont{Ma}}, \bibinfo{journal}{Phys.
  Rev.} \textbf{\bibinfo{volume}{D73}}, \bibinfo{pages}{077301}
  (\bibinfo{year}{2006}), \eprint{hep-ph/0601225}.

\bibitem[{\citenamefont{Kubo et~al.}(2006)\citenamefont{Kubo, Ma, and
  Suematsu}}]{Kubo:2006yx}
\bibinfo{author}{\bibfnamefont{J.}~\bibnamefont{Kubo}},
  \bibinfo{author}{\bibfnamefont{E.}~\bibnamefont{Ma}}, \bibnamefont{and}
  \bibinfo{author}{\bibfnamefont{D.}~\bibnamefont{Suematsu}},
  \bibinfo{journal}{Phys. Lett.} \textbf{\bibinfo{volume}{B642}},
  \bibinfo{pages}{18} (\bibinfo{year}{2006}), \eprint{hep-ph/0604114}.

\bibitem[{\citenamefont{Garny et~al.}(2015)\citenamefont{Garny, Ibarra, and
  Vogl}}]{Garny:2015wea}
\bibinfo{author}{\bibfnamefont{M.}~\bibnamefont{Garny}},
  \bibinfo{author}{\bibfnamefont{A.}~\bibnamefont{Ibarra}}, \bibnamefont{and}
  \bibinfo{author}{\bibfnamefont{S.}~\bibnamefont{Vogl}},
  \bibinfo{journal}{Int. J. Mod. Phys.} \textbf{\bibinfo{volume}{D24}},
  \bibinfo{pages}{1530019} (\bibinfo{year}{2015}), \eprint{1503.01500}.

\bibitem[{\citenamefont{Silveira and Zee}(1985)}]{Silveira:1985rk}
\bibinfo{author}{\bibfnamefont{V.}~\bibnamefont{Silveira}} \bibnamefont{and}
  \bibinfo{author}{\bibfnamefont{A.}~\bibnamefont{Zee}},
  \bibinfo{journal}{Phys. Lett.} \textbf{\bibinfo{volume}{B161}},
  \bibinfo{pages}{136} (\bibinfo{year}{1985}).

\bibitem[{\citenamefont{McDonald}(1994)}]{McDonald:1993ex}
\bibinfo{author}{\bibfnamefont{J.}~\bibnamefont{McDonald}},
  \bibinfo{journal}{Phys. Rev.} \textbf{\bibinfo{volume}{D50}},
  \bibinfo{pages}{3637} (\bibinfo{year}{1994}), \eprint{hep-ph/0702143}.

\bibitem[{\citenamefont{Duerr et~al.}(2016{\natexlab{b}})\citenamefont{Duerr,
  Fileviez~Pérez, and Smirnov}}]{Duerr:2015aka}
\bibinfo{author}{\bibfnamefont{M.}~\bibnamefont{Duerr}},
  \bibinfo{author}{\bibfnamefont{P.}~\bibnamefont{Fileviez~Pérez}},
  \bibnamefont{and} \bibinfo{author}{\bibfnamefont{J.}~\bibnamefont{Smirnov}},
  \bibinfo{journal}{JHEP} \textbf{\bibinfo{volume}{06}}, \bibinfo{pages}{152}
  (\bibinfo{year}{2016}{\natexlab{b}}), \eprint{1509.04282}.

\bibitem[{\citenamefont{Gunion et~al.}(2000)\citenamefont{Gunion, Haber, Kane,
  and Dawson}}]{Gunion:1989we}
\bibinfo{author}{\bibfnamefont{J.~F.} \bibnamefont{Gunion}},
  \bibinfo{author}{\bibfnamefont{H.~E.} \bibnamefont{Haber}},
  \bibinfo{author}{\bibfnamefont{G.~L.} \bibnamefont{Kane}}, \bibnamefont{and}
  \bibinfo{author}{\bibfnamefont{S.}~\bibnamefont{Dawson}},
  \bibinfo{journal}{Front. Phys.} \textbf{\bibinfo{volume}{80}},
  \bibinfo{pages}{1} (\bibinfo{year}{2000}).

\bibitem[{\citenamefont{Gustafsson et~al.}(2007)\citenamefont{Gustafsson,
  Lundstrom, Bergstrom, and Edsjo}}]{Gustafsson:2007pc}
\bibinfo{author}{\bibfnamefont{M.}~\bibnamefont{Gustafsson}},
  \bibinfo{author}{\bibfnamefont{E.}~\bibnamefont{Lundstrom}},
  \bibinfo{author}{\bibfnamefont{L.}~\bibnamefont{Bergstrom}},
  \bibnamefont{and} \bibinfo{author}{\bibfnamefont{J.}~\bibnamefont{Edsjo}},
  \bibinfo{journal}{Phys. Rev. Lett.} \textbf{\bibinfo{volume}{99}},
  \bibinfo{pages}{041301} (\bibinfo{year}{2007}), \eprint{astro-ph/0703512}.

\bibitem[{\citenamefont{Garcia~Cely}(2014-08-06)}]{GarciaCely:2014jha}
\bibinfo{author}{\bibfnamefont{C.~A.} \bibnamefont{Garcia~Cely}}, Ph.D. thesis,
  \bibinfo{school}{Munich, Tech. U.} (\bibinfo{year}{2014-08-06}),
  \urlprefix\url{http://mediatum.ub.tum.de?id=1224968}.

\bibitem[{\citenamefont{Garcia-Cely et~al.}(2016)\citenamefont{Garcia-Cely,
  Gustafsson, and Ibarra}}]{Garcia-Cely:2015khw}
\bibinfo{author}{\bibfnamefont{C.}~\bibnamefont{Garcia-Cely}},
  \bibinfo{author}{\bibfnamefont{M.}~\bibnamefont{Gustafsson}},
  \bibnamefont{and} \bibinfo{author}{\bibfnamefont{A.}~\bibnamefont{Ibarra}},
  \bibinfo{journal}{JCAP} \textbf{\bibinfo{volume}{1602}}, \bibinfo{pages}{043}
  (\bibinfo{year}{2016}), \eprint{1512.02801}.

\bibitem[{\citenamefont{D'Eramo}(2007)}]{D'Eramo:2007ga}
\bibinfo{author}{\bibfnamefont{F.}~\bibnamefont{D'Eramo}},
  \bibinfo{journal}{Phys.Rev.} \textbf{\bibinfo{volume}{D76}},
  \bibinfo{pages}{083522} (\bibinfo{year}{2007}), \eprint{0705.4493}.

\bibitem[{\citenamefont{Abe et~al.}(2014)\citenamefont{Abe, Kitano, and
  Sato}}]{Abe:2014gua}
\bibinfo{author}{\bibfnamefont{T.}~\bibnamefont{Abe}},
  \bibinfo{author}{\bibfnamefont{R.}~\bibnamefont{Kitano}}, \bibnamefont{and}
  \bibinfo{author}{\bibfnamefont{R.}~\bibnamefont{Sato}}
  (\bibinfo{year}{2014}), \eprint{1411.1335}.

\bibitem[{\citenamefont{Calibbi et~al.}(2015)\citenamefont{Calibbi, Mariotti,
  and Tziveloglou}}]{Calibbi:2015nha}
\bibinfo{author}{\bibfnamefont{L.}~\bibnamefont{Calibbi}},
  \bibinfo{author}{\bibfnamefont{A.}~\bibnamefont{Mariotti}}, \bibnamefont{and}
  \bibinfo{author}{\bibfnamefont{P.}~\bibnamefont{Tziveloglou}},
  \bibinfo{journal}{JHEP} \textbf{\bibinfo{volume}{10}}, \bibinfo{pages}{116}
  (\bibinfo{year}{2015}), \eprint{1505.03867}.

\bibitem[{\citenamefont{Restrepo et~al.}(2015)\citenamefont{Restrepo, Rivera,
  Sánchez-Peláez, Zapata, and Tangarife}}]{Restrepo:2015ura}
\bibinfo{author}{\bibfnamefont{D.}~\bibnamefont{Restrepo}},
  \bibinfo{author}{\bibfnamefont{A.}~\bibnamefont{Rivera}},
  \bibinfo{author}{\bibfnamefont{M.}~\bibnamefont{Sánchez-Peláez}},
  \bibinfo{author}{\bibfnamefont{O.}~\bibnamefont{Zapata}}, \bibnamefont{and}
  \bibinfo{author}{\bibfnamefont{W.}~\bibnamefont{Tangarife}},
  \bibinfo{journal}{Phys. Rev.} \textbf{\bibinfo{volume}{D92}},
  \bibinfo{pages}{013005} (\bibinfo{year}{2015}), \eprint{1504.07892}.

\bibitem[{\citenamefont{Binosi and Theussl}(2004)}]{Binosi:2003yf}
\bibinfo{author}{\bibfnamefont{D.}~\bibnamefont{Binosi}} \bibnamefont{and}
  \bibinfo{author}{\bibfnamefont{L.}~\bibnamefont{Theussl}},
  \bibinfo{journal}{Comput. Phys. Commun.} \textbf{\bibinfo{volume}{161}},
  \bibinfo{pages}{76} (\bibinfo{year}{2004}), \eprint{hep-ph/0309015}.

\bibitem[{\citenamefont{Moretti}(2015)}]{Moretti:2014rka}
\bibinfo{author}{\bibfnamefont{S.}~\bibnamefont{Moretti}},
  \bibinfo{journal}{Phys. Rev.} \textbf{\bibinfo{volume}{D91}},
  \bibinfo{pages}{014012} (\bibinfo{year}{2015}), \eprint{1407.3511}.

\bibitem[{\citenamefont{Landau}(1948)}]{Landau}
\bibinfo{author}{\bibfnamefont{L.~D.} \bibnamefont{Landau}},
  \bibinfo{journal}{Dokl. Akad. Nauk Ser. Fiz.} \textbf{\bibinfo{volume}{60}},
  \bibinfo{pages}{207} (\bibinfo{year}{1948}).

\bibitem[{\citenamefont{Yang}(1950)}]{Yang:1950rg}
\bibinfo{author}{\bibfnamefont{C.-N.} \bibnamefont{Yang}},
  \bibinfo{journal}{Phys. Rev.} \textbf{\bibinfo{volume}{77}},
  \bibinfo{pages}{242} (\bibinfo{year}{1950}).

\end{thebibliography}

\end{document}